\documentclass[aps,prd,amsmath,amssymb,showpacs]{revtex4}
\usepackage{graphicx}
\usepackage{epsfig,float}
\usepackage{dcolumn}
\usepackage{bm}
\usepackage{bbm}
\usepackage{morefloats}
\usepackage{color}
\usepackage{amsmath}
\usepackage{mathrsfs}
\usepackage{slashed}
\usepackage{multirow}
\usepackage{amssymb}
\usepackage{bbding}
\usepackage{hyperref}
\usepackage{appendix}
\usepackage{multirow}
\usepackage{graphicx}
\usepackage{caption}
\usepackage{subfigure}
\usepackage{subfloat}    
\usepackage{float}
\usepackage[font=footnotesize]{caption}
\usepackage{amssymb}

\makeatletter
\newcommand*{\rom}[1]{\expandafter\@slowromancap\romannumeral #1@}
\makeatother

\newcommand{\bs}[1]{\boldsymbol #1}

\begin{document}

\title{A unified approach for the hadronic weak decays of $\Lambda$ and $\Sigma^{\pm}\to N\pi$}

\author{Ye Cao$^{1,2,4}$}\email{caoye@impcas.ac.cn} 
\author{Ming-Xiao Duan$^{1,5}$}
\author{Tao Zhong$^{1,6}$}
\author{Qiang Zhao$^{1,2,3}$}\email{ zhaoq@ihep.ac.cn} 

\affiliation{
$^1$ Institute of High Energy Physics, Chinese Academy of Sciences, Beijing 100049, China \\
$^2$ University of Chinese Academy of Sciences, Beijing 100049, China \\
$^3$ Center for High Energy Physics, Henan Academy of Sciences, Zhengzhou 450046, China\\
$^4$ Southern Center for Nuclear-Science Theory,
Institute of Modern Physics, Chinese Academy of Sciences, Huizhou 516000, China\\
$^5$Department of Physics, Guizhou Normal University, Guiyang 550025, China\\
$^6$School of Physics and Mechatronic Engineering, Guizhou Minzu University, Guiyang 550025, China}

\begin{abstract}

We provide a unified approach for the two-body hadronic weak decays of hyperons with $S=-1$, i.e. $\Lambda$ and $\Sigma^\pm$, in the framework of the non-relativistic constitute quark model (NRCQM). A combined analysis shows that the branching ratios and asymmetry parameters of the decay channels $\Lambda\to p\pi^-$ and $\Sigma^\pm\to N\pi$ can be well described in the same framework with the direct pion emission, color suppressed internal $W$ emission, and pole terms included. However, the channel $\Lambda\to n\pi^0$  indicates significant deviations from the experimental data  based on these mentioned transition mechanism. We demonstrate that the final state interactions (FSIs) via the coupled-channel rescatterings play a crucial role in $\Lambda\to n\pi^0$. Namely, the dominant decay channel of $\Lambda\to p\pi^-$ can contribute to  $\Lambda\to n\pi^0$ via the $p\pi^-\to n\pi^0$ rescatterings. This is a leading correction effect for $\Lambda\to n\pi^0$ at the one-loop level. We find that such FSIs only become the leading effects in $\Lambda\to n\pi^0$, but contribute as subleading contributions in other channels. We also demonstrate that the pole terms are indispensable in these hyperon decays. In particular, for $\Sigma^-\to n\pi^-$ and $\Sigma^+\to n\pi^+$, it shows that $\Lambda(1405)$ as the intermediate state in the pole term amplitude is necessary for reproducing the experimental data. We also find that a dynamic selection rule forbids the radial excitation state $N(1710)$ of the quark model multiplet $|70, ^28,2,0^+,1/2^+\rangle$ from contribution. To some extent, the hyperon hadronic weak decays serve as a special probe for the underlying transition mechanisms and can provide some constraints on the intermediate $J^P=1/2^\pm$ baryon resonances.

\end{abstract}

\maketitle

\section{Introduction}

The hyperon weak decays serve as a unique probe not only for testing the unitarity of the CKM matrix~\cite{Cabibbo:2003ea} and exploring CP violation (CPV) phenomena in the light flavor sector~\cite{He:2022jjc,Guo:2025bfn}, but also for revealing the non-perturative property of strong interactions in the low-energy regime~\cite{Borasoy:2003rc}.

In the light sector, there still exist puzzling issues in the hyperon radiative weak decays~\cite{Shi:2025xkp} and hadronic weak decays. For instance, the experimental observations show a significant suppression of the $|\Delta I|=3/2$ amplitude relative to the $|\Delta I|=1/2$ amplitude in the hyperon hadronic weak decays.  Namely, the branching ratios of $\Lambda\to p\pi^-$ and $n\pi^0$ indicate that the $|\Delta I|=3/2$ amplitude contributes are less than 2$\%$ in $\Lambda$ decays~\cite{Overseth:1969bxc}. Despite extensive theoretical efforts, the physical mechanism behind this suppression still lacks a satisfactory explanation. The second issue is the $S/P$-wave puzzle. Hyperon hadronic decays involve both parity-conserving ($P$-wave) and parity-violating ($S$-wave) transition processes, and the existing theories fail to provide a self-consistent description of both the $S$-wave and $P$-wave amplitudes~\cite{Jenkins:1991bt,AbdEl-Hady:1999llb}. As follows, we will first review the experimental and theoretical progresses, and then explain our motivation of this combined analysis in a unified approach. 

As the primary decay mode of hyperons, the two-body hadronic weak decay $B\to B^{\prime}\pi$ has accumulated a wealth of experimental data and attracted extensive attention from theorists. In 1950s, the Lawrence Radiation Laboratory first accomplished the measurement of the branching ratio fraction for $\Lambda\to p\pi^-$, i.e. $R_{\Lambda}\equiv\Gamma(\Lambda\to p\pi^-)/[\Gamma(\Lambda\to p\pi^-)+\Gamma(\Lambda\to n\pi^0)]=\Gamma_1/(\Gamma_1+\Gamma_2)=0.0624\pm0.030$\cite{Crawford:1959zza}. Later, the result $R_{\Lambda}=0.646\pm0.008$ from the Brookhaven National Laboratory with higher statistics showed a significant discrepancy~\cite{Baltay:1971bg}. Actually, due to the presence of neutral particles in the final states, the branching ratio of $\Lambda\to n\pi^0$ was not measured until the early 1960s, when it was determined by two independent experiments~\cite{Chretien:1963zz} and \cite{Brown:1963zz}. In Ref.~\cite{Chretien:1963zz} $R_{\Lambda}=\Gamma_2/(\Gamma_1+\Gamma_2)=0.35\pm0.05$ was reported, while Ref.~\cite{Brown:1963zz} found $R_{\Lambda}=\Gamma_2/(\Gamma_1+\Gamma_2)=0.291\pm0.034$. 

The branching ratio fraction between $p\pi^-$ and $n\pi^0$ was regarded as one key piece of evidence supporting the $|\Delta I|=1/2$ selection rule. In Ref.~\cite{Marraffino:1980dj} the branching ratio $BR(\Sigma^+\to p\pi^0)=0.5172\pm0.0036$ and the ratio of the asymmetry parameter $\alpha_+(\Sigma^+\to n\pi^+)/\alpha_0(\Sigma^+\to p\pi^0)=-0.073\pm0.021$, were reported. In Ref.~\cite{Nowak:1978au} the branching ratio fraction $R_{\Sigma^+}=\Gamma(\Sigma^+\to n\pi^+)/[\Gamma(\Sigma^+\to n\pi^+)+\Gamma(\Sigma^+\to p\pi^0)]=0.488\pm 0.008$ was obtained, which was in good agreement with the previous results~\cite{Tovee:1971ga}. The most recent analysis by BESIII provides $\alpha_-=0.757\pm0.011\pm0.008$ for  $\Lambda\to p\pi^-$~\cite{BESIII:2021ypr}, and it is  consistent with the earlier result $\alpha_-=0.750\pm0.009\pm0.004$ from BESIII~\cite{BESIII:2018cnd}. In Ref.~\cite{BESIII:2018cnd} the first measurement of $\bar{\alpha}_0=-0.692\pm0.016\pm0.006$ for the $\bar{\Lambda}\to\bar{n}\pi^0$ was also reported.  For the $\Sigma^\pm$ decays early measurements of the asymmetry parameters $\alpha_-=-0.071\pm 0.012$, $\alpha_0=-0.999\pm 0.022$ and $\alpha_+/\alpha_0=-0.062\pm0.016$  for $\Sigma^-\to n\pi^-$, $\Sigma^+\to p\pi^0$ and $\Sigma^+\to n\pi^+$, respectively, can be found in Ref.~\cite{Bangerter:1969fta}.  More recently, the BESIII collaboration has updated these quantities with more accurate values, i.e. $\alpha_0=-0.998\pm 0.037\pm 0.0009$~\cite{BESIII:2020fqg}, $\alpha_+=0.0481\pm0.0031\pm0.0019$ \cite{BESIII:2023sgt} and $\alpha_+/\alpha_0=-0.049\pm0.0032\pm0.0021$ \cite{BESIII:2023sgt}. It should be noted that benefiting from the correlated production of hyperon-anti-hyperon pairs in the BESIII experiment, it is now possible to simultaneously measure the branching ratios and asymmetry parameters for the charge-conjugated channels within the same data sample. This capability is highly conducive to probing CP violation in hyperon decays~\cite{BESIII:2021ypr,BESIII:2023sgt,BESIII:2023drj}.

In theoretical studies, while the heavy quark expansion cannot be applied to light-flavor baryon decays and factorization approaches do not perform well, phenomenological models, effective theories, and symmetry-based analyses remain valuable for understanding hyperon weak decays. In Ref. \cite{Flores-Mendieta:2019lao}, an analysis of hadronic decays of light-flavor baryons was conducted within chiral perturbation theory (ChPT), incorporating large-$N_c$ expansion and SU(3) flavor symmetry breaking effects. Quite extensive studies of the hyperon hadronic decays can be found in the literature, e.g. pole model~\cite{Nardulli:1987ns}, constituent quark model~\cite{LeYaouanc:1977ys,Wu:1985yb}, QCD sum rule~\cite{Katuya:1978ps}, Skyrme soliton model~\cite{Donoghue:1985dya}, effective quark model with chiral U(3)$\times$U(3) symmetry~\cite{Berdnikov:2007zza}, broken SU(3) symmetry approach~\cite{Zenczykowski:2005cs}, etc. 

In Ref. \cite{Ivanov:2021huf}, both short-distance and long-distance effects were considered in the study of $\Lambda\to p\pi^-$ and $n\pi^0$ decays. The short-distance effects were described by five types of topological diagrams involving the external and internal $W$ emissions, and  $W$ exchanges, while long-distance contributions arose from pole diagrams incorporating both $1/2^+$ and $1/2^-$ baryon resonances. The study revealed that short-distance contributions are significantly suppressed in $\Lambda$ decays, non-factorizable contributions must be included, and pole diagrams play an essential role in ensuring consistency between theoretical predictions and experimental data. The crucial role played by the non-factorizable contributions is also supported by studies with the topological diagram approach (TDA) and the irreducible representation amplitude (IRA) method under SU(3) flavor symmetry~\cite{Xu:2020jfr,Wu:2025hnh}. These findings collectively indicate that non-perturbative QCD play an indispensable role in the hyperon weak decay processes. A combined analysis and coherent description of both $\Lambda$ and $\Sigma^\pm$ hadronic weak decays should be necessary for a better understanding of the hyperon weak decay mechanism, and be the best place for demonstrating the key role played by non-perturbative mechanisms. 

To further clarify our motivation, we outline several key issues about the hyperon weak decays here. First, note that $\Lambda$ and $\Sigma$ belong to the same ground-state baryon octet with strangeness $S=-1$. They should share similar internal structures and decay properties. Secondly, for the $\Lambda$ decays, the direct pion emission (DPE) can contribute to $\Lambda\to p\pi^-$, but is absent in $\Lambda\to n\pi^0$. Although the color-suppressed (CS) $W$ emission can contribute to both channels, it is not necessary that the partial decay widths of these two channels to be $\Gamma(\Lambda\to p\pi^-)\simeq 2\Gamma(\Lambda\to n\pi^0)$. Interestingly, the pole terms satisfy this relation if they are dominant in the $\Lambda$ hadronic weak decays~\cite{Richard:2016hac}, or if the pole terms are comparable with the DPE transition. Thirdly, note that only the decay $\Sigma^-\to n\pi^-$ involves the DPE process in the $\Sigma^\pm$ hadronic weak decays, while the decay of $\Sigma^+\to n\pi^+$ has contributions only from the pole terms. The comparable partial decay widths between these two channels indicate that the pole terms play a significant role in the hyperon hadronic decays. The last point is that since the threshold of the final states $N\pi$ and the initial baryon $\Lambda/\Sigma^{\pm}$ is not far away, it implies that the long-distance interaction, namely, the final-state interactions may also be important. We will show that such long-distance contributions indeed provide a significant enhancement effect for $\Lambda\to n\pi^0$ mode. However, the decay of $\Sigma^+\to N\pi$ is constrained by charge conservation and cannot obtain the rescattering contributions. In brief, although some of these aspects have been addressed by existing literature, a combined and unified analysis is not available. With the improved experimental data, a systematic investigation of the hyperon hadronic weak decays is needed and will allow us to gain deeper insights into the underlying dynamics. 

As follows, we will first introduce the framework of our approach in Sec.~\ref{framework-main}. Then, the calculation results for the partial widths and asymmetry parameters and discussions will be presented in Sec.~\ref{numerical-results}. A brief summary will be given in Sec.~\ref{sec-summary}.

\section{framework}\label{framework-main}
We adopt the NRCQM to describe the baryon states in the transition amplitudes. The detailed spin, flavor and spatial wavefunctions are presented in Appendix~\ref{wf-QM}. In the low energy regime, the constituent degrees of freedom become crucial and they are manifested by the wavefunction convolutions in the transition amplitudes. Meanwhile, different transition amplitudes can be connected to each other via the SU(3) flavor symmetry. An interesting and non-negligible feature is that the SU(3) flavor symmetry should be broken and the breaking effects can be accounted for via both constituent quark mass and the wavefunction convolutions. This feature is different from parametrization approaches at hadronic level where the SU(3) flavor symmetry effects are absorbed into the coupling constants, and connections and relative phases among different transition amplitudes are actually unclear. 

In the NRCQM transition amplitudes for the DPE and CS processes can be explicitly calculated without additional parameters introduced. Meanwhile, one notices that the the pion-nucleon final-state interactions (FSIs) may also play an important role. Since the masses of the initial $\Lambda$ and $\Sigma^\pm$ are not far away from the final $\pi N$ threshold, it implies that the PV channel may receive non-negligible contributions from the final-state pion and nucleon interactions. We mention in advance that indeed the FSIs contributions are necessary for obtaining the consistent results in the combined analysis of the partial widths and symmetry parameters. 

As follows, we present the detailed formalisms for these three transition mechanisms: $W$ emission processes (i.e. DPE and CS processes), pole terms, and FSIs.

\subsection{DPE and CS processes}

\begin{figure}
    \centering
    \subfigure[ \ DPE]{ 
    \label{fig: DE}    
    \includegraphics[width=4.cm]{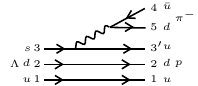}}
    \subfigure[ \ CS-1]{ 
    \label{fig: CS-1}    
    \includegraphics[width=4.cm]{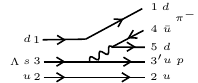}}
    \subfigure[ \ CS-1]{
    \label{fig: CS-1}    
    \includegraphics[width=4.cm]{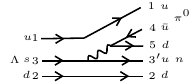}}
    \subfigure[ \ CS-2]{
    \label{fig: CS-2}    
    \includegraphics[width=4.cm]{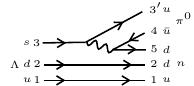}}
   \caption{Illustrations for the two-body hadronic weak decay of $\Lambda$ into $p\pi$ and $n\pi$ at the quark level. (a) Direct pion emission (DPE) processes, (b)-(d) Color suppressed (CS) pion emission processes.}
   \label{fig:Lambda DPE and CS}
  \end{figure}

\begin{figure}
    \centering
    \subfigure[ \ DPE]{ 
    \label{fig: DE}    
    \includegraphics[width=4.cm]{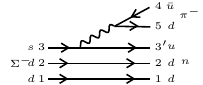}}
    \subfigure[ \ CS-1]{ 
    \label{fig: CS-1}    
    \includegraphics[width=4.cm]{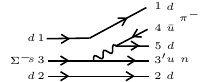}}
    \subfigure[ \ CS-1]{
    \label{fig: CS-1}    
    \includegraphics[width=4.cm]{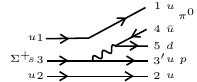}}
    \subfigure[ \ CS-2]{
    \label{fig: CS-2}    
    \includegraphics[width=4.cm]{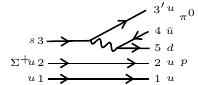}}
   \caption{Illustrations for the two-body hadronic weak decay of $\Sigma^-\to n\pi^-$ and $\Sigma^+\to p\pi^0$ at the quark level. (a) Direct pion emission (DPE) processes, (b)-(d) Color suppressed (CS) pion emission processes.}
   \label{fig:Sigma DPE and CS}
  \end{figure}

Both the DPE and CS processes are leading-order decay mechanisms. They are distinguished by whether the $W$ boson is emitted externally or internally, thereby a color suppression factor will be present in the CS amplitudes. As illustrated in Figs.~\ref{fig:Lambda DPE and CS} and \ref{fig:Sigma DPE and CS}, only the decays of $\Lambda\to p\pi^-$ and $\Sigma\to n\pi^-$ involve the DPE process. With the $s$ quark in the initial-state baryon converting to a $u$ quark by emitting a $W^-$ boson, the quark pair $d\bar{u}$ coupled to the $W^-$ will directly hadronize into the final-state meson $\pi^-$. In contrast, the CS process describes the hadronization of the final-state meson with a quark in the initial-state baryon combined with the antiquark created by the $W^-$ emission. Depending on how the antiquark from the $W$ emission is combined with one quark from the initial baryon,  the CS process can be categorized into two types: CS-1 and CS-2, as labeled in Fig. \ref{fig:Lambda DPE and CS} and \ref{fig:Sigma DPE and CS} for $\Lambda$ and $\Sigma^\pm$ decays, respectively.

\subsubsection{Operators for the $``1\to 3"$ weak transition}
It is generally assumed that the hadronic weak decay can be described by a current-current type Hamiltonian. According to the quark label adopted in Fig. \ref{fig:Lambda DPE and CS} and \ref{fig:Sigma DPE and CS}, the non-relativistic Hamiltonian for the $1\to 3$ transition process at the quark level gives,
\begin{align}
    \begin{split}
        H_{W,1\to 3}^{(\text{PC})}&=\frac{G_F}{\sqrt{2}}V_{ud}V_{us}\frac{\beta}{(2\pi)^3}\hat{\alpha}_3^{(+)}\hat{I}_{\pi}^{\prime}\delta^3(\boldsymbol{p}_3-\boldsymbol{p}_3^{\prime}-\boldsymbol{p}_4-\boldsymbol{p}_5)\\
        &\times\Bigg\{\Big[\langle s_3^{\prime}|I|s_3\rangle\langle s_5\bar{s}_4|\boldsymbol{\sigma}|0\rangle-\langle s_3^{\prime}|\boldsymbol{\sigma}|s_3\rangle\langle s_5\bar{s}_4|I|0\rangle\Big]\cdot\Big[\Big(\frac{\boldsymbol{p}_5}{2m_5}+\frac{\boldsymbol{p}_4}{2m_4}\Big)-\Big(\frac{\boldsymbol{p}_3^{\prime}}{2m_3^{\prime}}+\frac{\boldsymbol{p}_3}{2m_3}\Big)\Big]\\
        &+i\langle s_3^{\prime}|\boldsymbol{\sigma}|s_3\rangle\times\langle s_5\bar{s}_4|\boldsymbol{\sigma}|0\rangle\cdot\Big[\Big(\frac{\boldsymbol{p}_4}{2m_4}-\frac{\boldsymbol{p}_5}{2m_5}\Big)-\Big(\frac{\boldsymbol{p}_3}{2m_3}-\frac{\boldsymbol{p}_3^{\prime}}{2m_3^{\prime}}\Big)\Big]\Bigg\},\\
        H_{W,1\to 3}^{(\text{PV})}&=\frac{G_F}{\sqrt{2}}V_{ud}V_{us}\frac{\beta}{(2\pi)^3}\hat{\alpha}_3^{(+)}\hat{I}_{\pi}^{\prime}\delta^3(\boldsymbol{p}_3-\boldsymbol{p}_3^{\prime}-\boldsymbol{p}_4-\boldsymbol{p}_5)\\
        &\times\Big(-\langle s_3^{\prime}|I|s_3\rangle\langle s_5\bar{s}_4|I|0\rangle+\langle s_3^{\prime}|\boldsymbol{\sigma}|s_3\rangle\cdot\langle s_5\bar{s}_4|\boldsymbol{\sigma}|0\rangle\Big),
        \label{eq:1to3 H_W}
    \end{split}
\end{align}
where $\hat{\alpha}^{(+)}$ is the flavor-changing operator which transforms $s$ quark to $u$, and $\hat{I}_{\pi}^{\prime}$ is the isospin operator for the pion production process. It has the form of 
\begin{equation}
    \hat{I}_j^{\pi}=\left\{
    \begin{array}{ll}
        {-}b_{d,j}^{\dagger}b_{d,j} &\text{for}\ \pi^-,\\
        {-}\frac{1}{\sqrt{2}}b_{d,j}^{\dagger}b_{u,j} &\text{for}\ \pi^0,
    \end{array}
    \right.
\end{equation}
for the singly-Cabibbo-suppressed processes and will act on the $j$-th quark of the initial baryon after considering the pion flavor wavefunction. Combining the effects of flavor changing and pion production, the flavor operator for the $\Lambda\to N\pi$ and $\Sigma^{\pm}\to N\pi$ processes can be expressed as
\begin{equation}
    \hat{O}_{\text{flavor}}=\left\{
    \begin{array}{ll}
        -b_{u,3}^{\dagger}b_{s,3}, &\Lambda\to p\pi^-\ [\text{DPE}],\\
        (b_{u,3}^{\dagger}b_{s,3})(-b_{d,1}^{\dagger}b_{d,1}), &\Lambda\to p\pi^-\ [\text{CS-1}],\\
        (b_{u,3}^{\dagger}b_{s,3})(-\frac{1}{\sqrt{2}}b_{d,1}^{\dagger}b_{u,1}), &\Lambda\to n\pi^0\ \,[\text{CS-1}],\\
        (b_{u,3}^{\dagger}b_{s,3})(-\frac{1}{\sqrt{2}}b_{d,3}^{\dagger}b_{u,3}), &\Lambda\to n\pi^0\ \,[\text{CS-2}].
    \end{array}
    \right.
\end{equation}
\begin{equation}
    \hat{O}_{\text{flavor}}=\left\{
    \begin{array}{ll}
        -b_{u,3}^{\dagger}b_{s,3}, &\Sigma^-\to n\pi^-\ [\text{DPE}],\\
        (b_{u,3}^{\dagger}b_{s,3})(-b_{d,1}^{\dagger}b_{d,1}), &\Sigma^-\to n\pi^-\ [\text{CS-1}],\\
        (b_{u,3}^{\dagger}b_{s,3})(-\frac{1}{\sqrt{2}}b_{d,1}^{\dagger}b_{u,1}), &\Sigma^+\to p\pi^0\ \ \,[\text{CS-1}],\\
        (b_{u,3}^{\dagger}b_{s,3})(-\frac{1}{\sqrt{2}}b_{d,3}^{\dagger}b_{u,3}), &\Sigma^+\to p\pi^0\ \ \,[\text{CS-2}].
    \end{array}
    \right.
\end{equation}

Since the wavefunctions of baryons are completely symmetric in flavor, spin and space under interchange of any two quarks, so the trick $\langle\mathbb{B}_f|\sum_{j}H_{W,s_j\to u_j\bar{u}_4d_5}^{(\text{P})}|\mathbb{B}_i\rangle=\beta\langle\mathbb{B}_f|H_{W,s_3\to u_3\bar{u}_4d_5}^{(\text{P})}|\mathbb{B}_i\rangle$ simplifies the calculation. The $\beta$ is a symmetry factor and it takes a value of $A_3^1=3$ in the DPE and CS-2 processes, since in these two processes, the $ud$ in $\Lambda$ or the $dd$ in $\Sigma^-$ are the spectator quarks, while $A_3^2=6$ in the CS-1 processes since only one quark in $\Lambda$ or $\Sigma^{\pm}$ is the spectator quark. In Eq. (\ref{eq:1to3 H_W}), the Pauli operator $\boldsymbol{\sigma}$ and the momentum $\boldsymbol{p}_i$ are the spin operator and spatial operator acting on the spin wave function and spatial wave function, respectively.

\subsubsection{Amplitudes of the DPE and CS processes}

In the framework of the NRCQM, the transition amplitudes of DPE and CS shown in Fig.\ref{fig:Lambda DPE and CS} and \ref{fig:Sigma DPE and CS} can be expressed as
\begin{align}
    \begin{split}
        \mathcal{M}_{\text{DPE/CS}}^{J_f,J_f^z;J_i,J_i^z}&=\langle \mathbb{B}_{f}(\boldsymbol{P}_f,J_f,J_f^z)\mathbb{M}(\boldsymbol{k})|H_{W,1\to3}^{\text{(P)}}|\mathbb{B}_i(\boldsymbol{P}_i,J_i,J_i^z)\rangle\\
        &=\sum_{S_f^z,L_f^z;S_i^z,L_i^z}\langle \phi_f|\hat{\mathcal{O}}_{\text{flavor}}|\phi_i\rangle\langle \chi_{S_f,S_f^z}|\hat{\mathcal{O}}_{\text{spin}}|\chi_{S_i,S_i^z}\rangle\langle\psi_{\pi}(\boldsymbol{k})\psi_{N_fL_fL_f^z}(\boldsymbol{P}_f)|\hat{\mathcal{O}}_{\text{spatial}}(\boldsymbol{p_i})|\psi_{N_iL_iL_i^z}(\boldsymbol{P}_i)\rangle\\
        &=\sum_{S_f^z,L_f^z;S_i^z,L_i^z}\langle \phi_f|\hat{\mathcal{O}}_{\text{flavor}}|\phi_i\rangle\langle \chi_{S_f,S_f^z}|\hat{\mathcal{O}}_{\text{spin}}|\chi_{S_i,S_i^z}\rangle I_{\text{\text{DPE/CS}}}^{L_f,L_f^z;L_i,L_i^z}.
        \label{eq:Lambda/Sigma amp}
    \end{split}
\end{align}
Apart from the color factor, the different arrangements of quarks in the DPE and CS processes make a difference between these two processes via the spatial integrals. For the DPE process, the momentum conservation requires $\boldsymbol{P}_f=\boldsymbol{p}_1+\boldsymbol{p}_2+\boldsymbol{p}_3^{\prime}$ and $\boldsymbol{k}=\boldsymbol{p}_5+\boldsymbol{p}_4$, while $\boldsymbol{P}_f=\boldsymbol{p}_5+\boldsymbol{p}_2+\boldsymbol{p}_3^{\prime}$ and $\boldsymbol{k}=\boldsymbol{p}_1+\boldsymbol{p}_4$ for the CS-1 process; $\boldsymbol{P}_f=\boldsymbol{p}_1+\boldsymbol{p}_2+\boldsymbol{p}_5$ and $\boldsymbol{k}=\boldsymbol{p}_3^{\prime}+\boldsymbol{p}_5$ for the CS-2 process. They are guaranteed by the delta function in the following expressions of the spatial wavefunction convolution $I_{\text{DPE/CS}}^{L_f,L_f^z;L_i,L_i^z}$.

\begin{table}
    \centering
    \caption{The flavor matrix elements in the DPE and CS processes.}
    \label{tab:DPE and CS fla}
    \scalebox{1}{
    \begin{tabular}{cccccc}
        \hline\hline
        Decay channels                           &Processes       &$\langle\phi_{f}^{\lambda}|\hat{\mathcal{O}}_{\text{flavor}}|\phi_{i}^{\lambda}\rangle$      &$\langle\phi_{f}^{\lambda}|\hat{\mathcal{O}}_{\text{flavor}}|\phi_{i}^{\rho}\rangle$    &$\langle\phi_{f}^{\rho}|\hat{\mathcal{O}}_{\text{flavor}}|\phi_{i}^{\lambda}\rangle$    &$\langle\phi_{f}^{\rho}|\hat{\mathcal{O}}_{\text{flavor}}|\phi_{i}^{\rho}\rangle$\\
        \hline
        \multirow{2}{*}{$\Lambda\to p\pi^-$}     &DPE             &0&0&0&$-\frac{2}{\sqrt{6}}$\\
                                                 &CS-1             &0&$-\frac{1}{3\sqrt{2}}$&0&$-\frac{1}{\sqrt{6}}$\\
        \hline
        \multirow{2}{*}{$\Lambda\to n\pi^0$}     &CS-1             &0&$\frac13$&0&0\\
                                                 &CS-2             &0&0&0&$-\frac{1}{\sqrt{3}}$\\  
        \hline
        \multirow{2}{*}{$\Sigma^-\to n\pi^-$}    &DPE        &$-\frac23$&0&0&0\\
                                                 &CS-1        &$-\frac23$&0&0&0\\
        \hline
        \multirow{2}{*}{$\Sigma^+\to p\pi^0$}    &CS-1        &$-\frac{1}{3\sqrt{2}}$&0&$-\frac{1}{\sqrt{6}}$&0\\
                                                 &CS-2        &$\frac{2}{3\sqrt{2}}$&0&0&0\\
        \hline\hline 
    \end{tabular}}
\end{table}

\begin{table}
    \centering
    \caption{The spin matrix elements for the PC and PV transitions in the {DPE} process.}
    \label{tab:DPE spin}
    \scalebox{1}{
    \begin{tabular}{ccccc}
    \hline\hline
    $\mathcal{O}_{\text{spin}}^{\text{(PC)}}$     &$\langle \chi_{\frac12,-\frac12}^{\lambda}|\mathcal{O}_{\text{spin}}^{(\text{PC})}|\chi_{\frac12,-\frac12}^{\lambda}\rangle$    &$\langle \chi_{\frac12,-\frac12}^{\lambda}|\mathcal{O}_{\text{spin}}^{(\text{PC})}|\chi_{\frac12,-\frac12}^{\rho}\rangle$   &$\langle \chi_{\frac12,-\frac12}^{\rho}|\mathcal{O}_{\text{spin}}^{(\text{PC})}|\chi_{\frac12,-\frac12}^{\lambda}\rangle$ &$\langle \chi_{\frac12,-\frac12}^{\rho}|\mathcal{O}_{\text{spin}}^{(\text{PC})}|\chi_{\frac12,-\frac12}^{\rho}\rangle$\\
    \hline
    $\langle s_3^{\prime}|I|s_3\rangle\langle s_5\bar{s}_4|\bs{\sigma}|0\rangle$  &0&0&0&0\\
    $\langle s_3^{\prime}|\bs{\sigma}|s_3\rangle\langle s_5\bar{s}_4|I|0\rangle$  &$-\frac{\sqrt{2}}{3}\hat{\bs{k}}$&0&0&$\sqrt{2}\hat{\bs{k}}$\\
    $\langle s_3^{\prime}|\boldsymbol{\sigma}|s_3\rangle\times\langle s_5\bar{s}_4|\boldsymbol{\sigma}|0\rangle$   &0&0&0&0\\
    \hline
    $\mathcal{O}_{\text{spin}}^{\text{(PV)}}$     &$\langle \chi_{\frac12,-\frac12}^{\lambda}|\mathcal{O}_{\text{spin}}^{\text{(PV)}}|\chi_{\frac12,-\frac12}^{\lambda}\rangle$    &$\langle \chi_{\frac12,-\frac12}^{\lambda}|\mathcal{O}_{\text{spin}}^{(\text{PV})}|\chi_{\frac12,-\frac12}^{\rho}\rangle$   &$\langle \chi_{\frac12,-\frac12}^{\rho}|\mathcal{O}_{\text{spin}}^{(\text{PV})}|\chi_{\frac12,-\frac12}^{\lambda}\rangle$ &$\langle \chi_{\frac12,-\frac12}^{\rho}|\mathcal{O}_{\text{spin}}^{\text{(PV)}}|\chi_{\frac12,-\frac12}^{\rho}\rangle$\\
    \hline
    $\langle s_3^{\prime}|I|s_3\rangle\langle s_5\bar{s}_4|I|0\rangle$  &$-\sqrt{2}$&0&0&$-\sqrt{2}$\\
    $\langle s_3^{\prime}|\bs{\sigma}|s_3\rangle\cdot\langle s_5\bar{s}_4|\bs{\sigma}|0\rangle$  &0&0&0&0\\
    \hline\hline
    \end{tabular}}
\end{table}

\begin{table}
    \centering
    \caption{The spin matrix elements for the PC and PV transitions in the CS-1 process.}
    \label{tab:CS1 spin}
    \scalebox{1}{
    \begin{tabular}{ccccc}
    \hline\hline
    $\mathcal{O}_{\text{spin}}^{\text{(PC)}}$     &$\langle \chi_{\frac12,-\frac12}^{\lambda}|\mathcal{O}_{\text{spin}}^{\text{(PC)}}|\chi_{\frac12,-\frac12}^{\lambda}\rangle$    &$\langle \chi_{\frac12,-\frac12}^{\lambda}|\mathcal{O}_{\text{spin}}^{\text{(PC)}}|\chi_{\frac12,-\frac12}^{\rho}\rangle$   &$\langle \chi_{\frac12,-\frac12}^{\rho}|\mathcal{O}_{\text{spin}}^{\text{(PC)}}|\chi_{\frac12,-\frac12}^{\lambda}\rangle$ &$\langle \chi_{\frac12,-\frac12}^{\rho}|\mathcal{O}_{\text{spin}}^{\text{(PC)}}|\chi_{\frac12,-\frac12}^{\rho}\rangle$\\
    \hline
    $\langle s_3^{\prime}|I|s_3\rangle\langle s_5\bar{s}_4|\bs{\sigma}|0\rangle$  &$\frac{2}{3\sqrt{2}}\hat{\bs{k}}$&$-\frac{1}{\sqrt{6}}\hat{\bs{k}}$&$-\frac{1}{\sqrt{6}}\hat{\bs{k}}$&0\\
    $\langle s_3^{\prime}|\bs{\sigma}|s_3\rangle\langle s_5\bar{s}_4|I|0\rangle$  &$-\frac{1}{3\sqrt{2}}\hat{\bs{k}}$&0&0&$\frac{1}{\sqrt{2}}\hat{\bs{k}}$\\
    $\langle s_3^{\prime}|\boldsymbol{\sigma}|s_3\rangle\times\langle s_5\bar{s}_4|\boldsymbol{\sigma}|0\rangle$   &0&$\frac{2}{\sqrt{6}}i\hat{\bs{k}}$&$-\frac{2}{\sqrt{6}}i\hat{\bs{k}}$&0\\
    \hline
    $\mathcal{O}_{\text{spin}}^{\text{(PV)}}$     &$\langle \chi_{\frac12,-\frac12}^{\lambda}|\mathcal{O}_{\text{spin}}^{(\text{PV})}|\chi_{\frac12,-\frac12}^{\lambda}\rangle$    &$\langle \chi_{\frac12,-\frac12}^{\lambda}|\mathcal{O}_{\text{spin}}^{(\text{PV})}|\chi_{\frac12,-\frac12}^{\rho}\rangle$   &$\langle \chi_{\frac12,-\frac12}^{\rho}|\mathcal{O}_{\text{spin}}^{\text{(PV)}}|\chi_{\frac12,-\frac12}^{\lambda}\rangle$ &$\langle \chi_{\frac12,-\frac12}^{\rho}|\mathcal{O}_{\text{spin}}^{\text{(PV)}}|\chi_{\frac12,-\frac12}^{\rho}\rangle$\\
    \hline
    $\langle s_3^{\prime}|I|s_3\rangle\langle s_5\bar{s}_4|I|0\rangle$  &$-\frac{1}{\sqrt{2}}$&0&0&$-\frac{1}{\sqrt{2}}$\\
    $\langle s_3^{\prime}|\bs{\sigma}|s_3\rangle\cdot\langle s_5\bar{s}_4|\bs{\sigma}|0\rangle$  &$\frac{2}{\sqrt{2}}$&$\frac{3}{\sqrt{6}}$&$\frac{3}{\sqrt{6}}$&0\\
    \hline\hline
    \end{tabular}}
\end{table}

\begin{table}
    \centering
    \caption{The spin matrix elements for the PC and PV transitions in the CS-2 process.}
    \label{tab:CS2 spin}
    \scalebox{1}{
    \begin{tabular}{ccccc}
    \hline\hline
    $\mathcal{O}_{\text{spin}}^{\text{(PC)}}$     &$\langle \chi_{\frac12,-\frac12}^{\lambda}|\mathcal{O}_{\text{spin}}^{\text{(PC)}}|\chi_{\frac12,-\frac12}^{\lambda}\rangle$    &$\langle \chi_{\frac12,-\frac12}^{\lambda}|\mathcal{O}_{\text{spin}}^{\text{(PC)}}|\chi_{\frac12,-\frac12}^{\rho}\rangle$   &$\langle \chi_{\frac12,-\frac12}^{\rho}|\mathcal{O}_{\text{spin}}^{\text{(PC)}}|\chi_{\frac12,-\frac12}^{\lambda}\rangle$ &$\langle \chi_{\frac12,-\frac12}^{\rho}|\mathcal{O}_{\text{spin}}^{\text{(PC)}}|\chi_{\frac12,-\frac12}^{\rho}\rangle$\\
    \hline
    $\langle s_3^{\prime}|I|s_3\rangle\langle s_5\bar{s}_4|\bs{\sigma}|0\rangle$  &$-\frac{1}{3\sqrt{2}}\hat{\bs{k}}$&0&0&$\frac{1}{\sqrt{2}}\hat{\bs{k}}$\\
    $\langle s_3^{\prime}|\bs{\sigma}|s_3\rangle\langle s_5\bar{s}_4|I|0\rangle$  &$-\frac{1}{3\sqrt{2}}\hat{\bs{k}}$&0&0&$\frac{1}{\sqrt{2}}\hat{\bs{k}}$\\
    $\langle s_3^{\prime}|\boldsymbol{\sigma}|s_3\rangle\times\langle s_5\bar{s}_4|\boldsymbol{\sigma}|0\rangle$&$\frac{2}{3\sqrt{2}}i\hat{\bs{k}}$&0&0&$-\sqrt{2}i\hat{\bs{k}}$\\
    \hline
    $\mathcal{O}_{\text{spin}}^{\text{(PV)}}$     &$\langle \chi_{\frac12,-\frac12}^{\lambda}|\mathcal{O}_{\text{spin}}^{(\text{PV})}|\chi_{\frac12,-\frac12}^{\lambda}\rangle$    &$\langle \chi_{\frac12,-\frac12}^{\lambda}|\mathcal{O}_{\text{spin}}^{(\text{PV})}|\chi_{\frac12,-\frac12}^{\rho}\rangle$   &$\langle \chi_{\frac12,-\frac12}^{\rho}|\mathcal{O}_{\text{spin}}^{\text{(PV)}}|\chi_{\frac12,-\frac12}^{\lambda}\rangle$ &$\langle \chi_{\frac12,-\frac12}^{\rho}|\mathcal{O}_{\text{spin}}^{\text{(PV)}}|\chi_{\frac12,-\frac12}^{\rho}\rangle$\\
    \hline
    $\langle s_3^{\prime}|I|s_3\rangle\langle s_5\bar{s}_4|I|0\rangle$  &$-\frac{1}{\sqrt{2}}$&0&0&$-\frac{1}{\sqrt{2}}$\\
    $\langle s_3^{\prime}|\bs{\sigma}|s_3\rangle\langle s_5\bar{s}_4|\bs{\sigma}|0\rangle$  &$-\frac{3}{\sqrt{2}}$&0&0&$-\frac{3}{\sqrt{2}}$\\
    \hline\hline
    \end{tabular}}
\end{table}

The flavor matrix elements and spin matrix elements in Eq. (\ref{eq:Lambda/Sigma amp}) have different results for the DPE and CS processes. They are listed  separately in Tabs.~\ref{tab:DPE and CS fla}-, \ref{tab:DPE spin}, \ref{tab:CS1 spin}, and \ref{tab:CS2 spin}. 
Moreover, the spatial matrix elements can be described by the overlap integral of the initial-state wave function acting under the spatial operator with the final-state wave function. Taking the CS-1 process as an example, the expression is:
\begin{align}
    \begin{split}
        I_{\text{CS-1}}^{L_f,L_f^z;L_i,L_i^z}&=\langle \psi_{\pi}(\boldsymbol{k})\psi_{N_fL_fL_f^z}({\bf P}_f)|\hat{\mathcal{O}}_{W,1\to 3}^{\text{spatial}}(\boldsymbol{p}_i)|\psi_{N_iL_iL_i^z}({\bf P}_i)\rangle\\
        &=\int d\boldsymbol{p}_1d\boldsymbol{p}_2d\boldsymbol{p}_3d\boldsymbol{p}_3^{\prime}d\boldsymbol{p}_4d\boldsymbol{p}_5\psi_{\pi}^*(\boldsymbol{p}_1,\boldsymbol{p}_4)\delta^3(\boldsymbol{k}-\boldsymbol{p}_1-\boldsymbol{p}_4)\psi_{N_fL_fL_f^z}^*(\boldsymbol{p}_5,\boldsymbol{p}_2,\boldsymbol{p}_3^{\prime})\\&\times\delta^3({\bf P}_f-\boldsymbol{p}_5-\boldsymbol{p}_2-\boldsymbol{p}_3^{\prime})\hat{\mathcal{O}}_{W,1\to 3}^{\text{spatial}}(\boldsymbol{p}_i)\psi_{N_iL_iL_i^z}(\boldsymbol{p}_1,\boldsymbol{p}_2,\boldsymbol{p}_3)\delta^3({\bf P}_i-\boldsymbol{p}_1-\boldsymbol{p}_2-\boldsymbol{p}_3)\\&\times\delta^3(\boldsymbol{p}_3-\boldsymbol{p}_3^{\prime}-\boldsymbol{p}_4-\boldsymbol{p}_5),
    \end{split}
\end{align}
where $\hat{\mathcal{O}}_{W,1\to3}^{\text{spatial}}(\boldsymbol{p}_i)$ is a function of the quark momentum $\boldsymbol{p}_i$, such as $\boldsymbol{p}_5/(2m_5)+\boldsymbol{p}_4/(2m_4)$ or just 1 for $H_{W,1\to 3}^{(\text{PV})}$. The significant difference between the two integral functions \(I_{\text{DPE}}^{L_f,L_f^z;L_i,L_i^z}\) and \(I_{\text{CS}}^{L_f,L_f^z;L_i,L_i^z}\) is the different momentum conservation conditions. It indicates that, for the DPE and CS processes, apart from the color suppression factor of $1/3$, there are also differences in the convolution of their spatial wave functions. This is a dynamic feature with the quark model approach. The detailed expressions for the amplitudes of the DPE and CS processes in each hyperon decay channel can be found in Appendix \ref{app:tree amp}.

\subsection{Pole terms via the internal conversions}

\begin{figure}
    \centering
    \subfigure[ \ type-A]{ 
    \label{fig: A}    
    \includegraphics[width=3.4cm]{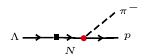}}
    \subfigure[ \ type-B]{ 
    \label{fig: B}    
    \includegraphics[width=3.4cm]{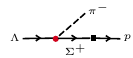}}
    \subfigure[ \ type-A]{
    \label{fig: A}    
    \includegraphics[width=3.4cm]{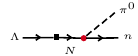}}
    \subfigure[ \ type-B]{
    \label{fig: B}    
    \includegraphics[width=3.4cm]{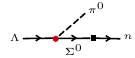}}
   \caption{Illustrations of the type-A and type-B pole terms for the initial baryon $\Lambda$ to hadronic final states $p\pi^-$ and $n\pi^0$. The intermediate states are $N$ and $\Sigma$ with quantum numbers of $\frac12^{\pm}$ and which could be off-shell (the intermediate state cannot be $\Lambda$ due to isospin violation). Black squares and red dots represent weak and strong vertices respectively.}
   \label{fig:Lambda pole terms}
  \end{figure}

\begin{figure}
    \centering
    \subfigure[ \ type-A]{ 
    \label{fig: A}    
    \includegraphics[width=3.3cm]{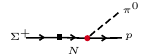}}
    \subfigure[ \ type-B]{
    \label{fig: B}    
    \includegraphics[width=3.3cm]{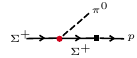}}
    \subfigure[ \ type-A]{
    \label{fig: A}    
    \includegraphics[width=3.3cm]{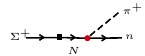}}
    \subfigure[ \ type-B]{
    \label{fig: B}    
    \includegraphics[width=3.3cm]{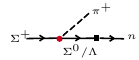}}
    \subfigure[ \ type-B]{ 
    \label{fig: B}    
    \includegraphics[width=3.2cm]{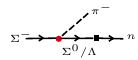}}
   \caption{Illustrations of the type-A and type-B pole terms for the initial baryon $\Sigma^\pm$ to hadronic final states $n\pi^\pm$ and $p\pi^0$. The intermediate states are $N$ and $\Sigma/\Lambda$ with quantum numbers of $\frac12^{\pm}$ and which could be off-shell. Only type-B pole terms contribute to the decay $\Sigma^-\to n\pi^-$. Black squares and red dots represent weak and strong vertices respectively.}
   \label{fig:Sigma pole terms}
  \end{figure}

The hyperon two-body hadronic weak decays also involve a typical transition processes known as the pole terms, as illustrated in Figs. \ref{fig:Lambda pole terms} and \ref{fig:Sigma pole terms}. It is noteworthy that although the schematic diagrams of the pole terms are described at the hadronic level, our calculation of the amplitudes for such processes remains within the framework of the NRCQM. Both the weak interaction operator describing internal flavor transitions and the strong interaction operator describing the pion emission are deduced at the quark level. Moreover, the initial and final hadron wavefunctions are fully symmetrized for the spin-flavor and spatial degrees of freedom. This implies that even if certain processes involve multiple diagrams at the quark level, we only need to consider one. For example, Fig. \ref{fig:Lambda pole terms} (a) corresponds to two diagrams at the quark level because the $\pi^-$ can be emitted from either $d$ quark in the neutral nucleon. However, in our symmetrized calculations, all possibilities are consistently accounted for. Therefore,  our schematic diagrams of the pole terms is described at the hadron level.

Pole terms are two-step processes with the weak transition either preceding or following the strong pion emission. We label (a) and (c) in Figs. \ref{fig:Lambda pole terms} (as well as (a) and (c) in Fig. \ref{fig:Sigma pole terms}) as type-A pole terms, while (b) and (d) (as well as (b), (d), and (e) in Fig. \ref{fig:Sigma pole terms}) are labeled as type-B pole terms. The amplitude of the pole term is expressed as
\begin{align}
    \begin{split}
        \mathcal{M}_{\text{Pole,(PC)}}^{J_f,J_f^z;J_i,J_i^z}&=\mathcal{M}_{\text{Pole,A(PC)}}^{J_f,J_f^z;J_i,J_i^z}+\mathcal{M}_{\text{Pole,B(PC)}}^{J_f,J_f^z;J_i,J_i^z},\\
        \mathcal{M}_{\text{Pole,(PV)}}^{J_f,J_f^z;J_i,J_i^z}&=\mathcal{M}_{\text{Pole,A(PV)}}^{J_f,J_f^z;J_i,J_i^z}+\mathcal{M}_{\text{Pole,B(PV)}}^{J_f,J_f^z;J_i,J_i^z},
    \end{split}
\end{align}
where 
\begin{align}
    \begin{split}
        &\mathcal{M}_{\text{Pole,A(PC)}}^{J_f,J_f^z;J_i,J_i^z}\\
        &\ \ \equiv\sum_{\mathbb{B}_m}\langle\mathbb{B}_f(\boldsymbol{P}_f;J_f,J_f^z)|H_{\pi}\frac{i}{\slashed{p}_{\mathbb{B}_m}-m_{\mathbb{B}_m}+i\frac{\Gamma_{\mathbb{B}_m}}{2}}|\mathbb{B}_m(\boldsymbol{P}_i;J_i,J_i^z)\rangle\langle \mathbb{B}_m(\boldsymbol{P}_i;J_i,J_i^z)|H_{W,2\to2}^{\text{(PC)}}|\mathbb{B}_i(\boldsymbol{P}_i;J_i,J_i^z)\rangle,\\
        &\mathcal{M}_{\text{Pole,A(PV)}}^{J_f,J_f^z;J_i,J_i^z}\\
        &\ \ \equiv\sum_{\mathbb{B}_m^{\prime}}\langle\mathbb{B}_f(\boldsymbol{P}_f;J_f,J_f^z)|H_{\pi}\frac{i}{\slashed{p}_{\mathbb{B}_m^{\prime}}-m_{\mathbb{B}_m^{\prime}}+i\frac{\Gamma_{\mathbb{B}_m^{\prime}}}{2}}|\mathbb{B}_m^{\prime}(\boldsymbol{P}_i;J_i,J_i^z)\rangle\langle \mathbb{B}_m^{\prime}(\boldsymbol{P}_i;J_i,J_i^z)|H_{W,2\to2}^{\text{(PV)}}|\mathbb{B}_i(\boldsymbol{P}_i;J_i,J_i^z)\rangle,\\
        &\mathcal{M}_{\text{Pole,B(PC)}}^{J_f,J_f^z;J_i,J_i^z}\\
        &\ \ \equiv\sum_{\mathbb{B}_m}\langle\mathbb{B}_f(\boldsymbol{P}_f;J_f,J_f^z)|H_{W,2\to2}^{\text{(PC)}}\frac{i}{\slashed{p}_{\mathbb{B}_m}-m_{\mathbb{B}_m}+i\frac{\Gamma_{\mathbb{B}_m}}{2}}|\mathbb{B}_m(\boldsymbol{P}_f;J_f,J_f^z)\rangle\langle\mathbb{B}_m(\boldsymbol{P}_f;J_f,J_f^z)|H_{\pi}|\mathbb{B}_i(\boldsymbol{P}_i;J_i,J_i^z)\rangle,\\
        &\mathcal{M}_{\text{Pole,B(PV)}}^{J_f,J_f^z;J_i,J_i^z}\\
        &\ \ \equiv\sum_{\mathbb{B}_m^{\prime}}\langle\mathbb{B}_f(\boldsymbol{P}_f;J_f,J_f^z)|H_{W,2\to2}^{\text{(PV)}}\frac{i}{\slashed{p}_{\mathbb{B}_m^{\prime}}-m_{\mathbb{B}_m^{\prime}}+i\frac{\Gamma_{\mathbb{B}_m^{\prime}}}{2}}|\mathbb{B}_m^{\prime}(\boldsymbol{P}_f;J_f,J_f^z)\rangle\langle\mathbb{B}_m^{\prime}(\boldsymbol{P}_f;J_f,J_f^z)|H_{\pi}|\mathbb{B}_i(\boldsymbol{P}_i;J_i,J_i^z)\rangle,
    \end{split}
\end{align}
in which $|\mathbb{B}_m(\boldsymbol{P}_{i/f};J_{i/f},J_{i/f}^z)\rangle$ and $|\mathbb{B}_m^{\prime}(\boldsymbol{P}_{i/f};J_{i/f},J_{i/f}^z)\rangle$ denote the intermediate baryon states of $J^P=1/2^+$ and $1/2^-$, respectively. In principle, all possible intermediate baryons should be included as the intermediate pole contributions for both PC and PV processes. However, due to the excessive suppression effects from the propagator, the primary contributions will come from those low-lying states. Therefore, in this study, we only consider the ground states and the first orbital/radial excited states.

For the intermediate baryons, the non-relativistic form for their propagators is applied:
\begin{align}
    \begin{split}
        \frac{1}{\slashed{p}-m+i\Gamma/2}\approx\frac{2m}{p^2-m^2+i m\Gamma }.
    \end{split}
\end{align}
It should be cautioned that this treatment will bring uncertainties into the theoretical results since the intermediate states are generally off-shell. However, such uncertainties can be absorbed into the quark model parameters for which the range of the favored values by experimental data can be estimated. We list the values of the masses and widths of the intermediate baryon adopted in this calculation in Tab. \ref{tab:intermediate states}.

\begin{table}
    \centering
    \caption{The masses and widths taken of baryons adopted in this work.} 
    \begin{tabular}{cccccccccccccccccc}
        \hline\hline
        Particles       &$p$                    &$n$                     &$N(1440)$                &$N(1710)$                 &$N(1535)$             &$N(1650)$               &-                         &$\Lambda$              &$\Lambda(1600)$           &$\Lambda(1810)$\\
        \hline
        $I(J^P)$        &$\frac12(\frac12^+)$   &$\frac12(\frac12^+)$    &$\frac12(\frac12^+)$     &$\frac12(\frac12^+)$      &$\frac12(\frac12^-)$  &$\frac12(\frac12^-)$    &-                         &$\frac12(\frac12^+)$   &$\frac12(\frac12^+)$      &$\frac12(\frac12^+)$\\
        Mass(GeV)       &0.938                  &0.94                   &1.44                    &1.71                     &1.54                 &1.665                   &-                         &1.116                  &1.57                       &1.79\\
        Width(GeV)      &-                      &-                       &0.35                     &0.14                      &0.12                  &0.1                   &-                         &$2.50\times10^{-15}$   &0.2                       &0.11\\               
        \hline
        Particles       &$\Sigma^+$             &$\Sigma^0$              &$\Sigma^-$               &$\Sigma(1660)$            &$\Sigma(1880)$        &$\Sigma(1620)$          &$\Sigma(1750)$            &$\Lambda(1670)$        &$\Lambda(1800)$           &$\Lambda(1405)$\\
        \hline
        $I(J^P)$        &$1(\frac12^+)$         &$1(\frac12^+)$          &$1(\frac12^+)$           &$1(\frac12^+)$            &$1(\frac12^+)$        &$1(\frac12^-)$          &$1(\frac12^-)$            &$\frac12(\frac12^-)$   &$\frac12(\frac12^-)$      &$\frac12(\frac12^-)$\\    
        Mass(GeV)       &1.189                  &1.193                   &1.2                     &1.68                     &1.82                 &1.62                   &1.8                     &1.674                  &1.8                       &1.405\\
        Width(GeV)      &$8.21\times10^{-15}$   &$8.89\times10^{-6}$     &$4.45\times10^{-15}$     &0.2                      &0.3                  &0.04                    &0.05                      &0.03                   &0.15                       &0.05\\
        \hline\hline
    \end{tabular}
    \label{tab:intermediate states}
\end{table}

\subsubsection{Operators of the weak $``2\to2"$ internal conversions}

The non-relativistic Hamiltonian describing the $``2\to2"$ internal conversion (IC) process at the quark level is:
\begin{align}
    \begin{split}
        H_{W,2\to 2}^{(\text{PC})}&=\frac{G_F}{\sqrt{2}}V_{q_1q_1^{\prime}}V_{q_2q_2^{\prime}}\frac{1}{(2\pi)^3}\sum_{i\neq j}\hat{\alpha}_i^{(-)}\hat{\beta}_j^{(+)}\delta^3(\boldsymbol{p}_i^{\prime}+\boldsymbol{p}_j^{\prime}-\boldsymbol{p}_i-\boldsymbol{p}_j)\\
        &\times\Big(\langle{s_i}^{\prime}|I|{s_i}\rangle\langle{s_j^{\prime}}|I|{s_j}\rangle-\langle{s_i}^{\prime}|\boldsymbol{\sigma}_i|{s_i}\rangle\cdot\langle{s_j^{\prime}}|\boldsymbol{\sigma}_j|{s_j}\rangle\Big),\\
        H_{W,2\to 2}^{(\text{PV})}&=\frac{G_F}{\sqrt{2}}V_{q_1q_1^{\prime}}V_{q_2q_2^{\prime}}\frac{1}{(2\pi)^3}\sum_{i\neq j}\hat{\alpha}_i^{(-)}\hat{\beta}_j^{(+)}\delta^3(\boldsymbol{p}_i^{\prime}+\boldsymbol{p}_j^{\prime}-\boldsymbol{p}_i-\boldsymbol{p}_j)\\
        &\times \Bigg\{\Big(\langle{s_i}^{\prime}|I|{s_i}\rangle\langle{s_j^{\prime}}|\boldsymbol{\sigma}_j|{s_j}\rangle-\langle{s_i}^{\prime}|\boldsymbol{\sigma}_i|{s_i}\rangle\langle{s_j^{\prime}}|I|{s_j}\rangle\Big)\cdot\Big[\Big(\frac{\boldsymbol{p}_i}{2m_i}-\frac{\boldsymbol{p}_j}{2m_j}\Big)+\Big(\frac{\boldsymbol{p}_i^{\prime}}{2m_i^{\prime}}-\frac{\boldsymbol{p}_j^{\prime}}{2m_j^{\prime}}\Big)\Big]\\
        &+i\Big(\langle{s_i}^{\prime}|\boldsymbol{\sigma}_i|{s_i}\rangle\times\langle{s_j^{\prime}}|\boldsymbol{\sigma}_j|{s_j}\rangle\Big)\cdot\Big[\Big(\frac{\boldsymbol{p}_i}{2m_i}-\frac{\boldsymbol{p}_j}{2m_j}\Big)-\Big(\frac{\boldsymbol{p}_i^{\prime}}{2m_i^{\prime}}-\frac{\boldsymbol{p}_j^{\prime}}{2m_j^{\prime}}\Big)\Big]
        \Bigg\},
        \label{eq:2to2 H_W}
    \end{split}
\end{align}
where the subscripts $i$ and $j$ ($i,j=1,2,3$ and $i\neq j$) indicate the quarks experiencing the weak interaction; $\hat{\alpha}_i^{(-)}$ and $\hat{\beta}_j$ are the are the flavor-changing operators,namely, $\alpha_i^{(-)}u_j=\delta_{ij}d_i$, $\hat{\beta}_j^{(+)}s_i=\delta_{ij}u_j$. In the calculation, we can fix the subscripts $i$ and $j$ to be 2 and 3 due to the symmetry of the total wavefunction, then the waek transition matrix element for the IC process can be simplified by multiplying a symmetry factor $\beta=A_3^2=6$, that is, $\langle\mathbb{B}_f|\sum_{i\neq j}H_{W,u_is_j\to d_iu_j}^{(\text{P})}|\mathbb{B}_i\rangle=\beta\langle\mathbb{B}_f|H_{W,u_2s_3\to d_2u_3}^{(\text{P})}|\mathbb{B}_i\rangle$. The flavor and spin matrix elements under this quark label are summarized in Tab. \ref{tab:IC fla} and \ref{tab:IC spin}. Additionally, the convolution of the spatial wave functions is expressed as:
\begin{align}
    \begin{split}
        I_{\text{IC}}^{L_f,L_f^z;L_i,L_i^z}&=\langle \psi_{N_fL_fL_f^z}({\bf P}_f)|\hat{\mathcal{O}}_{W,2\to 2}^{\text{spatial}}({\bf p}_i)|\psi_{N_iL_iL_i^z}(\bf{P}_i)\rangle\\
        &=\int d\boldsymbol{p}_1d\boldsymbol{p}_2d\boldsymbol{p}_3\boldsymbol{p}_2^{\prime}\boldsymbol{p}_3^{\prime}\psi_{N_fL_fL_f^z}^{*}(\boldsymbol{p}_1,\boldsymbol{p}_2^{\prime},\boldsymbol{p}_3^{\prime})\delta^3({\bf P}_f-\boldsymbol{p}_1-\boldsymbol{p}_2^{\prime}-\boldsymbol{p}_3^{\prime})\\
        &\times\psi_{N_iL_iL_i^z}(\boldsymbol{p}_1,\boldsymbol{p}_2,\boldsymbol{p}_3)\delta^3(\boldsymbol{P}_i-\boldsymbol{p}_1-\boldsymbol{p}_2-\boldsymbol{p}_3)\delta^3(\boldsymbol{p}_2^{\prime}+\boldsymbol{p}_3^{\prime}-\boldsymbol{p}_2-\boldsymbol{p}_3),\\
    \end{split}
\end{align}
where $\hat{\mathcal{O}}_{W,2\to2}^{\text{spatial}}(\boldsymbol{p}_i)$ is the functions of quark momentum $\boldsymbol{p}_i$, such as $\boldsymbol{p}_2/(2m_2)-\boldsymbol{p}_3/(2m_3)$ or just 1 for $H_{W,2\to 2}^{(\text{PC})}$. 

From Eq. (\ref{eq:2to2 H_W}), we can see that the flavor-spin operator in the Hamiltonian describing the PC process is decoupled from the momentum part (the momentum operator is $I$). Particularly for the IC process of ground-state baryons $\mathbb{B}_i(\frac12^+)\to\mathbb{B}_f(\frac12^+)$, we can directly extract the spin-flavor matrix element and define it as $C_{(\mathbb{B}_i\to\mathbb{B}_f)}^{f-s(\text{PC})}\equiv\langle\mathbb{B}_f|\hat{\mathcal{O}}_{f-s}^{\text{(PC)}}|\mathbb{B}_i\rangle$, where $\hat{\mathcal{O}}_{f-s}^{(\text{PC})}=\hat{\alpha}_2^{(-)}\hat{\beta}_3^{(+)}(1-\boldsymbol{\sigma}_2\cdot\boldsymbol{\sigma_3})$. The numerical values of the spin-flavor matrix elements for different IC processes are listed in Tab. \ref{tab:C_f-s}. However, in the Hamiltonian describing the PV process, the spin is coupled to the momentum ($\boldsymbol{\sigma}\cdot\boldsymbol{p}$), so the flavor-spin part cannot be separated from the momentum spatial convolution, and we cannot express the flavor-spin matrix element in a unified form.

\begin{table}
    \centering
    \caption{{The flavor matrix elements in the IC processes. The subscripts in $\Lambda_8$ and $\Lambda_1$ represent the octet and singlet states, respectively.}}
    \label{tab:IC fla}
    \scalebox{1}{
    \begin{tabular}{ccccccc}
        \hline\hline
        Processes         &$\langle\phi_f^{\lambda}|\hat{\alpha}_2^{(-)}\hat{\beta}_3^{(+)}|\phi_i^{\lambda}\rangle$      &$\langle\phi_f^{\lambda}|\hat{\alpha}_2^{(-)}\hat{\beta}_3^{(+)}|\phi_i^{\rho}\rangle$    &$\langle\phi_f^{\rho}|\hat{\alpha}_2^{(-)}\hat{\beta}_3^{(+)}|\phi_i^{\lambda}\rangle$    &$\langle\phi_f^{\rho}|\hat{\alpha}_2^{(-)}\hat{\beta}_3^{(+)}|\phi_i^{\rho}\rangle$  &$\langle\phi_f^{\lambda}|\hat{\alpha}_2^{(-)}\hat{\beta}_3^{(+)}|\phi_i^{a}\rangle$    &$\langle\phi_f^{\rho}|\hat{\alpha}_2^{(-)}\hat{\beta}_3^{(+)}|\phi_i^{a}\rangle$\\
        \hline
        $\Lambda_8\to n$  &$0$&$\frac{2}{3\sqrt{2}}$&$0$&$0$&-&-\\
        $\Sigma^+\to p$   &$\frac13$&$0$&$-\frac{1}{\sqrt{3}}$&$0$&-&-\\
        $\Sigma^0\to n$   &$\frac{2}{3\sqrt{2}}$&$0$&$0$&$0$&-&-\\
        $\Lambda_1\to n$  &-&-&-&-&$\frac13$&$0$\\
        \hline\hline        
    \end{tabular}}
\end{table}

\begin{table}
    \centering
    \caption{Spin matrix elements for the PC and PV transitions in the IC process.}
    \label{tab:IC spin}
    \scalebox{1}{
    \begin{tabular}{ccccc}
        \hline\hline
        $\mathcal{O}_{\rm spin}^{(\rm PC)}$ & $\langle\chi_{\frac{1}{2},-\frac{1}{2}}^\lambda | \mathcal{O}_{\rm spin}^{(\rm PC)} | \chi_{\frac{1}{2},-\frac{1}{2}}^\lambda\rangle$ & $\langle\chi_{\frac{1}{2},-\frac{1}{2}}^\lambda | \mathcal{O}_{\rm spin}^{(\rm PC)} | \chi_{\frac{1}{2},-\frac{1}{2}}^\rho\rangle$ & $\langle\chi_{\frac{1}{2},-\frac{1}{2}}^\rho | \mathcal{O}_{\rm spin}^{(\rm PC)} | \chi_{\frac{1}{2},-\frac{1}{2}}^\lambda\rangle$ & $\langle\chi_{\frac{1}{2},-\frac{1}{2}}^\rho | \mathcal{O}_{\rm spin}^{(\rm PC)} | \chi_{\frac{1}{2},-\frac{1}{2}}^\rho\rangle$ \\
        \hline
        $\langle s_2^\prime | I | s_2 \rangle \langle s_3^\prime | I | s_3 \rangle$ & $1$ & $0$ & $0$ & $1$ \\
        $\langle s_2^\prime | {\bs\sigma} | s_2 \rangle \cdot \langle s_3^\prime | {\bs\sigma} | s_3 \rangle$ & $-2$ & $\sqrt3$ & $\sqrt3$ & $0$ \\
        \hline
        $I - {\bs\sigma}_2 \cdot {\bs\sigma}_3$ & $3$ & $-\sqrt3$ & $-\sqrt3$ & $1$ \\
        \hline
        $\mathcal{O}_{\rm spin}^{(\rm PV)}$ & $\langle\chi_{\frac{1}{2},-\frac{1}{2}}^\lambda | \mathcal{O}_{\rm spin}^{(\rm PV)} | \chi_{\frac{1}{2},-\frac{1}{2}}^\lambda\rangle$ & $\langle\chi_{\frac{1}{2},-\frac{1}{2}}^\lambda | \mathcal{O}_{\rm spin}^{(\rm PV)} | \chi_{\frac{1}{2},-\frac{1}{2}}^\rho\rangle$ & $\langle\chi_{\frac{1}{2},-\frac{1}{2}}^\rho | \mathcal{O}_{\rm spin}^{(\rm PV)} | \chi_{\frac{1}{2},-\frac{1}{2}}^\lambda\rangle$ & $\langle\chi_{\frac{1}{2},-\frac{1}{2}}^\rho | \mathcal{O}_{\rm spin}^{(\rm PV)} | \chi_{\frac{1}{2},-\frac{1}{2}}^\rho\rangle$ \\
        \hline
        $\langle s_2^\prime | I | s_2 \rangle \langle s_3^\prime | {\bs\sigma} | s_3 \rangle$ & $\frac{1}{3} \hat{\bs k}$ & ${\bs 0}$ & ${\bs 0}$ & $-\hat{\bs k}$ \\
        $\langle s_2^\prime | {\bs\sigma} | s_2 \rangle \langle s_3^\prime | I | s_3 \rangle$ & $-\frac{2}{3} \hat{\bs k}$ & $-\frac{1}{\sqrt3} \hat{\bs k}$ & $-\frac{1}{\sqrt3} \hat{\bs k}$ & ${\bs 0}$ \\
        $\langle s_2^\prime | {\bs\sigma} | s_2 \rangle \times \langle s_3^\prime | {\bs\sigma} | s_3 \rangle$ & ${\bs 0}$ & $-\frac{2i}{\sqrt3} \hat{\bs k}$ & $\frac{2i}{\sqrt3} \hat{\bs k}$ & ${\bs 0}$ \\
        \hline
        $I_2 {\bs\sigma}_3 - {\bs\sigma}_2 I_3$ & $\hat{\bs k}$ & $\frac{1}{\sqrt3} \hat{\bs k}$ & $\frac{1}{\sqrt3} \hat{\bs k}$ & $-\hat{\bs k}$ \\
        $i {\bs\sigma}_2 \times {\bs\sigma}_3$ & ${\bs 0}$ & $\frac{2}{\sqrt3} \hat{\bs k}$ & $-\frac{2}{\sqrt3} \hat{\bs k}$ & ${\bs 0}$ \\
        \hline
        $\mathcal{O}_{\rm spin}^{(\rm PV)}$ & $\langle\chi_{\frac{1}{2},\frac{1}{2}}^\lambda | \mathcal{O}_{\rm spin}^{(\rm PV)} | \chi_{\frac{1}{2},-\frac{1}{2}}^\lambda\rangle$ & $\langle\chi_{\frac{1}{2},\frac{1}{2}}^\lambda | \mathcal{O}_{\rm spin}^{(\rm PV)} | \chi_{\frac{1}{2},-\frac{1}{2}}^\rho\rangle$ & $\langle\chi_{\frac{1}{2},\frac{1}{2}}^\rho | \mathcal{O}_{\rm spin}^{(\rm PV)} | \chi_{\frac{1}{2},-\frac{1}{2}}^\lambda\rangle$ & $\langle\chi_{\frac{1}{2},\frac{1}{2}}^\rho | \mathcal{O}_{\rm spin}^{(\rm PV)} | \chi_{\frac{1}{2},-\frac{1}{2}}^\rho\rangle$ \\
        \hline
        $\langle s_2^\prime | I | s_2 \rangle \langle s_3^\prime | {\bs\sigma} | s_3 \rangle$ & $-\frac{1}{3}(\hat{\bs i} - i\hat{\bs j})$ & ${\bs 0}$ & ${\bs 0}$ & $\hat{\bs i} - i\hat{\bs j}$ \\
        $\langle s_2^\prime | {\bs\sigma} | s_2 \rangle \langle s_3^\prime | I | s_3 \rangle$ & $\frac{2}{3}(\hat{\bs i} - i\hat{\bs j})$ & $\frac{1}{\sqrt3}(\hat{\bs i} - i\hat{\bs j})$ & $\frac{1}{\sqrt3}(\hat{\bs i} - i\hat{\bs j})$ & ${\bs 0}$ \\
        $\langle s_2^\prime | {\bs\sigma} | s_2 \rangle \times \langle s_3^\prime | {\bs\sigma} | s_3 \rangle$ & ${\bs 0}$ & $\frac{2}{\sqrt3}(i\hat{\bs i} + \hat{\bs j})$ & $-\frac{2}{\sqrt3}(i\hat{\bs i} + \hat{\bs j})$ & ${\bs 0}$ \\
        \hline
        $I_2 {\bs\sigma}_3 - {\bs\sigma}_2 I_3$ & $-(\hat{\bs i} - i\hat{\bs j})$ & $-\frac{1}{\sqrt3}(\hat{\bs i} - i\hat{\bs j})$ & $-\frac{1}{\sqrt3}(\hat{\bs i} - i\hat{\bs j})$ & $\hat{\bs i} - i\hat{\bs j}$ \\
        $i {\bs\sigma}_2 \times {\bs\sigma}_3$ & ${\bs 0}$ & $-\frac{2}{\sqrt3}(\hat{\bs i} - i\hat{\bs j})$ & $\frac{2}{\sqrt3}(\hat{\bs i} - i\hat{\bs j})$ & ${\bs 0}$ \\
        \hline
        $\mathcal{O}_{\rm spin}^{(\rm PV)}$ & $\langle\chi_{\frac{1}{2},-\frac{1}{2}}^\lambda | \mathcal{O}_{\rm spin}^{(\rm PV)} | \chi_{\frac{1}{2},\frac{1}{2}}^\lambda\rangle$ & $\langle\chi_{\frac{1}{2},-\frac{1}{2}}^\lambda | \mathcal{O}_{\rm spin}^{(\rm PV)} | \chi_{\frac{1}{2},\frac{1}{2}}^\rho\rangle$ & $\langle\chi_{\frac{1}{2},-\frac{1}{2}}^\rho | \mathcal{O}_{\rm spin}^{(\rm PV)} | \chi_{\frac{1}{2},\frac{1}{2}}^\lambda\rangle$ & $\langle\chi_{\frac{1}{2},-\frac{1}{2}}^\rho | \mathcal{O}_{\rm spin}^{(\rm PV)} | \chi_{\frac{1}{2},\frac{1}{2}}^\rho\rangle$ \\
        \hline
        $\langle s_2^\prime | I | s_2 \rangle \langle s_3^\prime | {\bs\sigma} | s_3 \rangle$ & $-\frac{1}{3}(\hat{\bs i} + i\hat{\bs j})$ & ${\bs 0}$ & ${\bs 0}$ & $\hat{\bs i} + i\hat{\bs j}$ \\
        $\langle s_2^\prime | {\bs\sigma} | s_2 \rangle \langle s_3^\prime | I | s_3 \rangle$ & $\frac{2}{3}(\hat{\bs i} + i\hat{\bs j})$ & $\frac{1}{\sqrt3}(\hat{\bs i} + i\hat{\bs j})$ & $\frac{1}{\sqrt3}(\hat{\bs i} + i\hat{\bs j})$ & ${\bs 0}$ \\
        $\langle s_2^\prime | {\bs\sigma} | s_2 \rangle \times \langle s_3^\prime | {\bs\sigma} | s_3 \rangle$ & ${\bs 0}$ & $-\frac{2}{\sqrt3}(-i\hat{\bs i} + \hat{\bs j})$ & $\frac{2}{\sqrt3}(-i\hat{\bs i} + \hat{\bs j})$ & ${\bs 0}$ \\
        \hline
        $I_2 {\bs\sigma}_3 - {\bs\sigma}_2 I_3$ & $-(\hat{\bs i} + i\hat{\bs j})$ & $-\frac{1}{\sqrt3}(\hat{\bs i} + i\hat{\bs j})$ & $-\frac{1}{\sqrt3}(\hat{\bs i} + i\hat{\bs j})$ & $\hat{\bs i} + i\hat{\bs j}$ \\
        $i {\bs\sigma}_2 \times {\bs\sigma}_3$ & ${\bs 0}$ & $-\frac{2}{\sqrt3}(\hat{\bs i} + i\hat{\bs j})$ & $\frac{2}{\sqrt3}(\hat{\bs i} + i\hat{\bs j})$ & ${\bs 0}$ \\
        \hline
        $\mathcal{O}_{\rm spin}^{(\rm PV)}$ & $\langle\chi_{\frac{3}{2},\frac{1}{2}}^s | \mathcal{O}_{\rm spin}^{(\rm PV)} | \chi_{\frac{1}{2},-\frac{1}{2}}^\lambda\rangle$ & $\langle\chi_{\frac{3}{2},\frac{1}{2}}^s | \mathcal{O}_{\rm spin}^{(\rm PV)} | \chi_{\frac{1}{2},-\frac{1}{2}}^\rho\rangle$ & $\langle\chi_{\frac{3}{2},-\frac{1}{2}}^s | \mathcal{O}_{\rm spin}^{(\rm PV)} | \chi_{\frac{1}{2},-\frac{1}{2}}^\lambda\rangle$ & $\langle\chi_{\frac{3}{2},-\frac{1}{2}}^s | \mathcal{O}_{\rm spin}^{(\rm PV)} | \chi_{\frac{1}{2},-\frac{1}{2}}^\rho\rangle$ \\
        \hline
        $\langle s_2^\prime | I | s_2 \rangle \langle s_3^\prime | {\bs\sigma} | s_3 \rangle$ & $\frac{2}{3\sqrt2}(\hat{\bs i} - i\hat{\bs j})$ & ${\bs 0}$ & $-\frac{4}{3\sqrt2} \hat{\bs k}$ & ${\bs 0}$ \\
        $\langle s_2^\prime | {\bs\sigma} | s_2 \rangle \langle s_3^\prime | I | s_3 \rangle$ & $-\frac{1}{3\sqrt2}(\hat{\bs i} - i\hat{\bs j})$ & $\frac{1}{\sqrt6}(\hat{\bs i} - i\hat{\bs j})$ & $\frac{2}{3\sqrt2} \hat{\bs k}$ & $-\frac{2}{\sqrt6} \hat{\bs k}$ \\
        $\langle s_2^\prime | {\bs\sigma} | s_2 \rangle \times \langle s_3^\prime | {\bs\sigma} | s_3 \rangle$ & $\frac{1}{\sqrt2}(i\hat{\bs i} + \hat{\bs j})$ & $-\frac{1}{\sqrt6}(i\hat{\bs i} + \hat{\bs j})$ & $-\sqrt2 i \hat{\bs k}$ & $\frac{2}{\sqrt6} i\hat{\bs k}$ \\
        \hline
        $I_2 {\bs\sigma}_3 - {\bs\sigma}_2 I_3$ & $\frac{1}{\sqrt2}(\hat{\bs i} - i\hat{\bs j})$ & $-\frac{1}{\sqrt6}(\hat{\bs i} - i\hat{\bs j})$ & $-\sqrt2 \hat{\bs k}$ & $\frac{2}{\sqrt6} \hat{\bs k}$ \\
        $i {\bs\sigma}_2 \times {\bs\sigma}_3$ & $-\frac{1}{\sqrt2}(\hat{\bs i} - i\hat{\bs j})$ & $\frac{1}{\sqrt6}(\hat{\bs i} - i\hat{\bs j})$ & $\sqrt2 \hat{\bs k}$ & $-\frac{2}{\sqrt6} \hat{\bs k}$ \\
        \hline
        $\mathcal{O}_{\rm spin}^{(\rm PV)}$ & $\langle\chi_{\frac{1}{2},-\frac{1}{2}}^\lambda | \mathcal{O}_{\rm spin}^{(\rm PV)} | \chi_{\frac{3}{2},\frac{1}{2}}^s\rangle$ & $\langle\chi_{\frac{1}{2},-\frac{1}{2}}^\rho | \mathcal{O}_{\rm spin}^{(\rm PV)} | \chi_{\frac{3}{2},\frac{1}{2}}^s\rangle$ & $\langle\chi_{\frac{1}{2},-\frac{1}{2}}^\lambda | \mathcal{O}_{\rm spin}^{(\rm PV)} | \chi_{\frac{3}{2},-\frac{1}{2}}^s\rangle$ & $\langle\chi_{\frac{1}{2},-\frac{1}{2}}^\rho | \mathcal{O}_{\rm spin}^{(\rm PV)} | \chi_{\frac{3}{2},-\frac{1}{2}}^s\rangle$ \\
        \hline
        $\langle s_2^\prime | I | s_2 \rangle \langle s_3^\prime | {\bs\sigma} | s_3 \rangle$ & $\frac{2}{3\sqrt2}(\hat{\bs i} + i\hat{\bs j})$ & ${\bs 0}$ & $-\frac{4}{3\sqrt2} \hat{\bs k}$ & ${\bs 0}$ \\
        $\langle s_2^\prime | {\bs\sigma} | s_2 \rangle \langle s_3^\prime | I | s_3 \rangle$ & $-\frac{1}{3\sqrt2}(\hat{\bs i} + i\hat{\bs j})$ & $\frac{1}{\sqrt6}(\hat{\bs i} + i\hat{\bs j})$ & $\frac{2}{3\sqrt2} \hat{\bs k}$ & $-\frac{2}{\sqrt6} \hat{\bs k}$ \\
        $\langle s_2^\prime | {\bs\sigma} | s_2 \rangle \times \langle s_3^\prime | {\bs\sigma} | s_3 \rangle$ & $-\frac{1}{\sqrt2}(i\hat{\bs i} - \hat{\bs j})$ & $-\frac{1}{\sqrt6}(-i\hat{\bs i} + \hat{\bs j})$ & $\sqrt2 i\hat{\bs k}$ & $-\frac{2i}{\sqrt6} \hat{\bs k}$ \\
        \hline
        $I_2 {\bs\sigma}_3 - {\bs\sigma}_2 I_3$ & $\frac{1}{\sqrt2}(\hat{\bs i} + i\hat{\bs j})$ & $-\frac{1}{\sqrt6}(\hat{\bs i} + i\hat{\bs j})$ & $-\sqrt2 \hat{\bs k}$ & $\frac{2}{\sqrt6} \hat{\bs k}$ \\
        $i {\bs\sigma}_2 \times {\bs\sigma}_3$ & ${\frac{1}{\sqrt2}(\hat{\bs i} + i\hat{\bs j})}$ & $-\frac{1}{\sqrt6}(\hat{\bs i} + i\hat{\bs j})$ & $-\sqrt2 \hat{\bs k}$ & $\frac{2}{\sqrt6} \hat{\bs k}$ \\
        \hline
        $\mathcal{O}_{\rm spin}^{(\rm PV)}$ & $\langle\chi_{\frac{3}{2},-\frac{3}{2}}^s | \mathcal{O}_{\rm spin}^{(\rm PV)} | \chi_{\frac{1}{2},-\frac{1}{2}}^\lambda\rangle$ & $\langle\chi_{\frac{3}{2},-\frac{3}{2}}^s | \mathcal{O}_{\rm spin}^{(\rm PV)} | \chi_{\frac{1}{2},-\frac{1}{2}}^\rho\rangle$ & $\langle\chi_{\frac{1}{2},-\frac{1}{2}}^\lambda | \mathcal{O}_{\rm spin}^{(\rm PV)} | \chi_{\frac{3}{2},-\frac{3}{2}}^s\rangle$ & $\langle\chi_{\frac{1}{2},-\frac{1}{2}}^\rho | \mathcal{O}_{\rm spin}^{(\rm PV)} | \chi_{\frac{3}{2},-\frac{3}{2}}^s\rangle$ \\
        \hline
        $\langle s_2^\prime | I | s_2 \rangle \langle s_3^\prime | {\bs\sigma} | s_3 \rangle$ & $-\frac{2}{\sqrt6}(\hat{\bs i} + i\hat{\bs j})$ & ${\bs 0}$ & $-\frac{2}{\sqrt6}(\hat{\bs i} - i\hat{\bs j})$ & ${\bs 0}$ \\
        $\langle s_2^\prime | {\bs\sigma} | s_2 \rangle \langle s_3^\prime | I | s_3 \rangle$ & $\frac{1}{\sqrt6}(\hat{\bs i} + i\hat{\bs j})$ & $-\frac{1}{\sqrt2}(\hat{\bs i} + i\hat{\bs j})$ & $\frac{1}{\sqrt6}(\hat{\bs i} - i\hat{\bs j})$ & $-\frac{1}{\sqrt2}(\hat{\bs i} - i\hat{\bs j})$ \\
        $\langle s_2^\prime | {\bs\sigma} | s_2 \rangle \times \langle s_3^\prime | {\bs\sigma} | s_3 \rangle$ & $\frac{3}{\sqrt6}(-i\hat{\bs i} + \hat{\bs j})$ & $-\frac{1}{\sqrt2}(-i\hat{\bs i} + \hat{\bs j})$ & $\frac{3}{\sqrt6}(i\hat{\bs i} + \hat{\bs j})$ & $-\frac{1}{\sqrt2}(i\hat{\bs i} + \hat{\bs j})$ \\
        \hline
        $I_2 {\bs\sigma}_3 - {\bs\sigma}_2 I_3$ & $-\frac{3}{\sqrt6}(\hat{\bs i} + i\hat{\bs j})$ & $\frac{1}{\sqrt2}(\hat{\bs i} + i\hat{\bs j})$ & $-\frac{3}{\sqrt6}(\hat{\bs i} - i\hat{\bs j})$ & $\frac{1}{\sqrt2}(\hat{\bs i} - i\hat{\bs j})$ \\
        $i {\bs\sigma}_2 \times {\bs\sigma}_3$ & $\frac{3}{\sqrt6}(\hat{\bs i} + i\hat{\bs j})$ & $-\frac{1}{\sqrt2}(\hat{\bs i} + i\hat{\bs j})$ & $-\frac{3}{\sqrt6}(\hat{\bs i} - i\hat{\bs j})$ & $\frac{1}{\sqrt2}(\hat{\bs i} - i\hat{\bs j})$ \\
        \hline\hline
    \end{tabular}}
\end{table}

\begin{table}
    \centering
    \caption{The flavor-spin matrix elements $C_{(\mathbb{B}_i\to\mathbb{B}_f)}^{f-s\text{(PC)}}$ for the ground-state baryon conversions.}
    \label{tab:C_f-s}
    \begin{tabular}{ccc}    
        \hline\hline
        $\langle n|\hat{\mathcal{O}}_{f-s}^{(\text{PC})}|\Lambda\rangle$  &$\langle p|\hat{\mathcal{O}}_{f-s}^{(\text{PC})}|\Sigma^+ \rangle$  &$\langle n|\hat{\mathcal{O}}_{f-s}^{(\text{PC})}|\Sigma^0 \rangle$\\
        \hline   
        $-1/\sqrt{6}$   &$1$   &$1/\sqrt{2}$\\
        \hline\hline
    \end{tabular}
\end{table}

\subsubsection{Pion-emission couplings in the chiral quark model}

The pseudoscalar meson emission process can be studied within the chiral quark model (CQM), with the transition Hamiltonian given by:
\begin{align}
    \begin{split}
        H_{m}=\frac{1}{f_{\pi}}\sum_j\bar{\psi}_j\gamma_{\mu}\gamma_5\partial^{\mu}\phi_{\pi}\hat{I}_j^{\pi}\psi_j,
    \end{split}
\end{align}
where \( f_{m} \) is the decay constant of the pseudoscalar meson, and \( j \) denotes the \( j \)-th quark inside the baryon that interacts with the emitted pseudoscalar meson. In the momentum representation, the chiral Lagrangian for quark-pseudoscalar meson coupling reduces to the non-relativistic form up to \( 1/m \) order as:
\begin{align}
    \begin{split}
        H_m^{NR}&=\frac{1}{f_m\sqrt{(2\pi)^32\omega_m}}\sum_j\left[\frac{\omega_m}{2m_j}\boldsymbol{\sigma}_j\cdot\boldsymbol{p}_j+\frac{\omega_m}{2m_j^{\prime}}\boldsymbol{\sigma}_j\cdot\boldsymbol{p}_j^{\prime}-\boldsymbol{\sigma}_j\cdot\boldsymbol{k}\right]\hat{I}_j^m\delta^3(\boldsymbol{p}_j-\boldsymbol{p}_j^{\prime}-\boldsymbol{k}),
        \label{eq:CQM L}
    \end{split}
\end{align}
where \( \omega_m \) and \( \boldsymbol{k} \) are the energy and three-momentum of the final-state pseudoscalar meson in the initial-state rest frame; \( \boldsymbol{p}_j \) and \( \boldsymbol{p}_j^{\prime} \) are the three-momenta of the \( j \)-th quark before and after emitting the pseudoscalar meson. Considering the baryon wave function is fully symmetric with the asymmetric color wavefunction, the strong interaction matrix element can be simplified using \( \langle\mathbb{B}_f|H_{m}^{NR}(j=1,2,3)|\mathbb{B}_i\rangle=3\langle\mathbb{B}_f|H_{m}^{NR}(j=1)|\mathbb{B}_i\rangle \). For the \( \pi \) emission process, the corresponding isospin operator \( \hat{I}_j^{\pi} \) is
\begin{equation}
    \hat{I}_j^{\pi}=\left\{
    \begin{array}{ll}
        a_j^{\dagger}(d)a_j(u) &\text{for}\ \pi^+,\\
        {-}a_j^{\dagger}(u)a_j(d) &\text{for}\ \pi^-,\\
        {-}\frac{1}{\sqrt{2}}[a_j^{\dagger}(u)a_j(u)-a_j^{\dagger}(d)a_j(d)] &\text{for}\ \pi^0,
    \end{array}
    \right.
\end{equation}
which acts on the flavor wave function, with the corresponding matrix elements listed in Tab. \ref{tab:CQM fla}. The Pauli operator and momentum operator in Eq. (\ref{eq:CQM L}) are coupled together, and we handle them by defining ladder operators,
\begin{align}
    \begin{split}
        \boldsymbol{\sigma}_j\cdot\boldsymbol{p}_j=\sigma_j^+p_j^-+\sigma_j^-p_j^++\sigma_{jz}p_{jz},
    \end{split}
\end{align}
where
\begin{equation}
    \left\{
    \begin{array}{l}
        \sigma_j^+=\frac12(\sigma_{jx}+i\sigma_{jy})\\
        \sigma_j^{-}=\frac12(\sigma_{jx}-i\sigma_{jy})
    \end{array},
    \right.
    \ \text{and}\ 
    \left\{
    \begin{array}{l}
        p_j^{+}=p_{jx}+ip_{jy}\\
        p_j^{-}=p_{jx}-ip_{jy}
    \end{array}.
    \right.
\end{equation}
The Pauli operator acts on the spin wavefunction, and we list the corresponding spin matrix elements in Table \ref{tab:CQM spin}. Additionally, the convolution of the spatial wave functions involved in the spatial matrix elements can be written in the momentum representation as 
\begin{align}
    \begin{split}
        I_{m}^{L_f,L_f^z;L_i,L_i^z}&=\langle\psi_{N_fL_fL_f^z}({\bf{P}}_f)|\hat{\mathcal{O}}_{m}^{\text{spatial}}({\boldsymbol{p}}_i)|\psi_{N_iL_iL_i^z}(\boldsymbol{P}_i)\rangle\\
        &=\int d\boldsymbol{p}_1d\boldsymbol{p}_1^{\prime}d\boldsymbol{p}_2d\boldsymbol{p}_3\psi_{N_fL_fL_f^z}^*(\boldsymbol{p}_1^{\prime},\boldsymbol{p}_2,\boldsymbol{p}_3)\hat{\mathcal{O}}_{m}^{\text{spatial}}(\boldsymbol{p}_i)\psi_{N_iL_iL_i^z}(\boldsymbol{p}_1,\boldsymbol{p}_2,\boldsymbol{p}_3)\\
        &\times\delta^3(\boldsymbol{p}_1^{\prime}+\boldsymbol{p}_2+\boldsymbol{p}_3-{\bf{P}}_f)\delta^3(\boldsymbol{p}_1+\boldsymbol{p}_2+\boldsymbol{p}_3-{\bf{P}}_i)\delta^3(\boldsymbol{p}_1-\boldsymbol{p}_1^{\prime}-\boldsymbol{k}).\\
    \end{split}
\end{align}

\begin{table}
    \centering
    \caption{The flavor matrix elements in the strong pion emission processes.}
    \label{tab:CQM fla}
    \scalebox{1}{
    \begin{tabular}{ccccccc}
        \hline
        Processes                    &$\langle \phi_f^{\lambda}|\hat{I}_1^{\pi}|\phi_i^{\lambda}\rangle$      &$\langle \phi_f^{\lambda}|\hat{I}_1^{\pi}|\phi_i^{\rho}\rangle$    &$\langle \phi_f^{\rho}|\hat{I}_1^{\pi}|\phi_i^{\lambda}\rangle$   &$\langle \phi_f^{\rho}|\hat{I}_1^{\pi}|\phi_i^{\rho}\rangle$  &$\langle \phi_f^{a}|\hat{I}_1^{\pi}|\phi_i^{\lambda}\rangle$  &$\langle \phi_f^{a}|\hat{I}_1^{\pi}|\phi_i^{\rho}\rangle$\\
        \hline             
        $n\to p\pi^-$                &$-\frac23$ &$\frac{1}{\sqrt{3}}$ &$\frac{1}{\sqrt{3}}$ &0 &- &-\\
        $n\to n\pi^0$                &$\frac{2}{3\sqrt{2}}$ &$-\frac{1}{\sqrt{6}}$ &$-\frac{1}{\sqrt{6}}$ &0 &- &-\\
        $p\to n\pi^+$                &$\frac23$ &$-\frac{1}{\sqrt{3}}$ &$-\frac{1}{\sqrt{3}}$ &0 &- &-\\
        $p\to p\pi^0$                &$-\frac{2}{3\sqrt{2}}$ &$\frac{1}{\sqrt{6}}$ &$\frac{1}{\sqrt{6}}$ &0 &- &-\\
        $\Sigma^+\to\Sigma^+\pi^0$   &$-\frac{5}{6\sqrt{2}}$ &$\frac{1}{2\sqrt{6}}$ &$\frac{1}{2\sqrt{6}}$ &$-\frac{1}{2\sqrt{2}}$ &- &-\\
        $\Sigma^+\to\Sigma^0\pi^+$   &$\frac{5}{6\sqrt{2}}$ &$-\frac{1}{2\sqrt{6}}$ &$-\frac{1}{2\sqrt{6}}$ &$\frac{1}{2\sqrt{2}}$ &- &-\\
        $\Sigma^+\to\Lambda_8\pi^+$  &$-\frac{1}{2\sqrt{6}}$ &$\frac{1}{2\sqrt{2}}$ &$\frac{1}{2\sqrt{2}}$ &$\frac{1}{2\sqrt{6}}$ &- &-\\
        $\Sigma^+\to\Lambda_1\pi^+$  &- &- &- &- &$\frac12$ &-$\frac{1}{2\sqrt{3}}$\\
        $\Sigma^-\to\Sigma^-\pi^0$   &$\frac{5}{6\sqrt{2}}$ &$-\frac{1}{2\sqrt{6}}$ &$-\frac{1}{2\sqrt{6}}$ &$\frac{1}{2\sqrt{2}}$ &- &-\\
        $\Sigma^-\to\Sigma^0\pi^-$   &$-\frac{5}{6\sqrt{2}}$ &$\frac{1}{2\sqrt{6}}$ &$\frac{1}{2\sqrt{6}}$ &$-\frac{1}{2\sqrt{2}}$ &- &-\\
        $\Sigma^-\to\Lambda_8\pi^-$  &$-\frac{1}{2\sqrt{6}}$ &$\frac{1}{2\sqrt{2}}$ &$\frac{1}{2\sqrt{2}}$ &$\frac{1}{2\sqrt{6}}$ &- &-\\
        $\Sigma^-\to\Lambda_1\pi^-$  &- &- &- &- &$\frac12$ &$-\frac{1}{2\sqrt{3}}$\\ 
        $\Lambda_8\to\Sigma^+\pi^-$  &$\frac{1}{2\sqrt{6}}$ &$-\frac{1}{2\sqrt{2}}$ &$-\frac{1}{2\sqrt{2}}$ &$-\frac{1}{2\sqrt{6}}$ &- &- \\
        $\Lambda_8\to\Sigma^0\pi^0$  &$-\frac{1}{2\sqrt{6}}$ &$\frac{1}{2\sqrt{2}}$ &$\frac{1}{2\sqrt{2}}$ &$\frac{1}{2\sqrt{6}}$ &- &-\\
        $\Lambda_8\to\Sigma^-\pi^+$  &$\frac{1}{2\sqrt{6}}$ &$-\frac{1}{2\sqrt{2}}$ &$-\frac{1}{2\sqrt{2}}$ &$-\frac{1}{2\sqrt{6}}$ &- &- \\
        \hline
    \end{tabular}}
\end{table}

\begin{table}
    \centering
    \caption{The spin matrix elements in the strong pion emission processes.}
    \label{tab:CQM spin}
    \begin{tabular}{cccc}
        \hline
        $\langle\chi_{\frac12,-\frac12}^{\lambda}|\sigma_{1z}|\chi_{\frac12,-\frac12}^{\lambda}\rangle$  &$\langle\chi_{\frac12,-\frac12}^{\lambda}|\sigma_{1z}|\chi_{\frac12,-\frac12}^{\rho}\rangle$  &$\langle\chi_{\frac12,-\frac12}^{\rho}|\sigma_{1z}|\chi_{\frac12,-\frac12}^{\lambda}\rangle$  &$\langle\chi_{\frac12,-\frac12}^{\rho}|\sigma_{1z}|\chi_{\frac12,-\frac12}^{\rho}\rangle$\\
        \hline
        $-\frac23$ &$\frac{1}{\sqrt{3}}$ &$\frac{1}{\sqrt{3}}$ &0\\
        \hline
        $\langle\chi_{\frac32,-\frac12}^s|\sigma_{1z}|\chi_{\frac12,-\frac12}^{\lambda}\rangle$  &$\langle\chi_{\frac32,-\frac12}^s|\sigma_{1z}|\chi_{\frac12,-\frac12}^{\rho}\rangle$  &$\langle\chi_{\frac12,-\frac12}^{\lambda}|\sigma_{1z}|\chi_{\frac32,-\frac12}^s\rangle$  &$\langle\chi_{\frac12,-\frac12}^{\rho}|\sigma_{1z}|\chi_{\frac32,-\frac12}^s\rangle$\\
        \hline     
        $\frac{2}{3\sqrt{2}}$ &$\frac{2}{\sqrt{6}}$ &$\frac{2}{3\sqrt{2}}$ &$\frac{2}{\sqrt{6}}$\\
        \hline
        $\langle\chi_{\frac12,\frac12}^{\lambda}|\sigma_{1}^+|\chi_{\frac12,-\frac12}^{\lambda}\rangle$  &$\langle\chi_{\frac12,\frac12}^{\lambda}|\sigma_{1}^+|\chi_{\frac12,-\frac12}^{\rho}\rangle$  &$\langle\chi_{\frac12,\frac12}^{\rho}|\sigma_{1}^+|\chi_{\frac12,-\frac12}^{\lambda}\rangle$  &$\langle\chi_{\frac12,\frac12}^{\rho}|\sigma_{1}^+|\chi_{\frac12,-\frac12}^{\rho}\rangle$\\
        \hline
        $\frac23$ &$-\frac{1}{\sqrt{3}}$ &$-\frac{1}{\sqrt{3}}$ &0\\
        \hline
        $\langle\chi_{\frac12,-\frac12}^{\lambda}|\sigma_{1}^-|\chi_{\frac12,\frac12}^{\lambda}\rangle$  &$\langle\chi_{\frac12,-\frac12}^{\lambda}|\sigma_{1}^-|\chi_{\frac12,\frac12}^{\rho}\rangle$  &$\langle\chi_{\frac12,-\frac12}^{\rho}|\sigma_{1}^-|\chi_{\frac12,\frac12}^{\lambda}\rangle$  &$\langle\chi_{\frac12,-\frac12}^{\rho}|\sigma_{1}^-|\chi_{\frac12,\frac12}^{\rho}\rangle$\\
        \hline
        $\frac23$ &$-\frac{1}{\sqrt{3}}$ &$-\frac{1}{\sqrt{3}}$ &0\\
        \hline
        $\langle\chi_{\frac32,\frac12}^s|\sigma_{1}^+|\chi_{\frac12,-\frac12}^{\lambda}\rangle$  &$\langle\chi_{\frac32,\frac12}^s|\sigma_{1}^+|\chi_{\frac12,-\frac12}^{\rho}\rangle$  &$\langle\chi_{\frac12,-\frac12}^{\lambda}|\sigma_{1}^-|\chi_{\frac32,\frac12}^s\rangle$  &$\langle\chi_{\frac12,-\frac12}^{\rho}|\sigma_{1}^-|\chi_{\frac32,\frac12}^s\rangle$\\
        \hline
        $-\frac{1}{3\sqrt{2}}$ &$-\frac{1}{\sqrt{6}}$ &$-\frac{1}{3\sqrt{2}}$ &$-\frac{1}{\sqrt{6}}$\\
        \hline
        $\langle\chi_{\frac32,-\frac32}^s|\sigma_{1}^-|\chi_{\frac12,-\frac12}^{\lambda}\rangle$  &$\langle\chi_{\frac32,-\frac32}^s|\sigma_{1}^-|\chi_{\frac12,-\frac12}^{\rho}\rangle$  &$\langle\chi_{\frac12,-\frac12}^{\lambda}|\sigma_{1}^+|\chi_{\frac32,-\frac32}^s\rangle$  &$\langle\chi_{\frac12,-\frac12}^{\rho}|\sigma_{1}^+|\chi_{\frac32,-\frac32}^s\rangle$\\
        \hline
        $\frac{1}{\sqrt{6}}$ &$\frac{1}{\sqrt{2}}$ &$\frac{1}{\sqrt{6}}$ &$\frac{1}{\sqrt{2}}$\\
        \hline
    \end{tabular}
\end{table}

Specially for the ground-state baryon decay $\mathbb{B}_i(\frac12^+)\to\mathbb{B}_f(\frac12^+)+\pi$, the axial current conservation gives 
\begin{align}
    \begin{split}
        g_A(\mathbb{B}_i\mathbb{B}_f\pi)\equiv\frac{\langle \mathbb{B}_f|\sum_j\hat{I}_j^{\pi}\sigma_{jz}|\mathbb{B}_i\rangle}{\langle \mathbb{B}_f|\sigma_z^{tot}|\mathbb{B}_i\rangle}=\frac{3\langle\mathbb{B}_f|\hat{I}_1^{\pi}\sigma_{1z}|\mathbb{B}_i\rangle}{\langle\mathbb{B}_f|\sigma_z^{tot}|\mathbb{B}_i},
    \end{split}
\end{align}
where $\sigma_{jz}$ and $\sigma_z^{tot}$ are the quark and baryon spin operator projections to the z axis, respectively. The values for $g_A$ in the SU(6) CQM are listed in Tab. \ref{tab:gA}.

\begin{table}
    \centering
    \caption{Axial-vector couplings for the pion emission.}
    \label{tab:gA}
    \scalebox{1}{
    \begin{tabular}{cccccccc}
        \hline
        Processes             &$g_A$               &Processes                             &$g_A$              &Processes                            &$g_A$               &Processes                            &$g_A$\\
        \hline
        $n\to p\pi^-$   &$-5/3$              &$\Sigma^+\to \Sigma^+\pi^0$     &$-4/(3\sqrt{2})$   &$\Sigma^-\to \Sigma^-\pi^0$    &$4/(3\sqrt{2})$     &$\Lambda_8\to\Sigma^+\pi^-$    &$2/\sqrt{6}$\\
        $n\to n\pi^0$   &$5/(3\sqrt{2})$     &$\Sigma^+\to \Sigma^0\pi^+$     &$4/(3\sqrt{2})$    &$\Sigma^-\to \Sigma^0\pi^-$    &$-4/(3\sqrt{2})$    &$\Lambda_8\to\Sigma^0\pi^0$    &$-2/\sqrt{6}$\\
        $p\to n\pi^+$   &$5/3$               &$\Sigma^+\to \Lambda_8\pi^+$    &$-2/\sqrt{6}$      &$\Sigma^-\to \Lambda_8\pi^-$   &$-2/\sqrt{6}$       &$\Lambda_8\to\Sigma^-\pi^+$    &$2/\sqrt{6}$\\
        $p\to p\pi^0$   &$-5/(3\sqrt{2})$    &&&&\\
        \hline
    \end{tabular}}
\end{table}

\subsubsection{Amplitudes of the Pole terms}

The amplitude of the pole term connects the weak interaction matrix element for changing flavor and the strong interaction matrix element for emitting $\pi$ meson through the intermediate state as a propagator. For the PC transitions, if the intermediate state is the ground-state baryon, the amplitudes for both type-A and type-B of the pole terms are:
\begin{align}
    \begin{split}
        \mathcal{M}_{\text{Pole,A}}^{(\text{PC})}&=\mathcal{F}_{A}^{(\text{PC})}(\boldsymbol{k})\times\mathcal{G}_{A}^{(\text{PC})},\\
        \mathcal{M}_{\text{Pole,B}}^{(\text{PC})}&=\mathcal{F}_{B}^{(\text{PC})}(\boldsymbol{k})\times\mathcal{G}_{B}^{(\text{PC})},
        \label{eq:PT amp}
    \end{split}
\end{align}
where $\mathcal{F}_{A/B}^{(\text{PC})}$ are functions related to the convolution of spatial wave functions. Taking $\Lambda\to p\pi^-$ as an example, their expressions are:
\begin{align}
    \begin{split}
        \mathcal{F}_{A}^{(\text{PC})}(\boldsymbol{k})&=\frac{G_F}{\sqrt{2}}V_{ud}V_{us}48\left(\frac{\alpha^2\alpha_{\rho}\alpha_{\lambda}}{\sqrt{\pi}(4\alpha^2+\alpha_{\lambda}^2+3\alpha_{\rho}^2)}\right)^{3/2}\\
        &\times\frac{1}{f_{\pi}\sqrt{(2\pi)^32\omega_0}}|\boldsymbol{k}|\big(-1-\frac{\omega_0}{6m_q}\big)\text{exp}\left[-\frac{\boldsymbol{k}^2}{6\alpha^2}\right],\\
        \mathcal{F}_{B}^{(\text{PC})}(\boldsymbol{k})&=\frac{G_F}{\sqrt{2}}V_{ud}V_{us}48\left(\frac{\alpha^2\alpha_{\rho}\alpha_{\lambda}}{\sqrt{\pi}(4\alpha^2+\alpha_{\lambda}^2+3\alpha_{\rho}^2)}\right)^{3/2}\\
        &\times\frac{1}{f_{\pi}\sqrt{(2\pi)^32\omega_0}}|\boldsymbol{k}|\big(-1-\frac{\omega_0}{4m_q+2m_s}\big)\text{exp}\left[-\frac{\boldsymbol{k}^2}{8}(\frac{3m_s^2}{(2m_q+m_s)^2\alpha_{\lambda}^2}+\frac{1}{\alpha_{\rho}^2})\right],
    \end{split}
\end{align}
It is easy to see that in the SU(3) flavor symmetry limit, due to the constituent mass $m_s = m_q$ ($q=u,d$), the overlap of the spatial wavefunctions for type-A and type-B pole terms is the same. The $\mathcal{G}_{A/B}^{(\text{PC})}$ in Eq. (\ref{eq:PT amp}) is a function that contains spin-flavor information of the initial-state, final-state, and intermediate baryons. In the SU(3) flavor limit, it can be defined as:
\begin{align}
    \mathcal{G}^{\text{(PC)}}=\mathcal{G}_{A}^{(\text{PC})}+\mathcal{G}_{B}^{(\text{PC})}.
\end{align}
For the decay channels $\Lambda\to N\pi$ and $\Sigma^{\pm}\to N\pi$, the expressions for $\mathcal{G}^{\text{(PC)}}$ are given by:
\begin{align}
    \begin{split}
        &\mathcal{G}^{\text{(PC)}}(\Lambda\to p\pi^-)\equiv\left[\frac{(2M_n)g_A^{np\pi^-}C_{(\Lambda\to n)}^{f-s\text{(PC)}}}{M_{\Lambda}^2-M_n^2}+\frac{(2M_{\Sigma})g_A^{\Lambda\Sigma^+\pi^-}C_{(\Sigma^+\to p)}^{f-s\text{(PC)}}}{M_p^2-M_{\Sigma}^2}\right],\\
        &\mathcal{G}^{\text{(PC)}}(\Lambda\to n\pi^0)\equiv\left[\frac{(2M_n)g_A^{nn\pi^0}C_{(\Lambda\to n)}^{f-s\text{(PC)}}}{M_{\Lambda}^2-M_n^2}+\frac{(2M_{\Sigma})g_A^{\Lambda\Sigma^0\pi^0}C_{(\Sigma^0\to n)}^{f-s\text{(PC)}}}{M_n^2-M_{\Sigma}^2}\right],\\
        &\mathcal{G}^{\text{(PC)}}(\Sigma^-\to n\pi^-)=\mathcal{G}_B^{\text{(PC)}}(\Sigma^-\to n\pi^-)\equiv\left[\frac{(2M_{\Sigma^0})g_A^{\Sigma^-\Sigma^0\pi^-}C_{(\Sigma^0\to n)}^{f-s\text{(PC)}}}{M_n^2-M_{\Sigma^0}^2}+\frac{(2M_\Lambda)g_A^{\Sigma^-\Lambda\pi^-}C_{(\Lambda\to n)}^{f-s\text{(PC)}}}{M_n^2-M_{\Lambda}^2}\right],\\
        &\mathcal{G}^{\text{(PC)}}(\Sigma^+\to p\pi^0)\equiv\left[\frac{(2M_p)g_A^{pp\pi^0}C_{(\Sigma^+\to p)}^{f-s\text{(PC)}}}{M_{\Sigma^+}^2-M_p^2}+\frac{(2M_{\Sigma^+})g_A^{\Sigma^+\Sigma^+\pi^0}C_{(\Sigma^+\to p)}^{f-s\text{(PC)}}}{M_p^2-M_{\Sigma^+}^2}\right],\\
        &\mathcal{G}^{\text{(PC)}}(\Sigma^+\to n\pi^+)\equiv\left[\frac{(2M_p)g_A^{pn\pi^+}C_{(\Sigma^+\to p)}^{f-s\text{(PC)}}}{M_{\Sigma^+}^2-M_p^2}+\frac{(2M_{\Sigma^0})g_A^{\Sigma^+\Sigma^0\pi^+}C_{(\Sigma^0\to n)}^{f-s\text{(PC)}}}{M_n^2-M_{\Sigma^0}^2}+\frac{(2M_{\Lambda})g_A^{\Sigma^+\Lambda\pi^+}C_{(\Lambda\to n)}^{f-s\text{(PC)}}}{M_n^2-M_{\Lambda}^2}\right],
        \label{eq:G}
    \end{split}
\end{align}
In the above expression, $g_A$ is the axial coupling constant calculated in the NRCQM, and $C^{f-s\text{(PC)}}$ is the spin-flavor factor for the IC process of ground-state baryons, with their values for different processes listed in Tab. \ref{tab:gA} and \ref{tab:C_f-s}.

According to the Eq. (\ref{eq:G}), we can derive the following conclusion:
\begin{itemize}
    \item Due to the isospin symmetry of the $\pi$ mesons, we obtain the relation $\mathcal{G}^{\text{(PC)}}(\Lambda \to n\pi^0)/\mathcal{G}^{\text{(PC)}}(\Lambda \to p\pi^-) = -1/\sqrt{2}$~\cite{Richard:2016hac}. In these two decay channels $\Lambda \to p\pi^-$ and $n\pi^0$, the ground state nucleon and $\Sigma$ are intermediate states of type-A and type-B pole terms, respectively, and the signs of the PC amplitudes for the two terms are opposite. Namely, a cancellation exists between these two terms.
    \item For $\Sigma^-\to n\pi^-$, only the type-B pole term contributes. Furthermore, the intermediate states can be the isospin-vector baryon $\Sigma^0$ and the isospin-scalar baryon $\Lambda$, and their PC amplitudes have opposite signs.
    \item For $\Sigma^+\to p\pi^0$, the ground-state baryons $p$ and $\Sigma^+$ contribute to the type-A and type-B pole amplitude for the PC transitions, respectively, and the signs of the two are opposite.
    \item For $\Sigma^+\to n\pi^+$, the type-A amplitude contributed by the ground-state baryon $p$ has an opposite sign to the type-B amplitudes contributions by $\Sigma^0$ and $\Lambda$ for the PC transition. Namely, a cancellation also exists between the type-A and type-B amplitudes. However, in this case, these two type-B amplitudes have the same sign.
\end{itemize}

The above features with the pole terms are based on the flavor symmetry relations in the NRCQM, and one crucial point is that there are cancellations involved  among the pole terms. We note in advance that the exclusive contributions from either type-A or type-B amplitude is sizeable. Without cancellations, the decay widths will usually be over-estimated. It indicates that an elaborated treatment of the pole terms are necessary and essential for understanding the transition mechanisms for the hyperon decays. 

In the case of the PV transition, the expressions of the amplitudes will be more complex due to the entanglement of spin and momentum in the corresponding operators, as shown in Eq. (\ref{eq:2to2 H_W}). In addition, the intermediate states are the first orbital excitation states with $J^P=\frac12^-$. We skip the detailed deductions of the PV transition amplitudes, but collect them in Appendix \ref{app:tree amp}. Due to the higher mass of the first orbital excited states compared to the ground-state baryons, and the fact that they possess a non-zero width, the amplitudes of these pole terms will have an imaginary part, and their relative signs are not as apparent as those in the case of the PC transition. 

For the two decay channels $\Lambda \to n\pi^0$ and $\Lambda \to p\pi^-$, regardless of whether it is a PC transition or a PV transition, the pole term amplitude ratio between these two channels is $-1/\sqrt{2}$. Thus, considering only the contributions of the pole terms (PC and PV), it is evident that $R_{\Lambda}=\frac{\Gamma(\Lambda\to p\pi^-)}{\Gamma(\Lambda\to p\pi^-)+\Gamma(\Lambda\to n\pi^0)}\simeq\frac23$, taking into account the phase spaces for these two channels are nearly identical. However, as shown in Fig.~\ref{fig:Lambda DPE and CS}, the $p\pi^-$ mode contains the DPE process, which is generally the dominant contribution. In contrast, the $n\pi^0$ channel does not involve the DPE transition. This difference arising from the DPE and CS processes will violate the relation between the $n\pi^0$ and $p\pi^-$ channel for the pole terms. We need to understand why the branching ratio fraction of $BR(\Lambda\to p\pi^-)/BR(\Lambda\to n\pi^0)\simeq 2$ with the apparent differences from the DPE and CS transitions. 

For $\Sigma^+\to n\pi^+$, only the pole terms contribute, indicating that pole terms are crucial in the hadronic weak decay of hyperons. For $\Sigma^\pm\to n\pi^\pm$, we also consider the contribution of singlet $\Lambda(1405)$ to the type-B terms in the PV transition, with $J^P=1/2^-$ and an additional phase $\theta_{\Lambda}$. Although the exact composition of $\Lambda(1405)$ remains theoretically uncertain, it is classified as a flavor singlet in the quark model. In our calculations, we likewise treat it as a singlet and introduce an additional phase $\theta_{\Lambda}$ between the singlet and octet amplitudes to account for effects arising from other possible dynamics.

\subsection{Final state interactions (FSIs) in hyperon decays}

One should be aware of the possible role played by the FSIs. Note that the phase spaces for $\Lambda$ and $\Sigma$ decays are rather small, and the nucleon and pion scattering has large cross sections. It is hence natural to consider that the FSIs may have sizeable effects on the hyperon hadronic weak decays. 

Focusing on the hadronic weak decays of  $\Lambda$ and $\Sigma$, it is interesting to note several features which seem to fit the request for fixing the puzzle. For the $\Lambda$ decays, due to the presence of the 
dominant DPE process in $\Lambda\to p\pi^-$, the FSI of $p\pi^-$ and their rescattering can contribute to the $n\pi^0$ channel as the leading correction. This additional transition amplitude in $\Lambda\to n\pi^0$ can enhance the transition amplitude by interference and ``coincidentally" retain the ratio of $BR(\Lambda\to p\pi^-)/BR(\Lambda\to n\pi^0)\simeq 2$ as observed in experiment. In turn, the rescattering contributions from the $\Lambda\to n\pi^0$ to $\Lambda\to p\pi^-$ will be much smaller and subleading since we are talking about a FSI amplitude from a small tree-level amplitude. Thus, its correction to the much larger tree-level amplitudes in $\Lambda\to p\pi^-$ will be  insignificant.  

For the $\Sigma^+$ decay, due to the violation of charge conservation, there is no feed-down FSI contributions from relevant processes. For $\Sigma^-\to n\pi^-$, its only decay channel is $n\pi^-$. It cannot either feed in other channels or receive feed-down contributions from other channels. It suggests that the FSIs will contribute to the decay of $\Lambda\to n\pi^0$ as the leading correction, but only play a sub-leading role in other decay processes.

\begin{figure}
    \centering
    \subfigure[ \ $t$-channel]{ 
    \label{fig: A}    
    \includegraphics[width=0.35\linewidth]{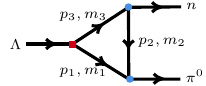}}
    \subfigure[ \ $s$-channel]{
    \label{fig: B}    
    \includegraphics[width=0.4\linewidth]{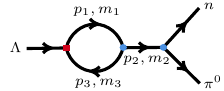}}
   \caption{Schematic diagram of the FSIs from $p\pi^-$ rescattering to $n\pi^0$. Black squares and red dots represent weak and strong vertices respectively.}
    \label{fig:loop}
  \end{figure}

As shown in Fig. \ref{fig:loop}, there are two types of schematic diagram of the FSIs. Specifically, Fig. \ref{fig:loop}(a) is the triangle loop diagram related to the $t$-channel, while (b) is the two-point bubble diagram related to the $s$ channel.

At the hadronic level, the PC amplitude and the PV amplitude can be described by pseudovector coupling and scalar coupling, respectively. We employ the effective field theory approach to calculate the loop-level amplitudes. The explicit Lagrangian forms for the above vertices can be written as
\begin{align}
    \begin{split}
        \mathcal{L}_{{B}{B}P}&=ig_{{B}{B}P}\bar{{B}}\gamma_{\mu}\gamma_5{B}\partial^{\mu}P,\\
        \mathcal{L}_{{B}{B}^*P}&=g_{{B}{B}^*P}\bar{{B}}{B}^*P,\\
        \mathcal{L}_{{B}{B}V}&=-g_{{B}{B}V}\bar{{B}}\gamma_{\mu}{B}V^{\mu}+\frac{f_{{B}{B}V}}{2M_{{B}}}\bar{{B}}\sigma^{\mu\nu}{B}\partial_{\nu}V_{\mu},\\
        \mathcal{L}_{VPP}&=ig_{VPP}Tr[(P\partial_{\mu}P-\partial_{\mu}PP)V^{\mu}],
    \end{split}
\end{align}
where $\sigma^{\mu\nu}=i[\gamma^{\mu},\gamma^{\nu}]/2$. ${B}$ and ${B}^*$ represent baryon fields with $J^P=\frac12^+$ and $J^P=\frac12^-$, respectively; $P$ and $V$ represent meson fields with $J^P=0^-$ and $J^P=1^-$, respectively. For the strong decay, the pseudo-vector couping is defined as $g_{\mathbb{B}\mathbb{B}P}={f_{\mathbb{B}\mathbb{B}P}}/{M_P}$. 

We use the notation $[t,(\text{P}),M1,M3,(M2)]$ to denote the FSIs via the triangle loop process shown in Fig. \ref{fig:loop}(a), where the intermediate particles $M1$ and $M3$ interact with each other by exchanging $M2$ for the PC or PV transitions. Similarly, we use the notation $[s,(\text{P}),M1,M3,(M2)]$ to denote the bubble process shown in Fig. \ref{fig:loop}(b), where particles $M1$ and $M3$ interact and rescatter into  intermediate states. The masses and 4-vector momenta of these internal particles are denoted by $(m_1,m_2,m_3)$ and $(p_1,p_2,p_3)$, respectively. The 4-vector momenta of the initial-state baryon $\Lambda$, final-state baryon $n$ and final-state meson $\pi^0$ are labeled as $p_{\Lambda}$, $p_n$ and $p_{\pi}$, respectively. As follow, we will write down the detailed amplitude for each loop transition.

\begin{itemize}
    \item $[t,(\text{PC}),\pi^-,p,(\rho^-)]$
    \begin{align}
        \begin{split}
            i\mathcal{M}&=\int\frac{d^4p_1}{(2\pi)^4}V_1D(\pi)V_{2\nu}D^{\alpha\nu}(\rho)V_{3\alpha}D(p)\mathcal{F}(p_i^2),
            \label{eq:loop}
        \end{split}
    \end{align}
    where the vertex functions are
    \begin{align}
        \begin{split}
            V_1&=-ig_{\Lambda p\pi}^{(\text{PC})}\gamma_{\mu}\gamma_5u(p_{\Lambda})p_1^{\mu},\\
            V_{2\nu}&=ig_{\rho\pi\pi}(p_1+p_{\pi})_{\nu},\\
            V_{3\alpha}&=-ig_{pn\rho}\bar{u}(p_n)\gamma_{\alpha}+\frac{f_{pn\rho}}{M_n+M_p}\bar{u}(p_n)\sigma_{\alpha\beta}p_2^{\beta}.
            \label{eq:vertex function}
        \end{split}
    \end{align}
    In Eq. (\ref{eq:loop}) functions $D(\pi)=i/(p_1^2-m_{\pi}^2+i\epsilon)$, $D^{\alpha\nu}(\rho)=-i(g^{\alpha\nu}-p_2^{\alpha}p_2^{\nu}/p_2^2)/(p_2^2-m_{\rho}^2+i\epsilon)$ and $D(p)=i/(\slashed{p}_3-m_p)$ are the propagators for pseudoscalar meson $\pi$, vector mseon $\rho$ and baryon $p$, respectively. $u(p_{\Lambda})$ and $u(p_n)$ are the spinors of initial-state baryon $\Lambda$ and final-state baryon $n$, respectively. We adopt the monopole form factor $\mathcal{F}(p_i)=(\Lambda_i^2-m_i^2)/(\Lambda_i^2-p_i^2)$ (where $\Lambda_i=m_i+\beta\Lambda_{\text{QCD}}$) to cut off UV divergence, where $m_i$ and $p_i$ are the mass and 4-vector momentum of the exchanged particle.

    \item $[t,(\text{PV}),\pi^-,p,(\rho^-)]$
    \begin{align}
        \begin{split}
            i\mathcal{M}&=\int\frac{d^4p_1}{(2\pi)^4}V_1D(\pi)V_{2\nu}D^{\alpha\nu}(\rho)V_{3\alpha}D(p)\mathcal{F}(p_i^2),
        \end{split}
    \end{align}
    where the vertex functions $V_{2\nu}$ and $V_{3\alpha}$ have the same form as that in Eq. (\ref{eq:vertex function}) and the other function is:
    \begin{align}
        \begin{split}
            V_1&=ig_{\Lambda p\pi}^{(\text{PV})}u(p_{\Lambda}).\\
            \label{eq:vertex function-2}
        \end{split}
    \end{align}

    \item $[t,(\text{PC}),p,\pi^-,(p)]$
    \begin{align}
        \begin{split}
            i\mathcal{M}&=\int\frac{d^4p_1}{(2\pi)^4}V_1D(p)V_2D(p)V_3D(\pi)\mathcal{F}(p_i^2),
        \end{split}
    \end{align}
    where the vertex function $V_1$ have been given in Eq. (\ref{eq:vertex function}) and the the other two functions are
    \begin{align}
        \begin{split}
            V_2&=-ig_{pp\pi}\gamma_{\alpha}\gamma_5p_{\pi}^{\alpha},\\
            V_3&=ig_{pn\pi}\bar{u}(p_n)\gamma_{\beta}\gamma_5p_3^{\beta}.
            \label{eq:vertex function-3}
        \end{split}
    \end{align}

    \item $[t,(\text{PV}),p,\pi^-,(p)]$
    \begin{align}
        \begin{split}
            i\mathcal{M}&=\int\frac{d^4p_1}{(2\pi)^4}V_1D(p)V_2D(p)V_3D(\pi)\mathcal{F}(p_i^2),
        \end{split}
    \end{align}
    where the vertex function $V_1$ has been given in Eq. (\ref{eq:vertex function-2}) and the other $V_2$ and $V_3$  have been given in Eq. (\ref{eq:vertex function-3}).

    \item $[s,(\text{PC}),p,\pi^-,(N(\frac12^+))]$
        \begin{align}
            \begin{split}
                i\mathcal{M}=\int\frac{d^4p_1}{(2\pi)^4}V_1D(p)V_2D(\pi)V_3\mathbb{D}(N)\mathcal{F}(p_i^2),
            \end{split}
        \end{align}
        where the vertex functions are, 
        \begin{align}
            \begin{split}
                V_1&=ig_{\Lambda p\pi}^{(\text{PC})}\gamma_{\mu}\gamma_5u(p_{\Lambda})p_3^{\mu},\\
                V_2&=-ig_{Np\pi^-}\gamma_{\alpha}\gamma_5p_3^{\alpha},\\
                V_3&=-ig_{Nn\pi^0}\gamma_{\beta}\gamma_5p_{\pi}^{\beta},
            \end{split}
        \end{align}
        and the propagator of the intermediate state $N$ can be written as $\mathbb{D}(N)=i(\slashed{p}_2+m_2)/(p_{\Lambda}^2-m_2^2+im_2\Gamma_2)$.
    \item $[s,(\text{PV}),p,\pi^-,(N(\frac12^-))]$
    \begin{align}
        \begin{split}
            i\mathcal{M}=\int\frac{d^4p_1}{(2\pi)^4}V_1D(p)V_2D(\pi)V_3\mathbb{D}(N)\mathcal{F}(p_i^2),
        \end{split}
    \end{align}
    where all the vertex functions are
    \begin{align}
        \begin{split}
            V_1&=ig_{\Lambda p\pi}^{(\text{PV})}u(p_{\Lambda}),\\
            V_2&=ig_{Np\pi^-},\\
            V_3&=ig_{Nn\pi^0}\bar{u}(p_n),
        \end{split}
    \end{align}
    and propagators have been given earlier.

\end{itemize}
The coupling constants of each vertex in the above loops can be determined by the following strategy:
\begin{itemize}
    \item The weak couplings $g_{\Lambda p\pi}^{(\text{PC})}=8.65\times 10^{-7}\ \text{GeV}^{-1}$ and $g_{\Lambda p\pi}^{(\text{PV})}=3.79\times 10^{-7}$ can be extracted by the DPE process of the $\Lambda\to p\pi^-$ which are calculated in the NRCQM.
    
    \item The coupling $g_{\rho\pi\pi}=\sqrt{2}g_{VPP}\simeq 5.96$ can be extracted by the experimental data of the decay $\rho^0\to \pi^+\pi^-$.

    \item The vector coupling constant for the ground-state nucleons $NN$ and $\rho$ meson ranges from $g_{NN\rho^0}=2.97$~\cite{Rijken:1998yy} to 3.19~\cite{Machleidt:1987hj}. In this work, we choose the value of the latter.
    
    \item The pseudovector coupling constant for the ground-state nucleons and pion is obtained from the Goldberger-Treiman relation~\cite{Goldberger:1958tr}, which gives $f_{NN\pi^0}=0.952\pm 0.003$. In this work, we use the value $f_{NN\pi^0}=0.989$ from the analysis of $NN$ scattering data~\cite{Janssen:1996kx}.

    \item The coupling constant for $N^*N\pi$ where $N^*$ stands for the first radial or orbital excitation state,  can be extracted by the experimental data of $N^*\to N\pi$. In this work, the corresponding couplings $g_{N(1440)N\pi^0}=2.75 \ \text{GeV}^{-1}$, $g_{N(1535)N\pi^0}=0.66$ and $g_{N(1650)N\pi^0}=0.68$ are extracted with the data from PDG~\cite{ParticleDataGroup:2024cfk}. 
\end{itemize}

\section{width and asymmetry parameter}

In this work, the amplitudes of the DPE, CS, and pole terms are all calculated within the framework of the NRCQM, with mesons and baryons represented by mock states, respectively,
\begin{align}
    \begin{split}
        |\mathbb{M}({\bf P}_c;J,J_z)\rangle&=\sum_{S_z,L_z;c_i}\langle L L_z;S S_z|J J_z\rangle\int d\boldsymbol{p}_1d\boldsymbol{p}_2\delta^3(\boldsymbol{p}_1+\boldsymbol{p}_2-{\bf P}_c)\\
        &\times\psi_{NLL_z}(\boldsymbol{p}_1,\boldsymbol{p}_2)\chi_{S,S_z}^{s_1,s_2}\frac{\delta_{c_1c_2}}{\sqrt{3}}\phi_{i_1i_2}b_{c_1,i_1,s_1,\boldsymbol{p}_1}^{\dagger}d_{c_2,i_2,s_2,\boldsymbol{p}_2}^{\dagger}|0\rangle,\\
        |\mathbb{B}({\bf P}_c;J,J_z)\rangle&=\sum_{S_z,L_z;c_i}\langle L L_z;S S_z|J J_z\rangle\int d\boldsymbol{p}_1d\boldsymbol{p}_2d\boldsymbol{p}_3\delta^3(\boldsymbol{p}_1+\boldsymbol{p}_2+\boldsymbol{p}_3-{\bf P}_c)\\
        &\times\psi_{NLL_z}(\boldsymbol{p}_1,\boldsymbol{p}_2,\boldsymbol{p}_3)\chi_{S,S_z}^{s_1,s_2,s_3}\frac{\epsilon_{c_1c_2c_3}}{\sqrt{6}}\phi_{i_1i_2i_3}b_{c_1,i_1,s_1,\boldsymbol{p}_1}^{\dagger}b_{c_2,i_2,s_2,\boldsymbol{p}_2}^{\dagger}b_{c_3,i_3,s_3,\boldsymbol{p}_3}^{\dagger}|0\rangle,
        \label{Eq: mock states}
    \end{split}
\end{align}
where $c_j$, $s_j$, $i_j$ are color, spin, and flavor indexes, respectively. $\psi_{NLL_z}$ is the spatial wavefunction which is taken as an harmonic oscillator wavefunction. $\boldsymbol{p}_i$ denotes the single quark (antiquark) three-vector momentum, and $\boldsymbol{P}_c$ ($\boldsymbol{P}_c^{\prime}$) denotes the hadron momentum. $\chi_{S,S_z}$ is the spin wavefunction; $\phi$ is the flavor wavefunction, and $\delta_{c_1c_2}/\sqrt{3}$ and $\epsilon_{c_1c_2c_3}/\sqrt{6}$ are the color wavefunctions for the meson and baryon, respectively. The normalization condition for the mock states are:
\begin{align}
    \begin{split}
        \langle \mathbb{M}({\bf P}_c^{\prime};J,J_z)|\mathbb{M}({\bf P}_c;J,J_z)\rangle&=\delta^3({\bf P}_c^{\prime}-{\bf P}_c),\\
        \langle \mathbb{B}({\bf P}_c^{\prime};J,J_z)|\mathbb{B}({\bf P}_c;J,J_z)\rangle&=\delta^3({\bf P}_c^{\prime}-{\bf P}_c).
        \label{Eq: normalization condition for the mock states}
    \end{split}
\end{align}

In the above convention, for the two-body decay $A \to B + C$, the $S$-matrix is defined by
\begin{align}  
    \begin{split}  
        \mathcal{S}=\mathcal{I}-2\pi i\delta^4(P_A-P_B-P_C)\mathcal{\tilde{M}},  
    \end{split}  
\end{align} 
where 
\begin{align}  
    \begin{split}  
        \delta^3(\boldsymbol{P}_A-\boldsymbol{P}_B-\boldsymbol{P}_c)\mathcal{\tilde{M}}\equiv\langle BC|H_I|A\rangle.  
    \end{split}  
\end{align} 
By integrating over phase space, the decay width can be expressed as:
\begin{align}  
    \begin{split}  
        \Gamma(A\to B+C)&=\frac{8\pi^2|\boldsymbol{k|}E_{{B}}E_{{C}}}{M_A}\frac{1}{2J_A+1}\sum_{\text{spin}}|\mathcal{\tilde{M}}|^2, 
        \label{eq:Gamma-1}
    \end{split}  
\end{align}
where $\boldsymbol{k}$ is the three-vector momentum of the final-state meson in the initial state rest frame, and $J_A$ is the spin of the initial state. $E_B$ and $E_C$ are the energies of the final states $B$ and $C$, respectively. 

By redefining 
\begin{equation}
    \mathcal{{M}}\equiv 8\pi^{3/2} (M_AE_BE_C)^{1/2}\mathcal{\tilde{M}} \ ,
\end{equation}
where $\mathcal{{M}}$ is the transition matrix element defined at the hadronic level, we can unify all the transition amplitudes as follows:
\begin{align}
    \begin{split}
        \mathcal{S}&=1+i\mathcal{T}=1+(2\pi)^4i\delta^4(P_A-P_B-P_C)\mathcal{M} \ .
    \end{split}
\end{align}
In this convention, the expression for the decay width is:
\begin{align}
    \begin{split}
        \Gamma(A\to B+C)=\frac{|\boldsymbol{k|}}{8\pi M_A}\frac{1}{2J_A+1}\sum_{\text{spin}}|\mathcal{M}|^2 \ .
        \label{eq:Gamma-2}
    \end{split}
\end{align}

The total amplitude considering various processes can be expressed as follows
\begin{align}  
    \begin{split}  
        \mathcal{M}_T^{\text{(P)}}=\mathcal{M}_{\text{DPE}}^{\text{(P)}}+\mathcal{M}_{\text{CS1}}^{\text{(P)}}+\mathcal{M}_{\text{CS2}}^{\text{(P)}}+e^{i\theta_S}\mathcal{M}_{\text{PT}}^{\text{(P)}},  
    \end{split}  
\end{align}
where the superscript $(\text{P}) = (\text{PC})$ and $(\text{PV})$ denotes the parity of the amplitude. The introduction of the phase $\theta_S$ is due to the fact that the the inclusion of long-range strong interaction dynamics in the pole terms. Since both DPE and CS processes involve the $``1\to 3"$ weak transition at the quark level, their relative phase can be determined by SU(3) flavor symmetry. Additionally, for the decay $\Lambda \to n\pi^0$, due to the rescattering contribution from $p\pi^-$, its total amplitude can be expressed as
\begin{align}  
    \begin{split}          \mathcal{M}^{\text{(P)}}=\mathcal{M}_T^{\text{(P)}}+e^{i\theta_L}\mathcal{M}_L^{\text{(P)}},  
    \end{split}  
\end{align} 
where phase $\theta_L$ is a natural phase between the tree-level and loop-level diagrams. The phase of each vertex in the loop is determined by the SU(3) flavor symmetry. It is noteworthy that the SU(3) flavor symmetry plays a crucial role for constraining these amplitudes.

We can also calculate the asymmetry parameters in our model. For the interest of understanding the transition mechanisms, we will focus on 
\begin{align}
    \begin{split}
        \alpha=\frac{2\text{Re}(A^*B)}{|A|^2+|B|^2},
    \end{split}
\end{align}
where $A$ and $B$ represent the $S$-wave (PV) and $P$-wave (PC) amplitudes, respectively. It is obvious that $\alpha$ is bounded by $-1\leq\alpha\leq 1$ and characterizes the relative strength between the PC and PV transitions.

\section{numerical results and discussions}\label{numerical-results}

In this work, we use $m_q$ and $m_s$ to denote the masses of the light quarks $u \ (d)$ and the strange quark $s$, respectively. The constituent masses for the $u \ (d)$ and $s$ quarks are different. As a consequence, the effective couplings extracted from the quark-level interactions will contain the wavefunction convolutions. The difference arising from the constituent quark masses  will serve as one of the sources of the SU(3) flavor symmetry breaking. Considering that the hadrons involved are either ground states or first orbital/radial excitations, it is reasonable to describe their spatial distributions using harmonic oscillator (H.O.) wavefunctions. For the nucleon, the H.O. parameters for both modes are the same and denoted by $\alpha$. For the strange baryons, the H.O. parameters for the $\rho$-mode and $\lambda$-mode are denoted by $\alpha_{\rho}$ and $\alpha_{\lambda}$, respectively. However, these parameters are not independent. They satisfy the relation $\alpha_{\lambda} = \left( \frac{3m_s}{2m_q + m_s} \right)^{1/4} \alpha_{\rho}$. The H.O. parameter for the final-state pion is denoted by $R$. Additionally, the pole term involves strong interaction associated with pion emission. For the processes $N \to N\pi$, $\Lambda \to \Sigma\pi$ and $\Sigma \to \Sigma\pi$, we treat the pion as a fundamental field and employ the NRCQM to extract the strong coupling constants for these processes. However, considering the mass difference between the nucleon and the strange baryons, we introduce SU(3) breaking parameters $C_{NN\pi}$, $C_{\Lambda\Sigma\pi}$ and $C_{\Sigma\Sigma\pi}$ here. Given that the nucleon $N$ is a system composed of light quarks, and the $NN\pi$ coupling has matched to the physical value, we set $C_{NN\pi}=1$ and the the deviation of the other two parameters from 1 indicates the degree of SU(3) flavor symmetry breaking.

To proceed, we adopt the following numerical study strategies to reveal the underlying mechanisms. First, we mention that in total there are 10 parameters to be fitted by 10 data points, i.e. 5 branching ratios and 5 asymmetry parameters. These 10 parameters include 3 constituent quark masses ($m_u$, $m_d$, and $m_s$), 3 H.O. strengths ($\alpha$, $\alpha_\rho$, and $R$), 2 SU(3) flavor symmetry breaking parameters ($C_{\Lambda\Sigma\pi}$ and $C_{\Sigma\Sigma\pi}$), and 2 phase angles ($\theta_S$ and $\theta_{\Lambda}$). Note that $m_u=m_d$ is a good approximation.  Thus, we actually have 9 parameters to be fitted.

In fitting Scheme-I we leave these 9 parameters free and to be fitted by the experimental data. In Scheme-II we fixed the constituent quark masses with the values fitted in Scheme-I, and carry out the fitting by excluding the  $\Lambda\to n\pi^0$ channel. This will allow us to figure out that the decay mechanism of $\Lambda\to n\pi^0$ is indeed different from other processes. In Scheme-III we carry out the fitting with $\Lambda\to n\pi^0$ included and for which the FSIs is also considered.

\subsubsection{Scheme-I: Global fitting}

We first perform a fitting of all the hadronic weak decay channels with the  DPE, CS and pole terms. As discussed earlier, the combined analysis will reveal the correlations among these channels and also expose the deficit if there are important pieces of dynamic information missing here. 

In Tab.~\ref{tab:scheme-I parameters} the fitted parameters are listed. It shows that these fitted parameters have values well within the expectations of the NRCQM. The relatively larger values of $R$ for the pion wavefunction is consistent with what found in the literature~\cite{Kokoski:1985is}. The deviations of $C_{\Lambda\Sigma\pi}$ and $C_{\Sigma\Sigma\pi}$ from unity suggest the SU(3) flavor symmetry breaking effects. Note that these two coefficients are introduced for the pole terms where destructive interference occurs between type-A and type-B processes. The scale of deviations are still within the typical range of the SU(3) flavor symmetry breaking. Supposing that there are only local electro-weak transitions via Figs.~\ref{fig:Lambda DPE and CS} and \ref{fig:Sigma DPE and CS}, it has been found in the literature that the SU(3) flavor symmetry breaking effects are unexpectedly large and hard to understand. The parameters defined here, $C_{\Lambda\Sigma\pi}$ and $C_{\Sigma\Sigma\pi}$, will pick up the SU(3) flavor symmetry breaking effects arising from the type-A and type-B terms. Due to the cancellations among these two types of amplitudes, it is quite impossible to judge the SU(3) flavor symmetry breaking effects by directly comparing the partial widths among the $\Lambda$ and $\Sigma^\pm$ hadronic weak decays.

\begin{table}
    \centering
    \caption{Under the fitting scheme-I, the values of the parameters being fitted.}
    \begin{tabular}{c|cccccccccccccccccc}
        \hline\hline
         Parameters   &$m_{u/d}$        &$m_s$       &$\alpha$         &$\alpha_{\rho}$     &$R$              \\
        \hline
        Values(GeV)  &$0.3\pm 0.05$   &$0.55\pm 0.24$    &$0.498\pm 0.021$        &$0.444\pm 0.036$           &$0.616\pm 0.014$ \\
        \hline
        &fitted &fitted &fitted &fitted &fitted\\
        \hline\hline
        Parameters   &$C_{NN\pi}$    &$C_{\Lambda\Sigma\pi}$   &$C_{\Sigma\Sigma\pi}$      &$\theta_{S}$       &$\theta_{\Lambda}$\\
        \hline
        Values &1              &$1.3\pm 0.03$                      &$0.75\pm 0.06$                        &$218.0^{\circ}\pm2.63^{\circ}$      &$316.5^{\circ}\pm0.89^{\circ}$\\
        \hline
        &fixed &fitted &fitted &fitted &fitted\\
        \hline\hline
    \end{tabular}
    \label{tab:scheme-I parameters}
\end{table}

\begin{table}
    \centering
    \caption{Under the fitting scheme-I, the best fitting results of the BR and AP for the decays $\Sigma^{\pm}\to N\pi$ and $\Lambda\to N\pi$.}
    \scalebox{0.6}{
    \begin{tabular}{c|c|c|c|c|c|c}
        \hline\hline
                                       &                                             &$\Sigma^-\to n\pi^-$                                           &$\Sigma^+\to p\pi^0$                                  &$\Sigma^+\to n\pi^+$ &$\Lambda\to p\pi^-$ &$\Lambda\to n\pi^0$\\
        \hline
        \multirow{4}{*}{BR(in$\%$)}    & \multirow{2}{*}{Expt.}                                       &\multirow{2}{*}{$99.849\pm 0.005$\cite{ParticleDataGroup:2024cfk}}              &\multirow{2}{*}{$51.57\pm 0.3$\cite{ParticleDataGroup:2024cfk}}        &\multirow{2}{*}{$48.31\pm 0.3$\cite{ParticleDataGroup:2024cfk}} &\multirow{2}{*}{$64.1\pm 0.5$\cite{ParticleDataGroup:2024cfk}} &$35\pm 5$\cite{Brown:1963zz}\\
        &&&&&&$29.1\pm 3.4$\cite{Chretien:1963zz}\\
        &Ours ($\bs{without}$ $\Lambda(1405)$)&$66.443\pm 8$ &$50.76\pm 9$ &$26.33\pm 5$ &$65.78\pm 6$   &$6.94\pm 2.1$\\
                                       &Ours ($\bs{with}$ $\Lambda(1405)$)          &$99.857\pm 21$                                                       &$50.76\pm 9$               &$48.38\pm 6$ &$65.78\pm 6$   &$6.94\pm 2.1$\\
        \hline
        \multirow{4}{*}{AP}            &\multirow{2}{*}{Expt.}                       &\multirow{2}{*}{$\alpha_-=-0.068\pm0.008$~\cite{ParticleDataGroup:2024cfk}}                                            &$\alpha_0=-0.998\pm0.037\pm0.009$~\cite{BESIII:2020fqg}                 &\multirow{2}{*}{$\alpha_+=0.0481\pm0.0031\pm0.0019$~\cite{BESIII:2023sgt}} &\multirow{2}{*}{$\alpha_-=0.757\pm0.011\pm0.008$~\cite{BESIII:2021ypr}} &$\bar{\alpha}_0=-0.692\pm0.016\pm0.006$~\cite{BESIII:2018cnd}\\
                                       &                                             &                                                                                                                      &$\alpha_0=\alpha_+/(-0.049\pm0.0032\pm0.0021)$~\cite{BESIII:2023sgt}    & & &$\alpha_0=(1.000\pm 0.068)\alpha_-$~\cite{Olsen:1970vb}\\
                                       &Ours ($\bs{without}$ $\Lambda(1405)$) &$-0.227\pm 0.005$ &$-0.593\pm 0.021$ &$-0.891\pm 0.031$ &$0.676\pm 0.05$   &$-0.055\pm 0.4$\\
                                       &Ours ($\bs{with}$ $\Lambda(1405)$)          &$-0.020\pm 0.016$                                                       &$-0.593\pm 0.021$             &$0.0397\pm $0.025 &$0.676\pm 0.05$      &$-0.055\pm 0.4$\\
        \hline\hline
    \end{tabular}}  
    \label{tab:scheme-I BR and AP} 
\end{table}

In Tab.~\ref{tab:scheme-I BR and AP} the fitting results of Scheme-I are presented. A less determined issue is the role played by the $J^P=1/2^-$ states in the pole terms in the PV processes. In $\Sigma^\pm\to n\pi^\pm$, $\Lambda(1405)$ can contribute through the PV pole terms. Note that this state does not contribute to other processes. To some extent, the inclusion of the intermediate $J^P=1/2^-$ states in the pole terms will also provide some constraints on these resonances. This is a unique feature arising from the hyperon weak decays. For $\Sigma^\pm\to n\pi^\pm$ it shows that whether to include the contributions from $\Lambda(1405)$ can cause a difference. In $\Lambda\to n\pi^0$ the inclusion of the $J^P=1/2^-$ states does not improve the overall fitting. The theoretical values of the BR and asymmetry parameter both have underestimated the experimental data. This deficit cannot be accounted for within the reasonable ranges of the parameters. The fitting results in Scheme-I, on the one hand, indicate that the pole terms are essential for understanding the underlying mechanisms, on the other hand, reveals the absence of key mechanisms in $\Lambda\to n\pi^0$.

\subsubsection{Scheme-II: Fitting excluding $\Lambda\to n\pi^0$ }

In order to further clarify the situation of  $\Lambda\to n\pi^0$, we carry out the fitting by excluding  the channel of $\Lambda\to n\pi^0$. Because of the reduced number of data points, we fix the quark masses to be the central values from Scheme-I. Still, we fix $C_{NN\pi}=1$ for the non-strange coupling. 

The fitted parameters are listed in Tab.~\ref{tab:scheme-II parameters}. It shows that the fitted values for the rest parameters are not dramatically different from those determined in Scheme-I. It suggests that the channel $\Lambda\to n\pi^0$ should have some unique property which is not strongly correlated with other channels.

The best fitting results of the BR and AP for the four channels $\Sigma^{\pm}\to n\pi^\pm$, $\Sigma^+\to p\pi^0$, and $\Lambda\to p\pi^-$ are presented in Tab. \ref{tab:scheme-II BR and AP}. To highlight the impact of $\Lambda(1405)$ in the $\Sigma^\pm$ decays, we also present the results {\it without} and {\it with} the contributions from $\Lambda(1405)$ as a comparison. It is quite apparent that the inclusion of $\Lambda(1405)$ is necessary for obtaining a good fitting result.

For these four decay channels, it shows that with contributions from the DPE, CS, and pole terms, one can obtain a reasonable description of the BR and AP. In particular, the signs of all the APs have been correctly described. In Tab. \ref{tab:scheme-II BR and AP}, the experimental value of $\alpha_0$ is nearly equal to $-1$. In contrast, $\alpha_{\pm}$ is very small for $\Sigma^\pm$ decays. These values indicate that the angular momentum states of the final pion-nucleon  are dominated either by the $S$-wave or by the $P$-wave. 

To help understand the underlying dynamics, we list the $S$-wave (PV) and $P$-wave (PC) amplitudes of each diagrams for $\Sigma^\pm\to N\pi^\pm$ and $\Lambda\to N\pi$ in Tabs. \ref{tab:amplitudes for Sigma} and \ref{tab:amplitudes for Lambda}, respectively. 
As shown in Tab. \ref{tab:amplitudes for Sigma}, the pole term contributions are mainly from the ground-state and first orbital excited state baryons. In contrast, the contributions from the higher excited state decreases significantly, and become negligible as expected.

In Tab.~\ref{tab:amplitudes for Lambda} one notices that $N(1710)$, which belongs to the radial excitation state of representation $|70,^28\rangle$, does not contribute to the PC amplitude in the type-A transition. This is due to an exact cancellation between the two terms involving different spatial wavefunction convolutions. In Tabs. \ref{tab:IC fla} and \ref{tab:IC spin}, there are two non-zero spin-flavor terms: $\langle \phi_N^{\lambda}|\hat{\alpha}_2^{(-)}\hat{\beta}_3^{(+)}|\phi_{\Lambda}^{\rho}\rangle\langle\chi_{\frac12,-\frac12}^{\rho}|\mathcal{O}_{\text{spin}}^{\text{(PC)}}|\chi_{\frac12,-\frac12}^{\rho}\rangle$ and $\langle \phi_N^{\lambda}|\hat{\alpha}_2^{(-)}\hat{\beta}_3^{(+)}|\phi_{\Lambda}^{\rho}\rangle\langle\chi_{\frac12,-\frac12}^{\lambda}|\mathcal{O}_{\text{spin}}^{\text{(PC)}}|\chi_{\frac12,-\frac12}^{\rho}\rangle$. After including these two factors, it occurs that the respective spatial wavefunction convolutions of $\psi_{200}^{\rho}$ and $\psi_{200}^{\lambda}$ (from state $N(1710)$) with the wavefunction $\psi_{000}^s$ of $\Lambda$ have the same magnitude but with different signs. Thus, they cancel exactly in the NRCQM with the spin-flavor SU(6) symmetry. This manifests a dynamic selection rule on the basis of the NRCQM. It gives a good example that rich phenomenological information can be gained based on the unified approach at the quark level.

It should be clarified that in  Tab.~\ref{tab:amplitudes for Lambda} the exclusive PV amplitudes with $J^P=1/2^-$ from the quark model representations $|70,^28\rangle$ and $|70,^48\rangle$ are to be compared with the exclusive amplitudes from their mixings, i.e. the physical states, $N(1535)$ and $N(1650)$. The results taken into account the state mixing for $N(1535)$ and $N(1650)$ do not lead to significant changes although the change to each amplitude is significant. This is because of the destructive phase between these two amplitudes. With the state mixing defined as follows,
\begin{equation}
    \left( 
    \begin{array}{c}
    N(1535) \\
    N(1650)
    \end{array}\right)
    =
    \left(
    \begin{array}{cc}
    \cos\theta_N & -\sin\theta_N \\
    \sin\theta_N & \cos\theta_N 
    \end{array}\right)
    \left( 
    \begin{array}{c}
    |70,^28\rangle\\
    |70,^48\rangle
    \end{array}\right) \ ,
\end{equation}
where the mixing angle $\theta_N=26^\circ$ is determined by other processes (see e.g. Refs.~\cite{Zhong:2024mnt}), one finds that the sum of these two amplitudes turns out to be comparable. 

The values of the AP for $\Sigma^{\pm}\to n\pi^{\pm}$ are essentially zero. It implies that the angular momentum state of the final pion-nucleon is highly dominated by either $S$-wave or $P$-wave. Our calculation listed in Tab. \ref{tab:amplitudes for Sigma} shows that these two channels are dominated by the $S$-wave (PV) transition.  Particularly, it shows that the flavor singlet $\Lambda(1405)$ is crucial for producing the small absolute values of the AP in $\Sigma^{\pm}\to n\pi^{\pm}$ as a dominant amplitude in the PV terms. Note that there exists a sign difference between the AP values in $\Sigma^-\to n\pi^-$ and $\Sigma^+\to n\pi^+$. This can be well understood in the quark model framework. Since  the type-A $J^P=1/2^-$ amplitudes from the $N|70, ^28\rangle]$ and $N|70, ^48\rangle]$ states only contribute to the decay of $\Sigma^+\to n\pi^+$, but are forbidden in $\Sigma^-\to n\pi^-$, its interference in $\Sigma^+\to n\pi^+$ results in the change of the AP sign, but keeps the absolute values small in both cases.

\begin{table}
    \centering
    \caption{The values of the parameters for the fitting scheme-II in this work.}
    \begin{tabular}{c|cccccccccccccccccccc}
        \hline\hline
         Parameters        &$m_{u/d}$      &$m_s$       &$\alpha$         &$\alpha_{\rho}$     &$R$              \\
        \hline
        Values(GeV)   &$0.3$  &$0.55$    &$0.506\pm 0.003$        &$0.447\pm 0.001$           &$0.619\pm 0.002$ \\
        \hline
        &fixed &fixed &fitted &fitted &fitted\\
        \hline\hline
        Parameters   &$C_{NN\pi}$    &$C_{\Lambda\Sigma\pi}$   &$C_{\Sigma\Sigma\pi}$      &$\theta_{S}$       &$\theta_{\Lambda}$\\
        \hline
        Values &1              &$1.25\pm 0.008$                      &$0.7\pm 0.003$                        &$219.0^{\circ}\pm0.68^{\circ}$      &$316.6^{\circ}\pm0.32^{\circ}$\\
        \hline
        &fixed &fitted &fitted &fitted &fitted\\
        \hline\hline
    \end{tabular}
    \label{tab:scheme-II parameters}
\end{table}

\begin{table}
    \centering
    \caption{Under the fitting scheme-II, the best fitting results of the BR and AP for the decays $\Sigma^{\pm}\to N\pi$ and $\Lambda\to N\pi$.}
    \scalebox{0.75}{
    \begin{tabular}{c|c|c|c|c|c}
        \hline\hline
                                       &                                             &$\Sigma^-\to n\pi^-$                                           &$\Sigma^+\to p\pi^0$                                  &$\Sigma^+\to n\pi^+$ &$\Lambda\to p\pi^-$\\
        \hline
        \multirow{3}{*}{BR(in$\%$)}    & Expt.                                       &$99.849\pm 0.005$\cite{ParticleDataGroup:2024cfk}              &$51.57\pm 0.3$\cite{ParticleDataGroup:2024cfk}        &$48.31\pm 0.3$\cite{ParticleDataGroup:2024cfk}&$64.1\pm 0.5$\cite{ParticleDataGroup:2024cfk}\\
                                       &Ours ($\bs{without}$ $\Lambda(1405)$)        &$62.102\pm 2$                                                       &$51.43\pm 1.6$                                               &$30.28\pm 0.8$   &$62.85\pm 0.8$\\
                                       &Ours ($\bs{with}$ $\Lambda(1405)$)           &$99.834\pm 5$                                                       &$51.43\pm 1.6$                                               &$48.56\pm 1.2$    &$62.85\pm 0.8$\\
        \hline
        \multirow{4}{*}{AP}            &\multirow{2}{*}{Expt.}                       &\multirow{2}{*}{$\alpha_-=-0.068\pm0.008$\cite{ParticleDataGroup:2024cfk}}                                            &$\alpha_0=-0.998\pm0.037\pm0.009$\cite{BESIII:2020fqg}                 &\multirow{2}{*}{$\alpha_+=0.0481\pm0.0031\pm0.0019$\cite{BESIII:2023sgt}} &\multirow{2}{*}{$\alpha_-=0.757\pm0.011\pm0.008$\cite{BESIII:2021ypr}} \\
                                       &                                             &                                                                                                                      &$\alpha_0=\alpha_+/(-0.049\pm0.0032\pm0.0021)$\cite{BESIII:2023sgt}    &&\\
                                       &Ours ($\bs{without}$ $\Lambda(1405)$)        &$-0.256\pm 0.001$                                                       &$-0.711\pm 0.004$                                              &$-0.960\pm 0.003$    &$0.702\pm 0.006$\\
                                       &Ours ($\bs{with}$ $\Lambda(1405)$)           &$-0.057\pm 0.004$                                                       &$-0.711\pm 0.004$                                              &$0.0758\pm 0.007$    &$0.702\pm 0.006$\\
        \hline\hline
    \end{tabular}}  
    \label{tab:scheme-II BR and AP}  
\end{table}

\begin{table}
    \centering
    \caption{The amplitudes of DPE, CS and pole terms with $J_f^z=J_i^z=-1/2$ for $\Lambda\to N\pi$ and the unit is $10^{-9}\ \text{GeV}^{-1/2}$. The parity-violating intermediate states $\Sigma(1620)$ and $\Sigma(1750)$ are the first orbital excited states $|70,^28\rangle$ and $|70,^48\rangle$,respectively. The parity-violating intermediate states $N(1535)$ and $N(1650)$ also can be regarded as pure $|70,^28\rangle$ and $|70,^48\rangle$ states, or they are the mixing states of $|70,^28\rangle$ and $|70,^48\rangle$ with a widely used mixing angle $\theta_N=26^\circ$. As can be seen from this table, the two options have a relatively small impact on the total amplitude after addition. In this work, we adopt the former option. } 
    \begin{tabular}{ccccccccccccccc}
        \hline\hline
        (PC)                   &DPE       &CS1       &CS2        &A[$n(939)$]             &A[$N(1440)$]             &A[$N(1710)$]      &B[$\Sigma(1193)$]    &B[$\Sigma(1660)$]   &B[$\Sigma(1880)$]\\
        \hline
        $\Lambda\to p\pi^-$    &$10.11$    &$0.61$    &-          &$25.75$                 &$0.95+0.58i$             &0                 &$-32.39$             &$-1.39-0.24i$       &$0.014+0.003i$\\
        $\Lambda\to n\pi^0$    &-         &$-0.44$   &$-2.46$    &$-18.89$                &$-0.69-0.42i$            &0                 &$23.59$              &$1.02+0.18i$        &$-0.010-0.002i$\\
        \hline
        (PV)                   &DPE       &CS1       &CS2        &A[$N|70,^28\rangle$]    &A[$N|70,^48\rangle$]     &A[$N(1535)$]      &A[$N(1650)$]         &B[$\Sigma(1620)$]   &B[$\Sigma(1750)$]\\
        \hline
        $\Lambda\to p\pi^-$    &$-40.14$   &$-4.39$   &-          &$24.10+3.95i$          &$-18.52-2.02i$           &$28.36+4.65i$    &$-21.97-2.39i$       &$10.92+0.41i$       &$-32.46-1.23i$\\
        $\Lambda\to n\pi^0$    &-         &$3.10$    &$9.46$     &$-16.92-2.77i$          &$13.00+1.42i$            &$-19.91-3.26i$    &$15.42+1.68i$        &$-7.71-0.29i$      &$22.91+0.87i$\\
        \hline\hline
    \end{tabular}
    \label{tab:amplitudes for Lambda}
\end{table}

\begin{table}
    \centering
    \caption{The amplitudes of DPE, CS and pole terms with $J_f^z=J_i^z=-1/2$ for $\Sigma^{\pm}\to N\pi$ and the unit is $10^{-9}\ \text{GeV}^{-1/2}$. Using the quark model classification, the first orbital excited states $N(1535)/\Sigma(1620)/\Lambda(1670)$ and $N(1650)/\Sigma(1750)/\Lambda(1800)$ are octet baryons $|70,^28\rangle$ and $|70,^48\rangle$, respectively. $\Lambda(1405)$ is classified as a singlet baryon $|70,^21\rangle$.} 
    \scalebox{0.75}{
    \begin{tabular}{ccccccccccccccc}
        \hline\hline
        (PC)                   &DPE       &CS-1       &CS-2        &A[$n(939)$]             &A[$N(1440)$]             &A[$N(1710)$]      &B[$\Sigma(1193)$]    &B[$\Lambda(1116)$]  &B[$\Sigma(1660)$]    &B[$\Lambda(1600)$]    &B[$\Sigma(1880)$]    &B[$\Lambda(1810)$]\\
        \hline
        $\Sigma^-\to n\pi^-$   &$-5.18$     &$-0.94$     &-          &-                       &-                        &-                 &$23.98$                &$29.85$              &$1.52+0.26i$         &$-1.54-0.31i$         &$-0.04-0.01i$       &$0.32+0.03i$\\
        $\Sigma^+\to p\pi^0$   &-         &$-0.65$     &$-1.20$      &$-49.46$                  &$-3.59-2.75i$            &$0.70+0.11i$      &$33.83$                &-                   &$2.07+0.36i$         &-                     &$-0.05-0.01i$       &-\\
        $\Sigma^+\to n\pi^+$   &-         &-         &-          &$68.59$                   &$4.99+3.82i$             &$-0.98-0.15i$     &$-23.31$               &$-29.02$              &$-1.43-0.25i$        &$-1.45-0.29i$         &$0.20+0.04i$        &$0.31+0.03i$\\
        \hline
        (PV)                   &DPE       &CS-1       &CS-2        &A[$N(1535)$]            &A[$N(1650)$]             &B[$\Sigma(1620)$]      &B[$\Lambda(1670)$]    &B[$\Sigma(1750)$]   &B[$\Lambda(1800)$]  &B[$\Lambda(1405)$]\\
        \hline
        $\Sigma^-\to n\pi^-$   &$-32.18$    &$-10.65$    &-          &-                       &-                        &$-15.12-0.56i$         &$4.10+0.11i$          &$-8.99-0.34i$       &$-19.20-2.20i$      &$-42.61-2.74i$\\
        $\Sigma^+\to p\pi^0$   &-         &$-7.53$     &$-7.59$      &$-49.59-9.58i$          &$39.32+4.82i$            &$-21.08-0.78i$         &-                     &$-12.53-0.48i$      &-                   &-\\
        $\Sigma^+\to n\pi^+$   &-         &-         &-          &$70.33+13.58i$          &$-55.77-6.84i$           &$14.89+0.55i$          &$4.04+0.11i$          &$8.84+0.34i$        &$-18.90-2.17i$      &$-41.94-2.70i$\\
        \hline\hline
    \end{tabular}}
    \label{tab:amplitudes for Sigma}
\end{table}

\subsubsection{Scheme-III: Including the FSIs in $\Lambda\to n\pi^0$}

For the decay channel $\Lambda\to n\pi^0$, we find that the theoretical predictions still deviate significantly from the experimental results with the CS and pole terms considered. The large difference in branching ratios between $p\pi^-$ and $n\pi^0$ is mainly because of the contribution from the DPE process in the $p\pi^-$ channel. It plays a dominant role in $\Lambda \to p\pi^-$ if regarding the type-A and type-B pole terms cancel each other. As shown in Tab. \ref{tab:amplitudes for Lambda}, the DPE amplitude is much larger than the CS which leads to an imbalance in the branching ratios of $p\pi^-$ and $n\pi^0$. 

Much larger amplitude for $\Lambda \to p\pi^-$ than $\Lambda \to n\pi^0$, and the closeness of these two threshold suggests that the FSI contributions from $\Lambda \to p\pi^-\to n\pi^0$ can be significant. It is unnecessary to be the case for $\Lambda \to n\pi^0\to p\pi^-$ because of the loop suppression. Therefore, it is reasonable to include the FSIs in $\Lambda \to n\pi^0$ as the leading correction. It is interesting to note that such a mechanism seems to be only significant in $\Lambda \to n\pi^0$ for both $\Lambda$ and $\Sigma^\pm$ hadronic weak decays. To be specific, one sees that although the decay of $\Sigma^-\to n\pi^-$ involves the DPE process, it has no coupled-channels to feed in. For $\Sigma^+\to p\pi^0$ and $n\pi^+$, they do not involve the DPE process and thus have comparable amplitudes. As the consequence, their coupled-channel contributions become subleading to each other due to the loop suppression. To some extent, the FSI effects seem to be a unique feature for $\Lambda\to n\pi^0$ within the $\Lambda$ and $\Sigma^\pm$ hadronic weak decays, and our numerical calculations have confirmed it.

Taking the strategy of including the FSIs in $\Lambda\to n\pi^0$ as an additional mechanism which only significantly contributes to $\Lambda\to n\pi^0$, we list the calculated  helicity amplitudes of each hadronic loop in Tab. \ref{tab:loop_amp} with the cutoff parameter $\beta=1$, 2 and 2.5 adopted. Here, the form factor has the following form: $\mathcal{F}(p_i)=(\Lambda_i^2-m_i^2)/(\Lambda_i^2-p_i^2)$ and $\Lambda_i=m_i+\beta\Lambda_{\text{QCD}}$. The results in Tab. \ref{tab:loop_amp} show that the PV amplitude through exchanging $\rho^-$ in the triangle loop is dominant. Meanwhile, the PC amplitude given by the intermediate neutron in the two-point bubble loop is dominant. These two loop amplitudes are at the same order of magnitude as those tree-level amplitudes listed in Tab.~\ref{tab:amplitudes for Lambda}. 

\begin{figure}
    \centering
    \includegraphics[width=0.5\linewidth]{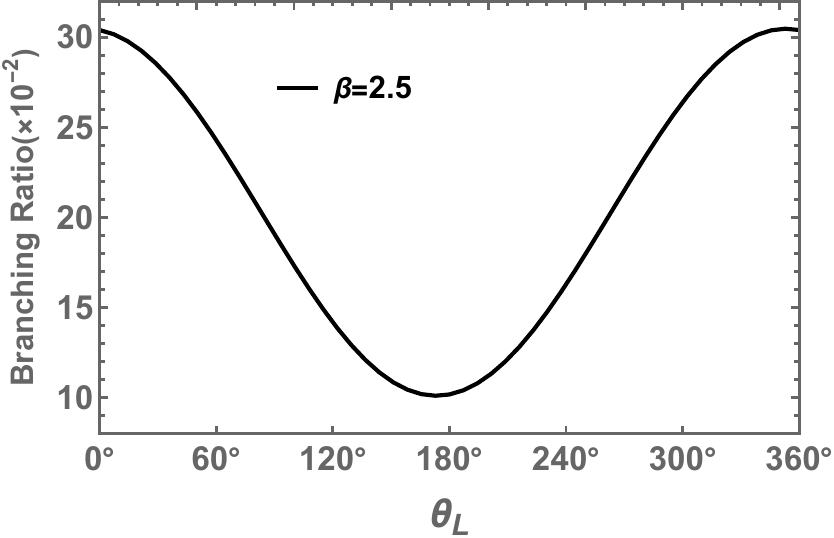}
    \caption{When $\beta=2.5$, the branching ratio of $\Lambda\to n\pi^0$ varies with the phase $\theta_L$ between the tree-level and loop-level amplitude.}
    \label{fig:theta_L}
\end{figure}

Figure \ref{fig:theta_L} indicates that the FSI contributions from the $p\pi^-$ channel can indeed enhance the decay branching ratio of $n\pi^0$. As a test of the range of the interference effects, we introduce a phase factor $e^{i\theta_L}$ for the FSI term via the $p\pi^-\to n\pi^0$ rescatterings, where $\theta_L$ is the phase angle. Interestingly, it shows that $\theta_L=0^\circ$, i.e. a natural phase, is favored, and can provide the largest interference. As shown in Tab. \ref{tab:BR and AP for Lambda}, after considering FSIs, the branching ratio and asymmetry parameter of $\Lambda\to n\pi^0$ agree well with the experimental measurements, and the results are quite insensitive to the cutoff parameter $\beta$.

We clarity that in this Scheme we do not carry out an overall fitting with the FSIs considered for $\Lambda\to n\pi^0$. One reason is due to the limited number of data. More importantly, as mentioned earlier, the other decay processes of $\Lambda$ and $\Sigma^\pm$ do not receive the FSI contributions via the coupled-channel rescatterings at the leading order. This feature actually makes Scheme-III useful for testing the role of the FSIs on the one hand, and on the other hand, justifies the mechanism as the last missing piece of information in the unified description of the $\Lambda\to n\pi^0$ partial width.

\begin{table}
    \centering
    \caption{The loop amplitudes with $J_f^z=J_i^z=-1/2$ for $\Lambda\to n\pi^0$ and the unit is $10^{-9}\ \text{GeV}^{-1/2}$.}
    \begin{tabular}{c|c|c|c}
          \hline\hline 
          $\mathcal{M}_L^{-\frac12,-\frac12}$ &$\beta=1$ &$\beta=2$ &$\beta=2.5$\\
          \hline
          $[t,(\text{PC}),\pi^-,p,(\rho^-)]$&$-0.46-0.18i$&$-1.13-0.28i$&$-1.64-0.31i$  \\
          $[t,(\text{PC}),p,\pi^-,(p)]$&$-0.03-0.04i$&$0.01-0.05i$&$0.08-0.05i$  \\
          $[s,(\text{PC}),p,\pi^-,(n)]$&$-2.25+0.58i$&$-8.76+0.58i$&$-13.56+0.58i$  \\
          $[s,(\text{PC}),p,\pi^-,(N(1440))]$&$-0.15-0.05i$&$-0.53-0.27i$&$-0.80-0.44i$  \\
         \hline
          $[t,(\text{PV}),\pi^-,p,(\rho^-)]$&$1.57+1.66i$&$3.63+2.52i$&$4.51+2.80i$  \\
          $[t,(\text{PV}),p,\pi^-,(p)]$&$0.26+0.20i$&$0.60+0.25i$&$0.77+0.27i$  \\
          $[s,(\text{PV}),p,\pi^-,(N(1535))]$&$0.42+0.56i$&$0.74+0.63i$&$0.86+0.65i$  \\
          $[s,(\text{PV}),p,\pi^-,(N(1650))]$&$0.36+0.42i$&$0.61+0.45i$&$0.71+0.47i$  \\
         \hline\hline
    \end{tabular}
    \label{tab:loop_amp}
\end{table}

\begin{table}
    \caption{For $\Lambda\to n\pi^0$, the results with FSIs corresponding to the cutoff parameter $\beta=2$ and $\beta=2.5$. $\bar{\alpha}_0$ is the asymmetry parameter of the charge conjugate decay of $\Lambda\to n\pi^0$.}
    \begin{tabular}{cc|c|c}
    \hline\hline
         & &BR($\%$) &AP  \\
    \hline
    \multirow{2}{*}{Expt.}  &   &$35\pm 5$\cite{Brown:1963zz}  &$\bar{\alpha}_0=-0.692\pm0.016\pm0.006$\cite{BESIII:2018cnd}\\
    & &$29.1\pm 3.4$\cite{Chretien:1963zz} &$\alpha_0=(1.000\pm 0.068)\alpha_-$\cite{Olsen:1970vb}\\
    \hline
    \multirow{2}{*}{Ours}   &$\beta=2$ &18.95 &0.737\\
                            &$\beta=2.5$ &30.41 &0.701\\
    \hline\hline
    \end{tabular} 
    \label{tab:BR and AP for Lambda}
\end{table}

\section{Summary}\label{sec-summary}

In this work we present a unified approach for the hadronic weak decays of $\Lambda$ and $\Sigma^\pm\to N\pi$. In the framework of the NRCQM, we obtain a self-consistent description of the external $W$ emissions (i.e. DPE), internal $W$ emissions (i.e. CS), and the internal $W$ conversions (i.e. the pole terms). One advantage is that these transition processes can be unified by the same set of quark model parameters with fixed relative phases. Meanwhile, the SU(3) flavor symmetry breaking effects can be recognized by the quark model wavefunction convolutions of which the quark mass differences have been included.  

The quark model framework allows us to figure out the role played by the pole terms in the internal conversion transitions. It shows that for the $\Lambda$ and $\Sigma^+$ decays there always exist cancellations among the pole terms for both PC and PV transition amplitudes. In particular, the cancellation between the type-A and type-B terms in the PC amplitude for $\Lambda$ is well known. Interestingly, the decay of $\Sigma^-\to n\pi^-$ does not involve type-A pole terms and has only contributions from the type-B transitions. It provides a clear evidence for the role played by the pole terms. Actually, without the pole terms, one cannot reproduce the experimental data with only contributions from the DPE and CS transitions. We also identify the crucial role played by the $J^P=1/2^-$ intermediate state $\Lambda(1405)$ in both $\Sigma^-\to n\pi^-$ and $\Sigma^+\to n\pi^+$. This seems to provide some constraints on $\Lambda(1405)$ from a special circumstance. Another interesting result is that we find that the radial excitation state $N(1710)$ of the SU(6)$\otimes$O(3) representation $|70, ^28, 2, 0, 1/2^+\rangle$ does not contribute to the type-A PC amplitudes because of a dynamic selection rule, which have not been identified before. Although the SU(6)$\otimes$O(3) symmetry should be broken, the leading approximation still provide important information about the structure of the intermediate baryons.

This combined analysis also allows us to identify the FSIs via the coupled-channel rescatterings serve as a leading correction to the transition amplitudes in $\Lambda\to n\pi^0$. Namely, the dominant decay channel of $\Lambda\to p\pi^-$ can contribute to  $\Lambda\to n\pi^0$ via the $p\pi^-\to n\pi^0$ rescatterings. This is a unique feature with FSIs are either subleading and negligible, or absent from the underlying dynamics. This explains the puzzling issue that the BR of $\Lambda\to n\pi^0$ is comparable with that of $\Lambda\to p\pi^-$. The approximate rate of $BR(\Lambda\to p\pi^-)\simeq 2\times BR(\Lambda\to n\pi^0)$ may be a coincident. However, the crucial role played by the pole terms and the unique contributions from the FSIs in $\Lambda\to n\pi^0$ indicate that some ``fine-tuning" mechanisms exist in the $\Lambda$ dynamics. 

In brief, based on the unified approach and systematic studies of the $\Lambda$ and $\Sigma^\pm$ hadronic weak decays, we have succeeded in understanding the details of the transition mechanisms for these hyperon states. By disentangling the common features and differences of these decay processes, we identify several key issues with the underlying dynamics, and provide an answer for the puzzling BR of $\Lambda\to n\pi^0$. This analysis will be useful for a better understanding of the role played by the hyperons in several hot topics, such as the hyperon interactions with nucleons and properties of hyper-nuclei, and the search for CP violation in hyperon decays.

\begin{acknowledgements}
This work is supported by the National Natural Science Foundation of China (Grant No. 12235018).
\end{acknowledgements}

\appendix
\begin{appendix}

\section{Quark model classification and the total wavefunctions}\label{wf-QM}

For a light baryon system, the color wavefunction $\mathcal{\psi}_c$ should be a color singlet under SU(3) symmetry. The flavor wavefunction $\phi$ can be constructed by the light $u$, $d$, $s$ quarks, which follows a flavor SU(3) symmetry. The spin wavefunction $\chi$ can be constructed as the eigen states of the quark spin and its projection along e.g. $z$ axis $s_z$, which follows the SU(2) symmetry. Combining together the spin and flavor symmetry, the spin-flavor wavefunctions follow an SU(6) symmetry and usually denoted by $|N_6,^{2\bm{S}+1}N_3\rangle$ in the literature, where $N_6$ and $N_3$ represent the dimensions of the SU(6) and SU(3) representations, respectively, and $\bm{S}$ stands for the quantum number of the total spin of a baryon state. The configurations of the spin-flavor wave functions under different symmetries are all listed in Table \ref{tab:SU6 wf}.

The spatial wavefunctions satisfy the $O(3)$ symmetry under a rotation transformation. For a baryon system containing three quarks, the spatial wavefunction $\psi_{NLL_z}^{\sigma}=[\psi_{n_{\rho l_{\rho}m_{\rho}}}(\boldsymbol{p}_{\rho})\otimes\psi_{n_{\lambda l_{\lambda}m_{\lambda}}}(\boldsymbol{p}_{\lambda})]_{NLL_z}$ is composed of $\rho$- and $\lambda$- mode spatial wavefunctions. The quantum numbers $n_{\rho}$, $l_{\rho}$ and $m_{\rho}$ [or $n_{\lambda}$, $l_{\lambda}$, $m_{\lambda}$] stand for that for the radial excitation, relative orbital angular momentum and its $z$ component for the $\rho$-mode [or $\lambda$-mode] wavefunction, respectively. While $N$, $L$ and $L_z$ stand for the principal quantum number, the quantum numbers of total orbital angular momentum and its $z$ component, respectively. They are defined by $N=2(n_{\rho}+n_{\lambda})+l_{\rho}+l_{\lambda}$, $|l_{\rho}-l_{\lambda}|\leq L\leq l_{\rho}+l_{\lambda}$ and $L_z=m_{\rho}+m_{\lambda}$. Furthermore, the superscript $\sigma$ appearing in wavefunctions represents their permutation symmetries.

\begin{table}[H]
  \centering
  \caption{The SU(6) spin-flavor wavefunctions of light baryons. $\chi$ and $\phi$ stand for the spin and flavor wavefunctions, respectively. The superscripts $s$, $a$, $\lambda$ and $\rho$ stand for the wavefunctions are symmetric, antisymmetric, mixed symmetric, and mixed antisymmetric, respectively.}
  \label{tab:SU6 wf}
  \begin{tabular}{cc|cc|cc}
      \hline\hline
      $|N_6,^{2\bm{S}+1}N_3\rangle$  &wave function                                                                 &$|N_6,^{2\bm{S}+1}N_3\rangle$    &wave function               &$|N_6,^{2\bm{S}+1}N_3\rangle$     &wave function\\
      \hline
      $|56,^28\rangle^s$             &$\frac{1}{\sqrt{2}}(\phi^{\rho}\chi^{\rho}+\phi^{\lambda}\chi^{\lambda})$     &$|70,^48\rangle^{\rho}$          &$\phi^{\rho}\chi^{s}$       &$|70,^21\rangle^{\rho}$           &$\phi^{a}\chi^{\lambda}$\\
      $|56,^410\rangle^s$            &$\phi^{s}\chi^{s}$                                                            &$|70,^48\rangle^{\lambda}$       &$\phi^{\lambda}\chi^{s}$    &$|70,^21\rangle^{\lambda}$        &$\phi^{a}\chi^{\rho}$\\
      $|70,^28\rangle^{\rho}$        &$\frac{1}{\sqrt{2}}(\phi^{\rho}\chi^{\lambda}+\phi^{\lambda}\chi^{\rho})$     &$|70,^210\rangle^{\rho}$         &$\phi^{s}\chi^{\rho}$       &$|20,^28\rangle^{a}$              &$\frac{1}{\sqrt{2}}(\phi^{\rho}\chi^{\lambda}-\phi^{\lambda}\chi^{\rho})$\\
      $|70,^28\rangle^{\lambda}$     &$\frac{1}{\sqrt{2}}(\phi^{\rho}\chi^{\rho}-\phi^{\lambda}\chi^{\lambda})$     &$|70,^210\rangle^{\lambda}$      &$\phi^{s}\chi^{\lambda}$    &$|20,^41\rangle^a$                &$\phi^{a}\chi^{s}$\\
      \hline\hline
  \end{tabular}
\end{table}

We adopt the simple harmonic oscillator (SHO) function to mimic the spatial wavefunction. And the $\rho$- (or $\lambda$-) mode spatial wavefunction can be written as
\begin{align}
    \begin{split}
        \psi_{nlm}(\boldsymbol{p})=(-1)^n(-i)^l\sqrt{\frac{2n!}{(n+l+1/2)!}}\frac{1}{\alpha^{l+3/2}}e^{-\frac{\boldsymbol{p}^2}{2\alpha^2}}L_n^{l+1/2}(\frac{\boldsymbol{p}^2}{\alpha^2})y_{lm}(\boldsymbol{p}),
    \end{split}
\end{align}
where $\alpha=\sqrt{m\omega}$ is a physical quantity related to the frequency of the harmonic oscillator, which can describe the spatial distribution of the wavefunction. The expressions for the spatial wavefunctions of baryons with $N\leq 2$ shells are shown in Table \ref{tab:baryon spatial wf}.

\begin{table}
  \centering
  \caption{The spatial wavefunctions of $N\leq 2$ shell $\psi_{NLL_z}^{\sigma}(\boldsymbol{p}_{\rho},\boldsymbol{p}_{\lambda})$ as the linear combination of $\psi_{n_{\rho}l_{\rho}m_{\rho}}(\boldsymbol{p}_{\rho})\otimes\psi_{n_{\lambda}l_{\lambda}m_{\lambda}}(\boldsymbol{p}_{\lambda})$. $\mathcal{N}=3^{3/4}$ is the normalization coefficient, due to the following normalization condition of the wavefunction: $\int_{-\infty}^{+\infty}d\boldsymbol{p}_1d\boldsymbol{p}_2d\boldsymbol{p}_3\delta^3(\boldsymbol{p}_1+\boldsymbol{p}_2+\boldsymbol{p}_3)|\psi(\boldsymbol{p}_1,\boldsymbol{p}_2,\boldsymbol{p}_3)|^2=1$, where $\psi(\boldsymbol{p}_1,\boldsymbol{p}_2,\boldsymbol{p}_3)=\mathcal{N}\psi_{n_{\rho}l_{\rho}m_{\rho}}(\boldsymbol{p}_{\rho})\psi_{n_{\lambda}l_{\lambda}m_{\lambda}}(\boldsymbol{p}_{\lambda})$.}
  \label{tab:baryon spatial wf}
  \begin{tabular}{cl}
      \hline\hline
      $\psi_{000}^s(\boldsymbol{p}_{\rho},\boldsymbol{p}_{\lambda})$             &$=\mathcal{N}\psi_{000}(\boldsymbol{p}_{\rho})\psi_{000}(\boldsymbol{p}_{\lambda})$\\
      $\psi_{11L_z}^{\lambda}(\boldsymbol{p}_{\rho},\boldsymbol{p}_{\lambda})$   &$=\mathcal{N}\psi_{000}(\boldsymbol{p}_{\rho})\psi_{01L_z}(\boldsymbol{p}_{\lambda})$\\
      $\psi_{11L_z}^{\rho}(\boldsymbol{p}_{\rho},\boldsymbol{p}_{\lambda})$      &$=\mathcal{N}\psi_{01L_z}(\boldsymbol{p}_{\rho})\psi_{000}(\boldsymbol{p}_{\lambda})$\\
      $\psi_{200}^s(\boldsymbol{p}_{\rho},\boldsymbol{p}_{\lambda})$             &$=\mathcal{N}\frac{1}{\sqrt{2}}[\psi_{100}(\boldsymbol{p}_{\rho})\psi_{000}(\boldsymbol{p}_{\lambda})+\psi_{000}(\boldsymbol{p}_{\rho})\psi_{100}(\boldsymbol{p}_{\lambda})]$\\
      $\psi_{200}^{\lambda}(\boldsymbol{p}_{\rho},\boldsymbol{p}_{\lambda})$     &$=\frac{\mathcal{N}}{\sqrt{2}}[-\psi_{100}(\boldsymbol{p}_{\rho})\psi_{000}(\boldsymbol{p}_{\lambda})+\psi_{000}(\boldsymbol{p}_{\rho})\psi_{100}(\boldsymbol{p}_{\lambda})]$\\
      $\psi_{200}^{\rho}(\boldsymbol{p}_{\rho},\boldsymbol{p}_{\lambda})$        &$=\frac{\mathcal{N}}{\sqrt{3}}[\psi_{011}(\boldsymbol{p}_{\rho})\psi_{01-1}(\boldsymbol{p}_{\lambda})-\psi_{010}(\boldsymbol{p}_{\rho})\psi_{010}(\boldsymbol{p}_{\lambda})+\psi_{01-1}(\boldsymbol{p}_{\rho})\psi_{011}(\boldsymbol{p}_{\lambda})]$\\
      $\psi_{21L_z}^{a}(\boldsymbol{p}_{\rho},\boldsymbol{p}_{\lambda})$         &$=\frac{\mathcal{N}}{\sqrt{2}}[\psi_{01m_{\rho}}(\boldsymbol{p}_{\rho})\psi_{01m_{\lambda}}(\boldsymbol{p}_{\lambda})-\psi_{01m_{\lambda}}(\boldsymbol{p}_{\rho})\psi_{01m_{\rho}}(\boldsymbol{p}_{\lambda})]$\\
      $\psi_{22L_z}^s(\boldsymbol{p}_{\rho},\boldsymbol{p}_{\lambda})$           &$=\frac{\mathcal{N}}{\sqrt{2}}[\psi_{02L_z}(\boldsymbol{p}_{\rho})\psi_{000}(\boldsymbol{p}_{\lambda})+\psi_{000}(\boldsymbol{p}_{\rho})\psi_{02L_z}(\boldsymbol{p}_{\lambda})]$\\
      $\psi_{22L_z}^{\lambda}(\boldsymbol{p}_{\rho},\boldsymbol{p}_{\lambda})$   &$=\frac{\mathcal{N}}{\sqrt{2}}[\psi_{02L_z}(\boldsymbol{p}_{\rho})\psi_{000}(\boldsymbol{p}_{\lambda})-\psi_{000}(\boldsymbol{p}_{\rho})\psi_{02L_z}(\boldsymbol{p}_{\lambda})]$\\
      $\psi_{222}^{\rho}(\boldsymbol{p}_{\rho},\boldsymbol{p}_{\lambda})$        &$=\mathcal{N}\psi_{011}(\boldsymbol{p}_{\rho})\psi_{011}(\boldsymbol{p}_{\lambda})$\\
      $\psi_{221}^{\rho}(\boldsymbol{p}_{\rho},\boldsymbol{p}_{\lambda})$        &$=\frac{\mathcal{N}}{\sqrt{2}}[\psi_{010}(\boldsymbol{p}_{\rho})\psi_{011}(\boldsymbol{p}_{\lambda})+\psi_{011}(\boldsymbol{p}_{\rho})\psi_{010}(\boldsymbol{p}_{\lambda})]$\\
      $\psi_{220}^{\rho}(\boldsymbol{p}_{\rho},\boldsymbol{p}_{\lambda})$        &$=\frac{\mathcal{N}}{\sqrt{6}}[\psi_{01-1}(\boldsymbol{p}_{\rho})\psi_{011}(\boldsymbol{p}_{\lambda})+2\psi_{010}(\boldsymbol{p}_{\rho})\psi_{010}(\boldsymbol{p}_{\lambda})+\psi_{011}(\boldsymbol{p}_{\rho})\psi_{01-1}(\boldsymbol{p}_{\lambda})]$\\
      $\psi_{22-1}^{\rho}(\boldsymbol{p}_{\rho},\boldsymbol{p}_{\lambda})$       &$=\frac{\mathcal{N}}{\sqrt{2}}[\psi_{01-1}(\boldsymbol{p}_{\rho})\psi_{010}(\boldsymbol{p}_{\lambda})+\psi_{010}(\boldsymbol{p}_{\rho})\psi_{01-1}(\boldsymbol{p}_{\lambda})]$\\
      $\psi_{22-2}^{\rho}(\boldsymbol{p}_{\rho},\boldsymbol{p}_{\lambda})$       &$=\mathcal{N}\psi_{01-1}(\boldsymbol{p}_{\rho})\psi_{01-1}(\boldsymbol{p}_{\lambda})$\\
      \hline\hline
  \end{tabular}
\end{table}

The total wavefunctions of a baryon can be expressed as $|qqq\rangle_A=|color\rangle_A\otimes|spin,flavor,spatial\rangle_S$. Based on the requirement of the SU(6)$\otimes$O(3) symmetry,one can obtain the configurations for the spin-flavor-spatial part. Specially, the total wavefunction with quantum number $\bm{J}^P=\frac12^{\pm}$ denoted by $|N_6,^{2\bm{S}+1}N_3,N,\bm{L}^P,\bm{J}^P\rangle$ can be constructed as follows:
\begin{align}
  \begin{split}
      |56,^28,0,0^+,\frac12^+\rangle&=|56,^28\rangle^s\psi_{000}^s(\boldsymbol{p}_{\rho},\boldsymbol{p}_{\lambda}),\\
      |56,^28,2,0^+,\frac12^+\rangle&=|56,^28\rangle^s\psi_{200}^s(\boldsymbol{p}_{\rho},\boldsymbol{p}_{\lambda}),\\
      |70,^28,2,0^+,\frac12^+\rangle&=\frac{1}{\sqrt{2}}\left(|70,^28\rangle^{\rho}\psi_{200}^{\rho}(\boldsymbol{p}_{\rho},\boldsymbol{p}_{\lambda})+|70,^28\rangle^{\lambda}\psi_{200}^{\lambda}(\boldsymbol{p}_{\rho},\boldsymbol{p}_{\lambda})\right),\\
      |70,^28,1,1^-,\frac12^-\rangle&=\sum_{L_z+S_z=J_z}\langle 1L_z;\frac12 S_z|\frac12 J_z\rangle\times \frac{1}{\sqrt{2}}\left(|70,^28\rangle^{\rho}\psi_{11L_z}^{\rho}(\boldsymbol{p}_{\rho},\boldsymbol{p}_{\lambda})+|70,^28\rangle^{\lambda}\psi_{11L_z}^{\lambda}(\boldsymbol{p}_{\rho},\boldsymbol{p}_{\lambda})\right),\\
      |70,^48,1,1^-,\frac12^-\rangle&=\sum_{L_z+S_z=J_z}\langle 1L_z;\frac32 S_z|\frac12 J_z\rangle\times\frac{1}{\sqrt{2}}\left(|70,^48\rangle^{\rho}\psi_{11L_z}^{\rho}(\boldsymbol{p}_{\rho},\boldsymbol{p}_{\lambda})+|70,^48\rangle^{\lambda}\psi_{11L_z}^{\lambda}(\boldsymbol{p}_{\rho},\boldsymbol{p}_{\lambda})\right),\\
      |70,^21,1,1^-,\frac12^-\rangle&=\sum_{L_z+S_z=J_z}\langle 1L_z;\frac12 S_z|\frac12 J_z\rangle\times\frac{1}{\sqrt{2}}\left(|70,^21\rangle^{\lambda}\psi_{11L_z}^{\lambda}(\boldsymbol{p}_{\rho},\boldsymbol{p}_{\lambda})-|70,^21\rangle^{\rho}\psi_{11L_z}^{\rho}(\boldsymbol{p}_{\rho},\boldsymbol{p}_{\lambda})\right).\\
  \end{split}
\end{align}

\begin{align}
    \begin{split}
        |56,^28,0,0^+,\frac12^+\rangle&=\frac{1}{\sqrt{2}}(\phi^{\rho}\chi^{\rho}+\phi^{\lambda}\chi^{\lambda})\psi_{000}^s(\boldsymbol{p}_{\rho},\boldsymbol{p}_{\lambda})
    \end{split}
\end{align}

The wavefunction of pion mesons is written as:
\begin{align}
    \begin{split}
        \Phi_{000}(\boldsymbol{p}_1,\boldsymbol{p}_2)=\delta^3(\boldsymbol{p}_1+\boldsymbol{p}_2-\boldsymbol{P})\phi_{\pi}\chi_{0,0}^a\psi_{000}^s(\boldsymbol{p}_1,\boldsymbol{p}_2),
    \end{split}
\end{align}
where $\chi_{0,0}^a$ is the spin wavefunction:
\begin{align}
    \begin{split}
        \chi_{0,0}^a=\frac{1}{\sqrt{2}}(\uparrow\downarrow-\downarrow\uparrow),
    \end{split}
\end{align}
and $\phi_{\pi}$ is the flavor wavefunction and we adopt the following convention:
\begin{align}
    \begin{split}
        \phi_{\pi^+}=u\bar{d},\ \ \ \ \ 
        \phi_{\pi^0}=-\frac{1}{\sqrt{2}}(u\bar{u}-d\bar{d}),\ \ \ \ \ 
        \phi_{\pi^-}=-d\bar{u}.
    \end{split}
\end{align}
The spatial wavefunction is expressed as:
\begin{align}
    \begin{split}
        \psi_{000}^s(\boldsymbol{p}_1,\boldsymbol{p}_2)=\frac{1}{\pi^{3/4}R^{3/2}}\text{exp}\left[-\frac{(\boldsymbol{p}_1-\boldsymbol{p}_2)^2}{8R^2}\right],
    \end{split}
\end{align}
where $R$ is the harmonic oscillator strength parameter of the pion meson.

\section{Transition amplitudes extracted in the quark model}
\label{app:tree amp}
    The transition amplitudes denotes by the baryon polarization quantum numbers are to be provided. In addition, $\mathcal{M}^{J_f,J_f^z;J_i,J_i^z}$ is shortened to $\mathcal{M}^{J_f^z,J_i^z}$ as the spin of initial and final states are all 1/2. We will provide the expressions for the amplitude of $\mathcal{M}^{-\frac12,-\frac12}$ for each process and the symmetry relation gives: $\mathcal{M}_{\text{PC}}^{\frac12,\frac12}=-\mathcal{M}_{\text{PC}}^{-\frac12,-\frac12}$ and $\mathcal{M}_{\text{PV}}^{\frac12,\frac12}=\mathcal{M}_{\text{PV}}^{-\frac12,-\frac12}$. The amplitudes of $\mathcal{M}^{\pm\frac12,\mp\frac12}$ are vanishing.

    The following functions are to be used later:
    \begin{align}
        \begin{split}
            \mathcal{F}_{\pi}(\boldsymbol{k})=\text{exp}\left[-\frac{\boldsymbol{k}^2}{6\alpha^2}\right],\ \ \ \ \ \mathcal{F}_{\pi}^{\prime}(\boldsymbol{k})=\text{exp}\left[-\frac{\boldsymbol{k}^2}{8}(\frac{3m_s^2}{(2m_q+m_s)^2\alpha_{\lambda}^2}+\frac{1}{\alpha_{\rho}^2})\right]
        \end{split}
    \end{align}
    where $m_q$ is the mass of the $u/d$ quarks and $m_s$ is the mass of the $s$ quark; $\boldsymbol{k}$ and $\omega_0$ denote the three-vector momentum and energy of the pion, respectively. In order to use the typical value of the harmonic oscillator strengths directly, all the amplitudes are expressed with the conventional the harmonic oscillator strengths. $\alpha_{\rho}$ and $\alpha_{\lambda}=(\frac{3m_s}{2m_q+m_s})^{1/4}$ are the harmonic oscillator strengths for the strange baryons and $\alpha=\alpha_{\rho}^{\prime}=\alpha_{\lambda}^{\prime}$ for the light baryons. For the pole terms, the propagator is noted with $\mathcal{P}(m_1,m_2)$ which is defined as
    \begin{align}
        \begin{split}
            \mathcal{P}(m_1,m_2)=\frac{2m_2}{m_1^2-m_2^2+im_2\Gamma_{m_2}},
        \end{split}
    \end{align}
    where $m_1$ is the mass of initial baryon or final baryon; $m_2$ and $\Gamma_{m_2}$ are the mass and width of intermediate baryon, respectively.

\begin{itemize}
    \item $\Lambda\to p\pi^-$

\begin{itemize}
    \item DPE process
    \begin{align}
                \begin{split}
                    \mathcal{M}_{\text{DPE,PC}}^{-\frac12,-\frac12}&=\frac{2G_FV_{ud}V_{us}}{\sqrt{3}\pi^{9/4}}\frac{\boldsymbol{k}(3m_s\alpha^2-m_q\alpha_{\lambda}^2+2m_s\alpha_{\lambda}^2)}{m_qm_s(\alpha^2+\alpha_{\lambda}^2)^{5/2}}\left(\frac{\alpha^2\alpha_{\rho}\alpha_{\lambda}R}{\alpha^2+\alpha_{\rho}^2}\right)^{3/2}\text{exp}\left[-\frac{\boldsymbol{k}^2}{3(\alpha^2+\alpha_{\lambda}^2)}\right],\\
                    \mathcal{M}_{\text{DPE,PV}}^{-\frac12,-\frac12}&=-\frac{2\sqrt{3}G_FV_{ud}V_{us}}{\pi^{9/4}}\left(\frac{\alpha^2\alpha_{\rho}\alpha_{\lambda}R}{(\alpha^2+\alpha_{\rho}^2)(\alpha^2+\alpha_{\lambda}^2)}\right)^{3/2}\text{exp}\left[-\frac{\boldsymbol{k}^2}{3(\alpha^2+\alpha_{\lambda}^2)}\right].
                \end{split}
            \end{align}
    \item CS process
    \begin{align}
                \begin{split}
                    &\mathcal{M}_{\text{CS-1,PC}}^{-\frac12,-\frac12}=\frac{6\sqrt{2}G_FV_{ud}V_{us}}{\pi^{9/4}}\\
                    &\times\frac{\boldsymbol{k}(\alpha^2\alpha_{\rho}\alpha_{\lambda}R)^{3/2}\bigg(m_q\alpha_{\lambda}^2(-2R^2+2\alpha^2+\alpha_{\rho}^2)+m_s\big(3\alpha_{\lambda}^2\alpha_{\rho}^2+\alpha^2(\alpha_{\lambda}^2+3\alpha_{\rho}^2)+2R^2(6\alpha^2+2\alpha_{\lambda}^2+3\alpha_{\rho}^2)\big)\bigg)}{m_qm_s(6\alpha_{\lambda}^2\alpha_{\rho}^2+2\alpha^2(\alpha_{\lambda}^2+3\alpha_{\rho}^2)+3R^2(4\alpha^2+\alpha_{\lambda}^2+3\alpha_{\rho}^2))^{5/2}}\\
                    &\times\text{exp}\left[-\frac{\boldsymbol{k}^2(24R^2+36\alpha^2+25\alpha_{\lambda}^2+3\alpha_{\rho}^2)}{24(6\alpha_{\lambda}^2\alpha_{\rho}^2+2\alpha^2(\alpha_{\lambda}^2+3\alpha_{\rho}^2)+3R^2(4\alpha^2+\alpha_{\lambda}^2+3\alpha_{\rho}^2))}\right],\\
                    &\mathcal{M}_{\text{CS-1,PV}}^{-\frac12,-\frac12}=-\frac{12\sqrt{2}G_FV_{ud}V_{us}}{\pi^{9/4}}\left[\frac{\alpha^2\alpha_{\rho}\alpha_{\lambda}R}{6\alpha_{\rho}^2\alpha_{\lambda}^2+2\alpha^2(\alpha_{\lambda}^2+2\alpha_{\rho}^2)+3R^2(4\alpha^2+\alpha_{\lambda}^2+2\alpha_{\rho}^2)}\right]^{3/2}\\
                    &\times\text{exp}\left[-\frac{\boldsymbol{k}^2(24R^2+36\alpha^2+25\alpha_{\lambda}^2+3\alpha_{\rho}^2)}{24(6\alpha_{\lambda}^2\alpha_{\rho}^2+2\alpha^2(\alpha_{\lambda}^2+3\alpha_{\rho}^2)+3R^2(4\alpha^2+\alpha_{\lambda}^2+3\alpha_{\rho}^2))}\right].
                \end{split}
            \end{align}
    \item{Pole terms}
    \begin{align}
                \begin{split}
                    &\mathcal{M}_{\text{Pole,A1,PC}}^{-\frac12,-\frac12}[n(939)\\
                    &=\left[-\frac{8\sqrt{3}G_FV_{ud}V_{us}}{\pi^{3/2}}\left(\frac{\alpha^2\alpha_{\lambda}\alpha_{\rho}}{4\alpha^2+\alpha_{\lambda}+3\alpha_{\rho}^2}\right)^{3/2}\right]\left[-\frac{5\boldsymbol{k}(6m_q+\omega_0)}{72\pi^{3/2}\sqrt{\omega_0}f_{\pi}m_q}\mathcal{F}_{\pi}(\boldsymbol{k})\right]\mathcal{P}(m_{\Lambda},m_n),\\
                    &\mathcal{M}_{\text{Pole,A2,PC}}^{-\frac12,-\frac12}[N(1440)\\
                    &=\left[-\frac{24G_FV_{ud}V_{us}}{\pi^{3/2}}\frac{(\alpha^2\alpha_{\lambda}\alpha_{\rho})^{3/2}(\alpha_{\lambda}^2+3\alpha_{\rho}^2)}{(4\alpha^2+\alpha_{\lambda}+3\alpha_{\rho}^2)^{5/2}}\right]\left[-\frac{5\boldsymbol{k}\big(-12\alpha^2\omega_0+\boldsymbol{k}^2(6m_q+\omega_0)\big)}{432\sqrt{3}\pi^{3/2}\alpha^2\sqrt{\omega_0}f_{\pi}m_q}\mathcal{F}_{\pi}(\boldsymbol{k})\right]\mathcal{P}(m_{\Lambda},m_{N(1440)}),\\
                    &\mathcal{M}_{\text{Pole,A3,PC}}^{-\frac12,-\frac12}[N(1710)]=0\times\mathcal{P}(m_{\Lambda},m_{N(1710)}),\\
                    &\mathcal{M}_{\text{Pole,A1,PV}}^{-\frac12,-\frac12}[N|70,^28\rangle]\\
                    &=\left[i\frac{8\sqrt{6}G_FV_{ud}V_{us}}{\pi^{3/2}}\frac{\alpha^4(\alpha_{\lambda}\alpha_{\rho})^{3/2}(2m_s\alpha^2-m_q\alpha_{\lambda}^2+3m_s\alpha_{\rho}^2)}{m_qm_s(4\alpha^2+\alpha_{\lambda}^2+3\alpha_{\rho}^2)^{5/2}}\right]\left[i\frac{\big(18\alpha^2\omega_0-\boldsymbol{k}^2(6m_q+\omega_0)\big)}{54\sqrt{2}\pi^{3/2}\alpha\sqrt{\omega_0}f_{\pi}m_q}\mathcal{F}_{\pi}(\boldsymbol{k})\right]\\
                    &\times\mathcal{P}(m_{\Lambda},m_{N|70,^28\rangle}),\\
                    &\mathcal{M}_{\text{Pole,A2,PV}}^{-\frac12,-\frac12}[N|70,^48\rangle]\\
                    &=\left[-i\frac{8\sqrt{6}G_FV_{ud}V_{us}}{\pi^{3/2}}\frac{\alpha^4(\alpha_{\lambda}\alpha_{\rho})^{3/2}}{m_q(4\alpha^2+\alpha_{\lambda}^2+3\alpha_{\rho}^2)^{3/2}}\right]\left[i\frac{\big(18\alpha^2\omega_0-\boldsymbol{k}^2(6m_q+\omega_0)\big)}{108\sqrt{2}\pi^{3/2}\alpha\sqrt{\omega_0}f_{\pi}m_q}\mathcal{F}_{\pi}(\boldsymbol{k})\right]\mathcal{P}(m_{\Lambda},m_{N|70,^48\rangle}).
                \end{split}
            \end{align}

            \begin{align}
                \begin{split}
                    &\mathcal{M}_{\text{Pole,B1,PC}}^{-\frac12,-\frac12}[\Sigma^+(1193)]\\
                    &=\left[\frac{\boldsymbol{k}(4m_q+2m_s+\omega_0)}{4\sqrt{6}\pi^{3/2}\sqrt{\omega_0}f_{\pi}(2m_q+m_s)}\mathcal{F}_{\pi}^{\prime}(\boldsymbol{k})\right]\left[\frac{24\sqrt{2}G_FV_{ud}V_{us}}{\pi^{3/2}}(\frac{\alpha^2\alpha_{\lambda}\alpha_{\rho}}{4\alpha^2+\alpha_{\lambda}^2+3\alpha_{\rho}^2})^{3/2}\right]\mathcal{P}(m_p,m_{\Sigma^+})\\
                    &\mathcal{M}_{\text{Pole,B2,PC}}^{-\frac12,-\frac12}[\Sigma^+(1660)]\\
                    &=\left[\frac{\boldsymbol{k}\bigg(8(m_q+m_s)(2m_q+m_s)^2\alpha_{\lambda}^2\alpha_{\rho}^2\omega_0+\boldsymbol{k}^2\big((2m_q+m_s)^2\alpha_{\lambda}^2+3m_s^2\alpha_{\rho}^2\big)(4m_q+2m_s+\omega_0)\bigg)}{96\sqrt{2}\pi^{3/2}\alpha_{\lambda}^2\alpha_{\rho}^2\sqrt{\omega_0}f_{\pi}(2m_q+m_s)^3}\mathcal{F}_{\pi}^{\prime}(\boldsymbol{k})\right]\\
                    &\times\left[\frac{96\sqrt{6}G_FV_{ud}V_{us}}{\pi^{3/2}}\frac{\alpha^5(\alpha_{\lambda}\alpha_{\rho})^{3/2}}{(4\alpha^2+\alpha_{\lambda}^2+3\alpha_{\rho}^2)^{5/2}}\right]\mathcal{P}(m_p,m_{\Sigma^+(1660)}),\\
                    &\mathcal{M}_{\text{Pole,B3,PC}}^{-\frac12,-\frac12}[\Sigma^+(1880)]\\
                    &=\Bigg[-\bigg(\boldsymbol{k}\Big(4(2m_q+m_s)^2\alpha_{\lambda}\alpha_{\rho}\big(2m_q\alpha_{\lambda}(\alpha_{\lambda}-\alpha_{\rho})+m_s(\alpha_{\lambda^2}+3\alpha_{\rho^2})\big)\omega_0+\boldsymbol{k}^2\big((2m_q+m_s)^2\alpha_{\lambda}^2\\
                    &-6m_s(2m_q+m_s)\alpha_{\lambda}\alpha_{\rho}-3m_s^2\alpha_{\rho}^2\big)(4m_q+2m_s+\omega_0)\Big)\bigg)\mathcal{F}_{\pi}^{\prime}(\boldsymbol{k})/\bigg(768\pi^{3/2}\alpha_{\lambda}^2\alpha_{\rho}^2\sqrt{\omega_0}f_{\pi}(2m_q+m_s)^3\bigg)\Bigg]\\
                    &\times\left[\frac{24\sqrt{3}G_FV_{ud}V_{us}}{\pi^{3/2}}\frac{(\alpha^2\alpha_{\lambda}\alpha_{\rho})^{3/2}(\alpha_{\lambda}-\alpha_{\rho})(\alpha_{\lambda}+3\alpha_{\rho})}{(4\alpha^2+\alpha_{\lambda}^2+3\alpha_{\rho}^2)^{5/2}}\right]\mathcal{P}(m_p,m_{\Sigma^+(1880)}),\\
                    &\mathcal{M}_{\text{Pole,B1,PV}}^{-\frac12,-\frac12}[\Sigma^+(1620)]\\
                    &=\left[i\frac{2(2m_q+m_s)^2\alpha_{\lambda}\alpha_{\rho}(\alpha_{\lambda}+3\alpha_{\rho})\omega_0+\boldsymbol{k}^2\big(2m_q\alpha_{\lambda}+m_s(\alpha_{\lambda}+\alpha_{\rho})\big)(4m_q+m_s+\omega_0)}{64\sqrt{3}\pi^{3/2}\alpha_{\lambda}\sqrt{\omega_0}f_{\pi}(2m_q+m_s)^2}\mathcal{F}_{\pi}^{\prime}(\boldsymbol{k})\right]\\
                    &\times\left[i\frac{8G_FV_{ud}V_{us}}{\pi^{3/2}}\frac{(\alpha^2\alpha_{\lambda}\alpha_{\rho})^{3/2}\bigg(4m_q\alpha^2\alpha_{\lambda}+3m_q\alpha_{\lambda}\alpha_{\rho}(\alpha_{\lambda}+\alpha_{\rho})+3m_s\alpha_{\lambda}\alpha_{\rho}(\alpha_{\lambda}+\alpha_{\rho})+2m_s\alpha^2(\alpha_{\lambda}+3\alpha_{\rho})\bigg)}{m_qm_s(4\alpha^2+\alpha_{\lambda}^2+3\alpha_{\rho}^2)^{5/2}}\right]\\
                    &\times\mathcal{P}(m_p,m_{\Sigma^+(1620)}),\\
                    &\mathcal{M}_{\text{Pole,B2,PV}}^{-\frac12,-\frac12}[\Sigma^+(1750)]\\
                    &=\left[i\frac{2(2m_q+m_s)^2\alpha_{\lambda}\alpha_{\rho}(\alpha_{\lambda}+3\alpha_{\rho})\omega_0+\boldsymbol{k}^2\big(2m_q\alpha_{\lambda}+m_s(\alpha_{\lambda}+\alpha_{\rho})\big)(4m_q+2m_s+\omega_0)}{144\sqrt{3}\pi^{3/2}\alpha_{\lambda}\sqrt{\omega_0}f_{\pi}m_q}\mathcal{F}_{\pi}^{\prime}(\boldsymbol{k})\right]\\
                    &\times\left[-i\frac{32G_FV_{ud}V_{us}}{\pi^{3/2}}\frac{\alpha^3\alpha_{\lambda}^{5/2}\alpha_{\rho}^{3/2}\big(2(2m_q+m_s)\alpha^2+3(m_q+m_s)\alpha_{\rho}^2\big)}{m_qm_s(4\alpha^2+\alpha_{\lambda}^2+3\alpha_{\rho}^2)^{5/2}}\right]\mathcal{P}(m_p,m_{\Sigma^+(1750)}).
                \end{split}
            \end{align}
        \end{itemize}

\item $\Lambda\to n\pi^0$
    \begin{itemize}
        \item CS process
        \begin{align}
                \begin{split}
                    \mathcal{M}_{\text{CS-1,PC}}^{-\frac12,-\frac12}(\Lambda\to n\pi^0)&=-\frac{1}{\sqrt{2}}\mathcal{M}_{\text{CS-1,PC}}^{-\frac12,-\frac12}(\Lambda\to p\pi^-),\\
                    \mathcal{M}_{\text{CS-1,PV}}^{-\frac12,-\frac12}(\Lambda\to n\pi^0)&=-\frac{1}{\sqrt{2}}\mathcal{M}_{\text{CS-1,PV}}^{-\frac12,-\frac12}(\Lambda\to p\pi^-).
                \end{split}
            \end{align}
            \begin{align}
                \begin{split}
                    \mathcal{M}_{\text{CS-2,PC}}^{-\frac12,-\frac12}(\Lambda\to n\pi^0)&=\frac{\sqrt{6}G_FV_{ud}V_{us}}{9\pi^{9/4}}\frac{\boldsymbol{k}(-3m_s\alpha^2+mq\alpha_{\lambda}^2-2m_s\alpha_{\lambda}^2)}{m_qm_s(\alpha^2+\alpha_{\lambda}^2)^{5/2}}\left(\frac{\alpha^2\alpha_{\rho}\alpha_{\lambda}R}{\alpha^2+\alpha_{\rho}^2}\right)^{3/2}\\
                    &\times\text{exp}\left[-\frac{\boldsymbol{k}^2}{3(\alpha^2+\alpha_{\lambda}^2)}\right]=-\frac{1}{3\sqrt{2}}\mathcal{M}_{\text{DPE,PC}}^{-\frac12,-\frac12}(\Lambda\to p\pi^-),\\
                    \mathcal{M}_{\text{CS-2,PV}}^{-\frac12,-\frac12}(\Lambda\to n\pi^0)&=\frac{\sqrt{6}G_FV_{ud}V_{us}}{3\pi^{9/4}}\left(\frac{\alpha^2\alpha_{\rho}\alpha_{\lambda}R}{(\alpha^2+\alpha_{\lambda}^2)(\alpha^2+\alpha_{\rho}^2)}\right)^{3/2}\text{exp}\left[-\frac{\boldsymbol{k}^2}{3(\alpha^2+\alpha_{\lambda}^2)}\right]\\
                    &=-\frac{1}{3\sqrt{2}}\mathcal{M}_{\text{DPE,PV}}^{-\frac12,-\frac12}(\Lambda\to p\pi^-).
                \end{split}
            \end{align}
            \item Pole terms
            \begin{align}
                \begin{split}
                    \mathcal{M}_{\text{Pole,A1,PC}}^{-\frac12,-\frac12}[n(939)](\Lambda\to n\pi^0)&=-\frac{1}{\sqrt{2}}\mathcal{M}_{\text{Pole,A1,PC}}^{-\frac12,-\frac12}[n(939)](\Lambda\to p\pi^-),\\
                    \mathcal{M}_{\text{Pole,A2,PC}}^{-\frac12,-\frac12}[N(1440)](\Lambda\to n\pi^0)&=-\frac{1}{\sqrt{2}}\mathcal{M}_{\text{Pole,A2,PC}}^{-\frac12,-\frac12}[N(1440)](\Lambda\to p\pi^-),\\
                    \mathcal{M}_{\text{Pole,A3,PC}}^{-\frac12,-\frac12}[N(1710)](\Lambda\to n\pi^0)&=-\frac{1}{\sqrt{2}}\mathcal{M}_{\text{Pole,A3,PC}}^{-\frac12,-\frac12}[N(1710)](\Lambda\to p\pi^-)=0,\\
                    \mathcal{M}_{\text{Pole,A1,PV}}^{-\frac12,-\frac12}[N|70,^28\rangle](\Lambda\to n\pi^0)&=-\frac{1}{\sqrt{2}}\mathcal{M}_{\text{Pole,A1,PV}}^{-\frac12,-\frac12}[N|70,^28\rangle](\Lambda\to p\pi^-),\\
                    \mathcal{M}_{\text{Pole,A2,PV}}^{-\frac12,-\frac12}[N|70,^48\rangle](\Lambda\to n\pi^0)&=-\frac{1}{\sqrt{2}}\mathcal{M}_{\text{Pole,A2,PV}}^{-\frac12,-\frac12}[N|70,^48\rangle](\Lambda\to p\pi^-),\\
                    \mathcal{M}_{\text{Pole,B1,PC}}^{-\frac12,-\frac12}[\Sigma^0(1193)](\Lambda\to n\pi^0)&=-\frac{1}{\sqrt{2}}\mathcal{M}_{\text{Pole,B1,PC}}^{-\frac12,-\frac12}[\Sigma^+(1193)](\Lambda\to p\pi^-),\\
                    \mathcal{M}_{\text{Pole,B2,PC}}^{-\frac12,-\frac12}[\Sigma^0(1660)](\Lambda\to n\pi^0)&=-\frac{1}{\sqrt{2}}\mathcal{M}_{\text{Pole,B2,PC}}^{-\frac12,-\frac12}[\Sigma^+(1660)](\Lambda\to p\pi^-),\\
                    \mathcal{M}_{\text{Pole,B3,PC}}^{-\frac12,-\frac12}[\Sigma^0(1880)](\Lambda\to n\pi^0)&=-\frac{1}{\sqrt{2}}\mathcal{M}_{\text{Pole,B3,PC}}^{-\frac12,-\frac12}[\Sigma^+(1880)](\Lambda\to p\pi^-),\\
                    \mathcal{M}_{\text{Pole,B1,PV}}^{-\frac12,-\frac12}[\Sigma^0(1620)](\Lambda\to n\pi^0)&=-\frac{1}{\sqrt{2}}\mathcal{M}_{\text{Pole,B1,PV}}^{-\frac12,-\frac12}[\Sigma^+(1620)](\Lambda\to p\pi^-),\\
                    \mathcal{M}_{\text{Pole,B2,PV}}^{-\frac12,-\frac12}[\Sigma^0(1750)](\Lambda\to n\pi^0)&=-\frac{1}{\sqrt{2}}\mathcal{M}_{\text{Pole,B2,PV}}^{-\frac12,-\frac12}[\Sigma^+(1750)](\Lambda\to p\pi^-).
                \end{split}
            \end{align}
        \end{itemize}

        \item $\Sigma^-\to n\pi^-$
            \begin{itemize}
                \item DPE process
                \begin{align}
        \begin{split}
            \mathcal{M}_{\text{DPE,PC}}^{-\frac12,-\frac12}&=-\frac{2\sqrt{2}G_FV_{ud}V_{us}}{9\pi^{9/4}}\frac{\boldsymbol{k}(3m_s\alpha^2-m_q\alpha_{\lambda}^2+2m_s\alpha_{\lambda}^2)}{m_qm_s(\alpha^2+\alpha_{\lambda}^2)^{5/2}}\left(\frac{\alpha^2\alpha_{\rho}\alpha_{\lambda}R}{\alpha^2+\alpha_{\rho}^2}\right)^{3/2}\text{exp}\left[-\frac{\boldsymbol{k}^2}{3(\alpha^2+\alpha_{\lambda}^2)}\right],\\
            \mathcal{M}_{\text{DPE,PV}}^{-\frac12,-\frac12}&=-\frac{2\sqrt{2}G_FV_{ud}V_{us}}{\pi^{9/4}}\left(\frac{\alpha^2\alpha_{\rho}\alpha_{\lambda}R}{(\alpha^2+\alpha_{\rho}^2)(\alpha^2+\alpha_{\lambda}^2)}\right)^{3/2}\text{exp}\left[-\frac{\boldsymbol{k}^2}{3(\alpha^2+\alpha_{\lambda}^2)}\right].
        \end{split}
    \end{align}
    \item CS process
    \begin{align}
        \begin{split}
            &\mathcal{M}_{\text{CS-1,PC}}^{-\frac12,-\frac12}=-\frac{4\sqrt{3}G_FV_{ud}V_{us}}{\pi^{9/4}}\\
            &\times\frac{\boldsymbol{k}(\alpha^2\alpha_{\rho}\alpha_{\lambda}R)^{3/2}\bigg(m_q\alpha_{\lambda}^2(-2R^2+2\alpha^2+\alpha_{\rho}^2)+m_s\big(3\alpha_{\lambda}^2\alpha_{\rho}^2+\alpha^2(\alpha_{\lambda}^2+3\alpha_{\rho}^2)+2R^2(6\alpha^2+2\alpha_{\lambda}^2+3\alpha_{\rho}^2)\big)\bigg)}{m_qm_s\big(6\alpha_{\lambda}^2\alpha_{\rho}^2+2\alpha^2(\alpha_{\lambda}^2+3\alpha_{\rho}^2)+3R^2(4\alpha^2+\alpha_{\lambda}^2+3\alpha_{\rho}^2)\big)^{5/2}}\\
            &\times\text{exp}\left[-\frac{\boldsymbol{k}^2(24R^2+36\alpha^2+25\alpha_{\lambda}^2+3\alpha_{\rho}^2)}{24\big(6\alpha_{\lambda}^2\alpha_{\rho}^2+2\alpha^2(\alpha_{\lambda}^2+3\alpha_{\rho}^2)+3R^2(4\alpha^2+\alpha_{\lambda}^2+3\alpha_{\rho}^2)\big)}\right],\\
            &\mathcal{M}_{\text{CS-1,PV}}^{-\frac12,-\frac12}=-\frac{24\sqrt{3}G_FV_{ud}V_{us}}{\pi^{9/4}}\left(\frac{\alpha^2\alpha_{\rho}\alpha_{\lambda}R}{6\alpha_{\rho}^2\alpha_{\lambda}^2+2\alpha^2(\alpha_{\lambda}^2+3\alpha_{\rho}^2)+3R^2(4\alpha^2+\alpha_{\lambda}^2+3\alpha_{\rho}^2)}\right)^{3/2}\\
            &\times\text{exp}\left[-\frac{\boldsymbol{k}^2(24R^2+36\alpha^2+25\alpha_{\lambda}^2+3\alpha_{\rho}^2)}{24\big(6\alpha_{\lambda}^2\alpha_{\rho}^2+2\alpha^2(\alpha_{\lambda}^2+3\alpha_{\rho}^2)+3R^2(4\alpha^2+\alpha_{\lambda}^2+3\alpha_{\rho}^2)\big)}\right].
        \end{split}
    \end{align}
    \item Pole terms
    \begin{align}
        \begin{split}
            &\mathcal{M}_{\text{Pole,B1,PC}}^{-\frac12,-\frac12}[\Sigma^0(1193)]\\&=\left[-\frac{\boldsymbol{k}(4m_q+2m_s+\omega_0)}{6\sqrt{2}\pi^{3/2}\sqrt{\omega_0}f_{\pi}(2m_q+m_s)}\mathcal{F}_{\pi}^{\prime}(\boldsymbol{k})\right]\left[\frac{24G_FV_{ud}V_{us}}{\pi^{3/2}}\left(\frac{\alpha^2\alpha_{\rho}\alpha_{\lambda}}{4\alpha^2+\alpha_{\lambda}^2+3\alpha_{\rho}^2}\right)^{3/2}\right]\mathcal{P}(m_{n},m_{\Sigma^0}),\\
            &\mathcal{M}_{\text{Pole,B2,PC}}^{-\frac12,-\frac12}[\Lambda_8(1116)]\\&=\left[-\frac{\boldsymbol{k}(4m_q+2m_s+\omega_0)}{4\sqrt{6}\pi^{3/2}\sqrt{\omega_0}f_{\pi}(2m_q+m_s)}\mathcal{F}_{\pi}^{\prime}(\boldsymbol{k})\right]\left[-\frac{8\sqrt{3}G_FV_{ud}V_{us}}{\pi^{3/2}}\left(\frac{\alpha^2\alpha_{\rho}\alpha_{\lambda}}{4\alpha^2+\alpha_{\lambda}^2+3\alpha_{\rho}^2}\right)^{3/2}\right]\mathcal{P}(m_{n},m_{\Lambda}),\\
            &\mathcal{M}_{\text{Pole,B3,PC}}^{-\frac12,-\frac12}[\Sigma^0(1660)]\\&=\left[-\frac{\boldsymbol{k}\bigg(8(m_q+m_s)(2m_q+m_s)^2\alpha_{\lambda}^2\alpha_{\rho}^2\omega_0+\boldsymbol{k}^2m_q\big((2m_q+m_s)^2\alpha_{\lambda}^2+3m_s^2\alpha_{\rho}^2\big)(4m_q+2m_s+\omega_0)\bigg)}{48\sqrt{6}\pi^{3/2}\alpha_{\rho}^2\alpha_{\lambda}^2\sqrt{\omega_0}f_{\pi}m_q(2m_q+m_s)^3}\mathcal{F}_{\pi}^{\prime}(\boldsymbol{k})\right]\\
            &\times\left[\frac{96\sqrt{3}G_FV_{ud}V_{us}}{\pi^{3/2}}\frac{\alpha^5(\alpha_{\lambda}\alpha_{\rho})^{3/2}}{(4\alpha^2+\alpha_{\lambda}^2+3\alpha_{\rho}^2)^{5/2}}\right]\mathcal{P}(m_{n},m_{\Sigma^0(1660)}),\\
            &\mathcal{M}_{\text{Pole,B4,PC}}^{-\frac12,-\frac12}[\Lambda_8(1600)]\\&=\left[-\frac{\boldsymbol{k}\bigg(8(m_q+m_s)(2m_q+m_s)^3\alpha_{\lambda}^2\alpha_{\rho}^2\omega_0+\boldsymbol{k}^2m_q\big((2m_q+m_s)^2\alpha_{\lambda}^2+3m_s^2\alpha_{\rho}^2\big)(4m_q+m_s+\omega_0)\bigg)}{96\sqrt{2}\pi^{3/2}\alpha_{\rho}^2\alpha_{\lambda}^2\sqrt{\omega_0}f_{\pi}m_q(2m_q+m_s)^3}\mathcal{F}_{\pi}^{\prime}(\boldsymbol{k})\right]\\
            &\times\left[-\frac{96G_FV_{ud}V_{us}}{\pi^{3/2}}\frac{\alpha^5(\alpha_{\lambda}\alpha_{\rho})^{3/2}}{(4\alpha^2+\alpha_{\lambda}^2+3\alpha_{\rho}^2)^{5/2}}\right]\mathcal{P}(m_{n},m_{\Lambda(1600)}),\\
            &\mathcal{M}_{\text{Pole,B5,PC}}^{-\frac12,-\frac12}[\Sigma^0(1880)]\\
            &\Bigg[\bigg(20\boldsymbol{k}(2m_q+m_s)^2\alpha_{\lambda}\alpha_{\rho}\big(2m_q\alpha_{\lambda}(\alpha_{\lambda}-\alpha_{\rho})+m_s(\alpha_{\lambda}^2+3\alpha_{\rho}^2)\omega_0\big)-5\boldsymbol{k}^3m_q\big((2m_q+m_s)^2\alpha_{\lambda}^2\\
            &-6m_s(2m_q+m_s)\alpha_{\lambda}\alpha_{\rho}-3m_s^2\alpha_{\rho}^2\big)(4m_q+2m_s+\omega_0)\bigg)\mathcal{F}_{\pi}^{\prime}(\boldsymbol{k})/\bigg(768\sqrt{3}\pi^{3/2}\alpha_{\rho}^2\alpha_{\lambda}^2\sqrt{\omega_0}f_{\pi}m_q(2m_q+m_s)^3\bigg)\Bigg]\\
            &\times\left[\frac{12\sqrt{6}G_FV_{ud}V_{us}}{\pi^{3/2}}\frac{\alpha^3\big(2(\alpha_{\lambda}\alpha_{\rho})^{5/2}+\alpha_{\lambda}^{7/2}\alpha_{\rho}^{3/2}-3\alpha_{\lambda}^{3/2}\alpha_{\rho}^{7/2}\big)}{(4\alpha^2+\alpha_{\lambda}^2+3\alpha_{\rho}^2)^{5/2}}\right]\mathcal{P}(m_{n},m_{\Sigma^0(1880)}),\\
            &\mathcal{M}_{\text{Pole,B6,PC}}^{-\frac12,-\frac12}[\Lambda_8(1810)]\\
            &\Bigg[\bigg(4\boldsymbol{k}(2m_q+m_s)^2\alpha_{\lambda}\alpha_{\rho}\bigg(2m_q\alpha_{\lambda}(\alpha_{\lambda}-\alpha_{\rho})+m_s(\alpha_{\lambda}^2+3\alpha_{\rho}^2)\omega_0\bigg)-\boldsymbol{k}^3m_q\bigg((2m_q+m_s)^2\alpha_{\lambda}^2\\
            &-6m_s(2m_q+m_s)\alpha_{\lambda}\alpha_{\rho}-3m_s^2\alpha_{\rho}^2\bigg)(4m_q+2m_s+\omega_0)\bigg)\mathcal{F}_{\pi}^{\prime}(\boldsymbol{k})/\bigg(768\sqrt{3}\pi^{3/2}\alpha_{\rho}^2\alpha_{\lambda}^2\sqrt{\omega_0}f_{\pi}m_q(2m_q+m_s)^3\bigg)\Bigg]\\
            &\times\left[\frac{12\sqrt{2}G_FV_{ud}V_{us}}{\pi^{3/2}}\frac{\alpha^3\big(-6(\alpha_{\lambda}\alpha_{\rho})^{5/2}+\alpha_{\lambda}^{7/2}\alpha_{\rho}^{3/2}-3\alpha_{\lambda}^{3/2}\alpha_{\rho}^{7/2}\big)}{(4\alpha^2+\alpha_{\lambda}^2+3\alpha_{\rho}^2)^{5/2}}\right]\mathcal{P}(m_{n},m_{\Lambda(1810)}).
        \end{split}
    \end{align}
    \begin{align}
        \begin{split}
            &\mathcal{M}_{\text{Pole,B1,PV}}^{-\frac12,-\frac12}[\Sigma^0(1620)]\\&=\left[-i\frac{5\bigg(2(2m_q+m_s)^2\alpha_{\lambda}\alpha_{\rho}(\alpha_{\lambda}+3\alpha_{\rho})\omega_0+\boldsymbol{k}^2m_q\big(2m_q\alpha_{\lambda}+m_s(\alpha_{\lambda}+\alpha_{\rho})\big)(4m_q+2m_s+\omega_0)\bigg)}{192\pi^{3/2}\alpha_{\lambda}\sqrt{\omega_0}f_{\pi}m_q(2m_q+m_s)^2}\mathcal{F}_{\pi}^{\prime}(\boldsymbol{k})\right]\\
            &\times\left[i\frac{4\sqrt{2}G_FV_{ud}V_{us}}{\pi^{3/2}}\frac{(\alpha^2\alpha_{\lambda}\alpha_{\rho})^{3/2}\bigg(4m_q\alpha^2\alpha_{\lambda}+3(m_q+m_s)\alpha_{\lambda}\alpha_{\rho}(\alpha_{\lambda}+\alpha_{\rho})+2m_s\alpha^2(\alpha_{\lambda}+3\alpha_{\rho})\bigg)}{m_qm_s(4\alpha^2+\alpha_{\lambda}^2+3\alpha_{\rho}^2)^{5/2}}\right]\\
            &\times\mathcal{P}(m_{n},m_{\Sigma^0(1620)}),\\
            &\mathcal{M}_{\text{Pole,B2,PV}}^{-\frac12,-\frac12}[\Lambda_8(1670)]\\&=\left[-i\frac{2(2m_q+m_s)^2\alpha_{\lambda}\alpha_{\rho}(\alpha_{\lambda}+3\alpha_{\rho})\omega_0+\boldsymbol{k}^2m_q\bigg(2m_q\alpha_{\lambda}+m_s(\alpha_{\lambda}+\alpha_{\rho})\bigg)(4m_q+2m_s+\omega_0)}{64\sqrt{3}\pi^{3/2}\alpha_{\lambda}\sqrt{\omega_0}f_{\pi}m_q(2m_q+m_s)^2}\mathcal{F}_{\pi}^{\prime}(\boldsymbol{k})\right]\\
            &\times\left[i\frac{4\sqrt{6}G_FV_{ud}V_{us}}{\pi^{3/2}}\frac{(\alpha^2\alpha_{\lambda}\alpha_{\rho})^{3/2}\bigg(-2(2m_q+m_s)\alpha^2\alpha_{\lambda}+\big(2m_s\alpha^2+(m_q+m_s)\alpha_{\lambda}^2\big)\alpha_{\rho}-3(m_q+m_s)\alpha_{\lambda}\alpha_{\rho}^2\bigg)}{m_qm_s(4\alpha^2+\alpha_{\lambda}^2+3\alpha_{\rho}^2)^{5/2}}\right]\\
            &\times\mathcal{P}(m_{n},m_{\Lambda(1670)}),\\
            &\mathcal{M}_{\text{Pole,B3,PV}}^{-\frac12,-\frac12}[\Sigma^0(1750)]\\&=\left[i\frac{2(2m_q+m_s)^2\alpha_{\lambda}\alpha_{\rho}(\alpha_{\lambda}+3\alpha_{\rho})\omega_0+\boldsymbol{k}^2\big(2m_q\alpha_{\lambda}+m_s(\alpha_{\lambda}+\alpha_{\rho})\big)(4m_q+2m_s+\omega_0)}{96\pi^{3/2}\alpha_{\lambda}\sqrt{\omega_0}f_{\pi}m_q(2m_q+m_s)^2}\mathcal{F}_{\pi}^{\prime}(\boldsymbol{k})\right]\\
            &\times\left[-i\frac{16\sqrt{2}G_FV_{ud}V_{us}}{\pi^{3/2}}\frac{\alpha^3\alpha_{\lambda}^{5/2}\alpha_{\rho}^{3/2}\big(2(2m_q+m_s)\alpha^2+3(m_q+m_s)\alpha_{\rho}^2\big)}{m_qm_s(4\alpha^2+\alpha_{\lambda}^2+3\alpha_{\rho}^2)^{5/2}}\right]\mathcal{P}(m_{n},m_{\Sigma^0(1750)}),\\
            &\mathcal{M}_{\text{Pole,B4,PV}}^{-\frac12,-\frac12}[\Lambda_8(1800)]\\&=\left[-i\frac{2(2m_q+m_s)^2\alpha_{\lambda}\alpha_{\rho}(\alpha_{\lambda}+3\alpha_{\rho})\omega_0+\boldsymbol{k}^2\big(2m_q\alpha_{\lambda}+m_s(\alpha_{\lambda}+\alpha_{\rho})\big)(4m_q+2m_s+\omega_0)}{32\sqrt{3}\pi^{3/2}\alpha_{\lambda}\sqrt{\omega_0}f_{\pi}m_q(2m_q+m_s)^2}\mathcal{F}_{\pi}^{\prime}(\boldsymbol{k})\right]\\
            &\times\left[i\frac{16\sqrt{6}G_FV_{ud}V_{us}}{\pi^{3/2}}\frac{\alpha^3\alpha_{\lambda}^{3/2}\alpha_{\rho}^{5/2}\big(2m_s\alpha^2+(m_q+m_s)\alpha_{\lambda}^2\big)}{m_qm_s(4\alpha^2+\alpha_{\lambda}^2+3\alpha_{\rho}^2)^{5/2}}\right]\mathcal{P}(m_{n},m_{\Lambda(1800)}),\\
            &\mathcal{M}_{\text{Pole,B5,PV}}^{-\frac12,-\frac12}[\Lambda_1(1405)]\\&=\left[i\frac{\sqrt{3}\bigg(2(2m_q+m_s)^2\alpha_{\lambda}\alpha_{\rho}(\alpha_{\lambda}+3\alpha_{\rho})\omega_0+\boldsymbol{k}^2m_q\big(2m_q\alpha_{\lambda}+m_s(\alpha_{\lambda}+\alpha_{\rho})\big)(4m_q+2m_s+\omega_0)\bigg)}{64\pi^{3/2}\alpha_{\lambda}\sqrt{\omega_0}f_{\pi}m_q(2m_q+m_s)^2}\mathcal{F}_{\pi}^{\prime}(\boldsymbol{k})\right]\\
            &\times\left[-i\frac{4\sqrt{6}G_FV_{ud}V_{us}}{\pi^{3/2}}\frac{\alpha^3(\alpha_{\lambda}\alpha_{\rho})^{3/2}\bigg(4m_q\alpha^2\alpha_{\lambda}+2m_s\alpha^2(\alpha_{\lambda}+\alpha_{\rho})+(m_q+m_s)\alpha_{\lambda}\alpha_{\rho}(\alpha_{\lambda}+3\alpha_{\rho})\bigg)}{m_qm_s(4\alpha^2+\alpha_{\lambda}^2+3\alpha_{\rho}^2)^{5/2}}\right]\\
            &\times\mathcal{P}(m_{n},m_{\Lambda_1(1405)}).
        \end{split}
    \end{align}
\end{itemize}

\item $\Sigma^+\to p\pi^0$
\begin{itemize}
    \item CS process
    \begin{align}
        \begin{split}
            \mathcal{M}_{\text{CS-1,PC}}^{-\frac12,-\frac12}(\Sigma^+\to p\pi^0)&=\frac{1}{\sqrt{2}}\mathcal{M}_{\text{CS-1,PC}}^{-\frac12,-\frac12}(\Sigma^-\to n\pi^-),\\
            \mathcal{M}_{\text{CS-1,PV}}^{-\frac12,-\frac12}(\Sigma^+\to p\pi^0)&=\frac{1}{\sqrt{2}}\mathcal{M}_{\text{CS-1,PV}}^{-\frac12,-\frac12}(\Sigma^-\to n\pi^-).
        \end{split}
    \end{align}
    \begin{align}
        \begin{split}
            \mathcal{M}_{\text{CS-2,PC}}^{-\frac12,-\frac12}(\Sigma^+\to p\pi^0)&=-\frac{2G_FV_{ud}V_{us}}{27\pi^{9/4}}\frac{\boldsymbol{k}(3m_s\alpha^2-mq\alpha_{\lambda}^2+2m_s\alpha_{\lambda}^2)}{m_qm_s(\alpha^2+\alpha_{\lambda}^2)^{5/2}}\left(\frac{\alpha^2\alpha_{\rho}\alpha_{\lambda}R}{\alpha^2+\alpha_{\rho}^2}\right)^{3/2}\\
            &\times\text{exp}\left[-\frac{\boldsymbol{k}^2}{3(\alpha^2+\alpha_{\lambda}^2)}\right]=\frac{1}{3\sqrt{2}}\mathcal{M}_{\text{DPE,PC}}^{-\frac12,-\frac12}(\Sigma^-\to n\pi^-),\\
            \mathcal{M}_{\text{CS-2,PV}}^{-\frac12,-\frac12}(\Sigma^+\to p\pi^0)&=-\frac{2G_FV_{ud}V_{us}}{3\pi^{9/4}}\left(\frac{\alpha^2\alpha_{\rho}\alpha_{\lambda}R}{(\alpha^2+\alpha_{\lambda}^2)(\alpha^2+\alpha_{\rho}^2)}\right)^{3/2}\text{exp}\left[-\frac{\boldsymbol{k}^2}{3(\alpha^2+\alpha_{\lambda}^2)}\right]\\
            &=\frac{1}{3\sqrt{2}}\mathcal{M}_{\text{DPE,PV}}^{-\frac12,-\frac12}(\Sigma^-\to n\pi^-).
        \end{split}
    \end{align}
    \item Pole terms
    \begin{align}
        \begin{split}
            &\mathcal{M}_{\text{Pole,A1,PC}}^{-\frac12,-\frac12}[p(938)]\\&=\left[\frac{24\sqrt{2}G_FV_{ud}V_{us}}{\pi^{3/2}}\left(\frac{\alpha^2\alpha_{\lambda}\alpha_{\rho}}{4\alpha^2+\alpha_{\lambda}^2+3\alpha_{\rho}^2}\right)^{3/2}\right]\left[-\frac{5\boldsymbol{k}(6m_q+\omega_0)}{72\sqrt{2}\pi^{3/2}\sqrt{\omega_0}f_{\pi}m_q}\mathcal{F}_{\pi}(\boldsymbol{k})\right]\mathcal{P}(m_{\Sigma^+},m_p),\\
            &\mathcal{M}_{\text{Pole,A2,PC}}^{-\frac12,-\frac12}[N(1440)]\\&=\left[\frac{24\sqrt{6}G_FV_{ud}V_{us}}{\pi^{3/2}}\frac{(\alpha^2\alpha_{\lambda}\alpha_{\rho})^{3/2}(\alpha_{\lambda}^2+3\alpha_{\rho}^2)}{(4\alpha^2+\alpha_{\lambda}^2+3\alpha_{\rho}^2)^{5/2}}\right]\left[-\frac{5\boldsymbol{k}\big(-12\alpha^2\omega_0+\boldsymbol{k}^2(6m_q+\omega_0)\big)}{432\sqrt{6}\pi^{3/2}\alpha^2\sqrt{\omega_0}f_{\pi}m_q}\mathcal{F}_{\pi}(\boldsymbol{k})\right]\\
            &\times\mathcal{P}(m_{\Sigma^+},m_{N(1440)}),\\
            &\mathcal{M}_{\text{Pole,A3,PC}}^{-\frac12,-\frac12}[N(1710)]\\&=\left[\frac{96\sqrt{3}G_FV_{ud}V_{us}}{\pi^{3/2}}\frac{\alpha^5(\alpha_{\lambda}\alpha_{\rho})^{3/2}}{(4\alpha^2+\alpha_{\lambda}^2+3\alpha_{\rho}^2)^{5/2}}\right]\left[\frac{\boldsymbol{k}\big(-12\alpha^2\omega_0+\boldsymbol{k}^2(6m_q+\omega_0)\big)}{432\sqrt{3}\pi^{3/2}\alpha^2\sqrt{\omega_0}f_{\pi}m_q}\mathcal{F}_{\pi}(\boldsymbol{k})\right]\times\mathcal{P}(m_{\Sigma^+},m_{N(1710)}).
        \end{split}
    \end{align}
    \begin{align}
        \begin{split}
            &\mathcal{M}_{\text{Pole,A1,PV}}^{-\frac12,-\frac12}[N|70,^28\rangle]\\&=\left[-i\frac{16G_FV_{ud}V_{us}}{\pi^{3/2}}\frac{\alpha^4\bigg((m_q+2m_s)\alpha_{\lambda}^{7/2}\alpha_{\rho}^{3/2}+m_s(6\alpha^2\alpha_{\lambda}^{3/2}\alpha_{\rho}^{3/2}+3\alpha_{\lambda}^{3/2}\alpha_{\rho}^{7/2})\bigg)}{m_qm_s(4\alpha^2+\alpha_{\lambda}^2+3\alpha_{\rho}^2)^{5/2}}\right]\\
            &\times\left[-i\frac{-18\alpha^2\omega_0+\boldsymbol{k}^2(6m_q+\omega_0)}{108\pi^{3/2}\alpha\sqrt{\omega_0}f_{\pi}m_q}\mathcal{F}_{\pi}(\boldsymbol{k})\right]\mathcal{P}(m_{\Sigma^+},m_{N|70,^28\rangle}),\\
            &\mathcal{M}_{\text{Pole,A2,PV}}^{-\frac12,-\frac12}[N|70,^48\rangle]\\&=\left[i\frac{48G_FV_{ud}V_{us}}{\pi^{3/2}}\frac{\alpha^4(\alpha_{\lambda}\alpha_{\rho})^{3/2}}{m_q(4\alpha^2+\alpha_{\lambda}^2+3\alpha_{\rho}^2)^{3/2}}\right]\left[-i\frac{-18\alpha^2\omega_0+\boldsymbol{k}^2(6m_q+\omega_0)}{216\pi^{3/2}\alpha\sqrt{\omega_0}f_{\pi}m_q}\mathcal{F}_{\pi}(\boldsymbol{k})\right]\mathcal{P}(m_{\Sigma^+},m_{N|70,^48\rangle}).
        \end{split}
    \end{align}
\begin{align}
        \begin{split}
            \mathcal{M}_{\text{Pole,B1,PC}}^{-\frac12,-\frac12}[\Sigma^+(1193)](\Sigma^+\to p\pi^0)&=\sqrt{2}\mathcal{M}_{\text{Pole,B1,PC}}^{-\frac12,-\frac12}[\Sigma^0(1193)](\Sigma^-\to n\pi^-),\\
            \mathcal{M}_{\text{Pole,B2,PC}}^{-\frac12,-\frac12}[\Sigma^+(1660)](\Sigma^+\to p\pi^0)&=\sqrt{2}\mathcal{M}_{\text{Pole,B3,PC}}^{-\frac12,-\frac12}[\Sigma^0(1660)](\Sigma^-\to n\pi^-),\\
            \mathcal{M}_{\text{Pole,B3,PC}}^{-\frac12,-\frac12}[\Sigma^+(1880)](\Sigma^+\to p\pi^0)&=\sqrt{2}\mathcal{M}_{\text{Pole,B5,PC}}^{-\frac12,-\frac12}[\Sigma^0(1880)](\Sigma^-\to n\pi^-),\\
            \mathcal{M}_{\text{Pole,B1,PV}}^{-\frac12,-\frac12}[\Sigma^+(1620)](\Sigma^+\to p\pi^0)&=\sqrt{2}\mathcal{M}_{\text{Pole,B1,PV}}^{-\frac12,-\frac12}[\Sigma^0(1620)](\Sigma^-\to n\pi^-),\\
            \mathcal{M}_{\text{Pole,B2,PV}}^{-\frac12,-\frac12}[\Sigma^+(1750)](\Sigma^+\to p\pi^0)&=\sqrt{2}\mathcal{M}_{\text{Pole,B3,PV}}^{-\frac12,-\frac12}[\Sigma^0(1750)](\Sigma^-\to n\pi^-).
        \end{split}
    \end{align}
\end{itemize}

\item $\Sigma^+\to n\pi^+$
    
\begin{itemize}
    \item Pole terms
    \begin{align}
        \begin{split}
            \mathcal{M}_{\text{Pole,A1,PC}}^{-\frac12,-\frac12}[p(938)](\Sigma^+\to n\pi^+)&=-\sqrt{2}\mathcal{M}_{\text{Pole,A1,PC}}^{-\frac12,-\frac12}[p(938)](\Sigma^+\to p\pi^0),\\
            \mathcal{M}_{\text{Pole,A2,PC}}^{-\frac12,-\frac12}[N(1440)](\Sigma^+\to n\pi^+)&=-\sqrt{2}\mathcal{M}_{\text{Pole,A2,PC}}^{-\frac12,-\frac12}[N(1440)](\Sigma^+\to p\pi^0),\\
            \mathcal{M}_{\text{Pole,A3,PC}}^{-\frac12,-\frac12}[N(1710)](\Sigma^+\to n\pi^+)&=-\sqrt{2}\mathcal{M}_{\text{Pole,A3,PC}}^{-\frac12,-\frac12}[N(1710)](\Sigma^+\to p\pi^0),\\
            \mathcal{M}_{\text{Pole,A1,PV}}^{-\frac12,-\frac12}[N|70,^28\rangle](\Sigma^+\to n\pi^+)&=-\sqrt{2}\mathcal{M}_{\text{Pole,A1,PV}}^{-\frac12,-\frac12}[N|70,^28\rangle](\Sigma^+\to p\pi^0),\\
            \mathcal{M}_{\text{Pole,A2,PV}}^{-\frac12,-\frac12}[N|70,^48\rangle](\Sigma^+\to n\pi^+)&=-\sqrt{2}\mathcal{M}_{\text{Pole,A2,PV}}^{-\frac12,-\frac12}[N|70,^48\rangle](\Sigma^+\to p\pi^0).
        \end{split}
    \end{align}
    \begin{align}
        \begin{split}
            \mathcal{M}_{\text{Pole,B1,PC}}^{-\frac12,-\frac12}[\Sigma^0(1193)](\Sigma^+\to n\pi^+)&=-\mathcal{M}_{\text{Pole,B1,PC}}^{-\frac12,-\frac12}[\Sigma^0(1193)](\Sigma^-\to n\pi^-)\\
            &=-\frac{1}{\sqrt{2}}\mathcal{M}_{\text{Pole,B1,PC}}^{-\frac12,-\frac12}[\Sigma^+(1193)](\Sigma^+\to p\pi^0),\\
            \mathcal{M}_{\text{Pole,B2,PC}}^{-\frac12,-\frac12}[\Lambda_8(1116)](\Sigma^+\to n\pi^+)&=+\mathcal{M}_{\text{Pole,B2,PC}}^{-\frac12,-\frac12}[\Lambda_8(1116)](\Sigma^-\to n\pi^-),\\
            \mathcal{M}_{\text{Pole,B3,PC}}^{-\frac12,-\frac12}[\Sigma^0(1660)](\Sigma^+\to n\pi^+)&=-\mathcal{M}_{\text{Pole,B3,PC}}^{-\frac12,-\frac12}[\Sigma^0(1660)](\Sigma^-\to n\pi^-)\\
            &=-\frac{1}{\sqrt{2}}\mathcal{M}_{\text{Pole,B2,PC}}^{-\frac12,-\frac12}[\Sigma^+(1660)](\Sigma^+\to p\pi^0),\\
            \mathcal{M}_{\text{Pole,B4,PC}}^{-\frac12,-\frac12}[\Lambda_8(1600)](\Sigma^+\to n\pi^+)&=+\mathcal{M}_{\text{Pole,B4,PC}}^{-\frac12,-\frac12}[\Lambda_8(1600)](\Sigma^-\to n\pi^-),\\
            \mathcal{M}_{\text{Pole,B5,PC}}^{-\frac12,-\frac12}[\Sigma^0(1880)](\Sigma^+\to n\pi^+)&=-\mathcal{M}_{\text{Pole,B5,PC}}^{-\frac12,-\frac12}[\Sigma^0(1880)](\Sigma^-\to n\pi^-)\\
            &=-\frac{1}{\sqrt{2}}\mathcal{M}_{\text{Pole,B3,PC}}^{-\frac12,-\frac12}[\Sigma^+(1880)](\Sigma^+\to p\pi^0),\\
            \mathcal{M}_{\text{Pole,B6,PC}}^{-\frac12,-\frac12}[\Lambda_8(1810)](\Sigma^+\to n\pi^+)&=+\mathcal{M}_{\text{Pole,B6,PC}}^{-\frac12,-\frac12}[\Lambda_8(1810)](\Sigma^-\to n\pi^-),\\            \mathcal{M}_{\text{Pole,B1,PV}}^{-\frac12,-\frac12}[\Sigma^0(1620)](\Sigma^+\to n\pi^+)&=-\mathcal{M}_{\text{Pole,B1,PV}}^{-\frac12,-\frac12}[\Sigma^0(1620)](\Sigma^-\to n\pi^-)\\
            &=-\frac{1}{\sqrt{2}}\mathcal{M}_{\text{Pole,B1,PV}}^{-\frac12,-\frac12}[\Sigma^+(1620)](\Sigma^+\to p\pi^0),\\
            \mathcal{M}_{\text{Pole,B2,PV}}^{-\frac12,-\frac12}[\Lambda_8(1670)](\Sigma^+\to n\pi^+)&=+\mathcal{M}_{\text{Pole,B2,PV}}^{-\frac12,-\frac12}[\Lambda_8(1670)](\Sigma^-\to n\pi^-),\\
            \mathcal{M}_{\text{Pole,B3,PV}}^{-\frac12,-\frac12}[\Sigma^0(1750)](\Sigma^+\to n\pi^+)&=-\mathcal{M}_{\text{Pole,B3,PV}}^{-\frac12,-\frac12}[\Sigma^0(1750)](\Sigma^-\to n\pi^-)\\
            &=-\frac{1}{\sqrt{2}}\mathcal{M}_{\text{Pole,B2,PV}}^{-\frac12,-\frac12}[\Sigma^+(1750)](\Sigma^+\to p\pi^0),\\
            \mathcal{M}_{\text{Pole,B4,PV}}^{-\frac12,-\frac12}[\Lambda_8(1800)](\Sigma^+\to n\pi^+)&=+\mathcal{M}_{\text{Pole,B4,PV}}^{-\frac12,-\frac12}[\Lambda_8(1800)](\Sigma^-\to n\pi^-),\\
            \mathcal{M}_{\text{Pole,B5,PV}}^{-\frac12,-\frac12}[\Lambda_1(1405)](\Sigma^+\to n\pi^+)&=+\mathcal{M}_{\text{Pole,B5,PV}}^{-\frac12,-\frac12}[\Lambda_1(1405)](\Sigma^-\to n\pi^-).
        \end{split}
    \end{align}
\end{itemize}

\end{itemize}

\section{The helicity amplitudes of loop diagrams}
\label{app:loop amp}
In this Appendix we provide the method for extracting the helicity amplitude of loop diagrams. 

In this work, the metric in four-dimensional Minkowski space is
\begin{align}
    g_{\mu\nu}=g^{\mu\nu}=\begin{pmatrix}
        1&0&0&0\\
        0&-1&0&0\\
        0&0&-1&0\\
        0&0&0&-1
    \end{pmatrix},
\end{align}
and the Einstein summation convention is used. The Dirac matrices $\gamma^{\mu}$ satisfy
\begin{align}
    \begin{split}
        \{\gamma^{\mu},\gamma^{\nu}\}=2g^{\mu\nu}.
    \end{split}
\end{align}
We choose the standard representation for the Dirac matrices:
\begin{align}
    \gamma^0=\begin{pmatrix}
        \bs{1}&0\\
        0&-\bs{1}
    \end{pmatrix},\ \ \ \gamma^1=\begin{pmatrix}
        0&\bs{\sigma}^1\\
        -\bs{\sigma}^1&0
    \end{pmatrix},\ \ \ \gamma^2=\begin{pmatrix}
        0&\bs{\sigma}^2\\
        -\bs{\sigma}^2&0
    \end{pmatrix},\ \ \ \gamma^3=\begin{pmatrix}
        0&\bs{\sigma}^3\\
        -\bs{\sigma}^3&0
    \end{pmatrix}
\end{align}
where $\bs{\sigma}^i$ ($i=1,2,3$) are the Pauli matrices, and 
\begin{align}
    \gamma_5\equiv i\gamma^0\gamma^1\gamma^2\gamma^3=\begin{pmatrix}
        0&\bs{1}\\
        \bs{1}&0
    \end{pmatrix}.
\end{align}

Dirac equations are
\begin{align}
    (\slashed{p}-m)u(\boldsymbol{p},s)=0,\ \ \ (\slashed{p}+m)v(\boldsymbol{p},s)=0,
\end{align}
where $u(\bs{p},s)$ and $v(\bs{p},s)$ are the  positive energy solution and Negative energy solution, respectively with its three-vector momentum $\boldsymbol{p}$ and spin $s$. Since the spin is not a good quantum number, its projection in the direction of three-momentum which is defined as helicity $\lambda$ is a conserved quantity. The state $\omega^{(\lambda})$ labeled by helicity satisfies the eigenvalue equation:
\begin{align}
    \begin{split}
        \frac12\frac{\boldsymbol{\sigma}\cdot\boldsymbol{p}}{|\boldsymbol{p}|}\omega^{(\lambda)}=\lambda\omega^{(\lambda)},
    \end{split}
\end{align}
and the forms of these two helicity eigenstates $\omega^{(\lambda)}$ is chosen to be
\begin{align}
    \begin{split}
    \omega^{(\lambda=+1/2)}=\begin{pmatrix}
            \text{e}^{-i\phi/2}\text{cos}\frac{\theta}{2}\\
            \text{e}^{i\phi/2}\text{sin}\frac{\theta}{2}
        \end{pmatrix},\ \ \ \ \ \omega^{(\lambda=-1/2)}=\begin{pmatrix}
            -\text{e}^{-i\phi/2}\text{sin}\frac{\theta}{2}\\
            \text{e}^{i\phi/2}\text{cos}\frac{\theta}{2}
        \end{pmatrix},
    \end{split}
\end{align}
where $\theta$ and $\phi$ are the polar and azimuthal angles of the momentum $\bs{p}$,respectively. It is obvious that $\omega^{(\lambda)}$ satisfies the normalization condition
\begin{align}
    \begin{split}
        \omega^{(\lambda)\dagger}\omega^{(\lambda^{\prime})}=\delta^{\lambda\lambda^{\prime}}.
    \end{split}
\end{align}
The Dirac spinors labelled by the three-momentum and helicity are
\begin{align}
    \begin{split}
    u(\bs{p},\lambda)=\sqrt{E+m}\begin{pmatrix}
        \omega^{(\lambda)}\\
        \frac{\boldsymbol{\sigma}\cdot\boldsymbol{p}}{E+m}\omega^{(\lambda)}
    \end{pmatrix},\ \ \ 
    v(\bs{p},\lambda)=\sqrt{E+m}\begin{pmatrix}
        \frac{\boldsymbol{\sigma}\cdot\boldsymbol{p}}{E+m}\omega^{(\lambda)}\\
        \omega^{(\lambda)}
    \end{pmatrix},
    \end{split}     
\end{align}
satisfying the normalization conditions
\begin{align}
    \begin{split}
        \bar{u}(\bs{p},\lambda)u(\bs{p},\lambda^{\prime})=\delta_{\lambda\lambda^{\prime}},\ \ \ \bar{v}(\bs{p},\lambda)v(\bs{p},\lambda^{\prime})=-\delta_{\lambda\lambda^{\prime}},
    \end{split}
\end{align}
and the completeness conditions
\begin{align}
    \begin{split}
        \sum_{\lambda}{u}(\bs{p},\lambda)\bar{u}(\bs{p},\lambda)=\slashed{p}+m,\ \ \ \sum_{\lambda}{v}(\bs{p},\lambda)\bar{v}(\bs{p},\lambda)=\slashed{p}-m.
    \end{split}
\end{align}

\end{appendix}

\bibliographystyle{unsrt}
\bibliography{ref.bib}

\end{document}